\documentclass[twoside,12pt]{article}
\usepackage{epsfig}

\def\p{\vec p}

\def\sax{\mbox{SAX~J1808.4-3658}}
\def\sgr0526{\mbox{SGR~0526-66}}
\def\sgreighteenosix{\mbox{SGR~1806-20}}
\def\sgrnineteenoo{\mbox{SGR~1900+14}}
\def\groj1744{\mbox{GRO~J1744-28}}
\def\rxj1856{\mbox{RX~J1856.5-3754}}
\def\0720{\mbox{RX~J0720.4-3125}}
\def\eu1728{\mbox{4U~1728-34}}
\def\3c58{\mbox{3C58}}
\def\1e1207{\mbox{1E~1207.4-5209}}
\def\p0943{\mbox{PSR~0943+10}}
\def\psrotwoofive{\mbox{PSR~0205+6449}}
\def\exo{\mbox{EXO~0748-1676}}
\def\ls{\lower0.5ex\hbox{$\; \buildrel < \over \sim \;$}}
\def\gs{\lower0.5ex\hbox{$\; \buildrel > \over \sim \;$}}
\def\dmmax{\mbox{${\dot{M}}_{\rm max}$}}
\def\dmmin{\mbox{${\dot{M}}_{\rm min}$}}
\def\fmax{\mbox{$F_{\rm max}$}}
\def\fmin{\mbox{$F_{\rm min}$}}
\def\slash#1{#1\!\!\!/}
\def\etal{{\it{et al.\/}}}

\newcommand{\be}{\begin{equation}}
\newcommand{\ee}{\end{equation}}
\newcommand{\bea}{\begin{eqnarray}}
\newcommand{\eea}{\end{eqnarray}}

\newcommand{\el}{{\rm el}}

\newcommand{\kFB}{{k_{F_B}}}

\newcommand{\msun}{{M_{\odot}}}
\newcommand{\eos}{equation of state~}
\newcommand{\eoss}{equations of state~}
\newcommand{\eossp}{equations of state}
\newcommand{\eosp}{equation of state}
\newcommand{\Eos}{Equation of state~}

\newcommand{\okgr}{{\Omega_{\rm K}}}

\newcommand{\fm}{{\rm fm}}

\newcommand{\fmmt}{{\rm fm}^{-3}}
\newcommand{\mevt}{{\rm MeV/fm}^3}
\newcommand{\gcmt}{{\rm g/cm}^3}

\newcommand{\eut}{{\rm UV}_{14}\!+\!{\rm TNI}}
\newcommand{\euu}{{\rm UV}_{14}\!+\!{\rm UVII}}
\newcommand{\eau}{{\rm AV}_{14}\!+\!{\rm UVII}}
\newcommand{\eaix}{{\rm A}_{18}\!+\!\delta v\!+\!{\rm UIX}}

\newcommand{\KBt}{{\rm G}^{\rm K300}_{\rm B180}}
\newcommand{\KM}{{\rm G}^{\rm K240}_{\rm M78}}

\newcommand{\bag}{B^{1/4}}
\newcommand{\edrip}{{\epsilon_{\rm drip}}}

\newcommand{\brho}{\mbox{\mbox{\boldmath$\rho$}}}

\newcommand{\btau}{\mbox{\mbox{\boldmath$\tau$}}}
\newcommand{\bpi}{\mbox{\mbox{\boldmath$\pi$}}}

\newcommand{\bcdot}{\mbox{\mbox{\boldmath$\cdot$}}}
\newcommand{\bfG}{{\bf G}}

\newcommand{\psiB}{{\psi_B}}

\newcommand{\bpsiB}{{\bar\psi_B}}
\newcommand{\kFBt}{{k_{F_B}^3}}
\newcommand{\kFL}{{k_{F_L}}}

\newcommand{\kFft}{{k_{F_f}^3}}
\newcommand{\kFf}{{k_{F_f}}}
\newcommand{\pFf}{{p_{F_f}}}
\newcommand{\epswd}{{\epsilon_{\rm wd}}}
\newcommand{\ecrusti}{{\epsilon_{\rm crust}}}
\newcommand{\cm}{{\rm cm}}
\newcommand{\cmmt}{{\rm cm}^{-3}}
\newcommand{\rms}{{\rm s}}
\newcommand{\ergs}{{\rm erg/s}}
\newcommand{\mev}{{\rm MeV}}
\newcommand{\tev}{{\rm TeV}}
\newcommand{\gev}{{\rm GeV}}
\newcommand{\mH}{{m_{\rm H}}}
\newcommand{\pkgr}{{P_{\,\rm K}}}
\newcommand{\const}{{\rm const}}
\newcommand{\rcrust}{{R_{\rm crust}}}
\newcommand{\icrust}{{I_{\rm crust}}}
\newcommand{\itotal}{{I_{\rm total}}}
\newcommand{\nth}{{$n$th~}}
\newcommand{\secm}{{\rm s}^{-1}}

\newcommand{\dfq}{{{d^4q}\over{(2\pi)^4}}}
\newcommand{\kFo}{k_{F_0}}
\newcommand{\kFosq}{{k_{F_0}^2}}

\begin{document}

\title{\bf Strange Quark Matter and Compact Stars}

\author{F. Weber \\ \\ Department of Physics \\ San Diego State
University \\ 5500 Campanile Drive, San Diego\\ 
California 92182, USA}

\maketitle

\vskip 2truecm

\begin{abstract}
Astrophysicists distinguish between three different types of compact
stars. These are white dwarfs, neutron stars, and black holes. The
former contain matter in one of the densest forms found in the
Universe which, together with the unprecedented progress in
observational astronomy, make such stars superb astrophysical
laboratories for a broad range of most striking physical phenomena.
These range from nuclear processes on the stellar surface to processes
in electron degenerate matter at subnuclear densities to boson
condensates and the existence of new states of baryonic matter--like
color superconducting quark matter--at supernuclear densities. More
than that, according to the strange matter hypothesis strange quark
matter could be more stable than nuclear matter, in which case neutron
stars should be largely composed of pure quark matter possibly
enveloped in thin nuclear crusts. Another remarkable implication of
the hypothesis is the possible existence of a new class of white
dwarfs.  This article aims at giving an overview of all these striking
physical possibilities, with an emphasis on the astrophysical
phenomenology of strange quark matter. Possible observational
signatures associated with the theoretically proposed states of matter
inside compact stars are discussed as well. They will provide most
valuable information about the phase diagram of superdense nuclear
matter at high baryon number density but low temperature, which is not
accessible to relativistic heavy ion collision experiments.
\end{abstract}



\eject \tableofcontents

\clearpage

\goodbreak
\section{Introduction}\label{sec:intro}

It is often stressed that there has never been a more exciting time in
the overlapping areas of nuclear physics, particle physics and
relativistic astrophysics than today.  This comes at a time where new
orbiting observatories such as the Hubble Space Telescope, Rossi X-ray
Timing Explorer (RXTE), Chandra X-ray satellite, and the X-ray Multi
Mirror Mission (XMM) have extended our vision tremendously, allowing
us to see vistas with an unprecedented clarity and angular resolution
that previously were only imagined, enabling astrophysicists for the
first time ever to perform detailed studies of large samples of
galactic and extragalactic objects. On the Earth, radio telescopes
(e.g., Arecibo, Green Bank, Parkes, VLA) and instruments using
adaptive optics and other revolutionary techniques have exceeded
previous expectations of what can be accomplished from the ground. The
gravitational wave detectors LIGO, LISA, VIRGO, and Geo-600 are
opening up a window for the detection of gravitational waves emitted
from compact stellar objects such as neutron stars and black holes.

\begin{figure}[tb] 
\begin{center}
\epsfig{figure=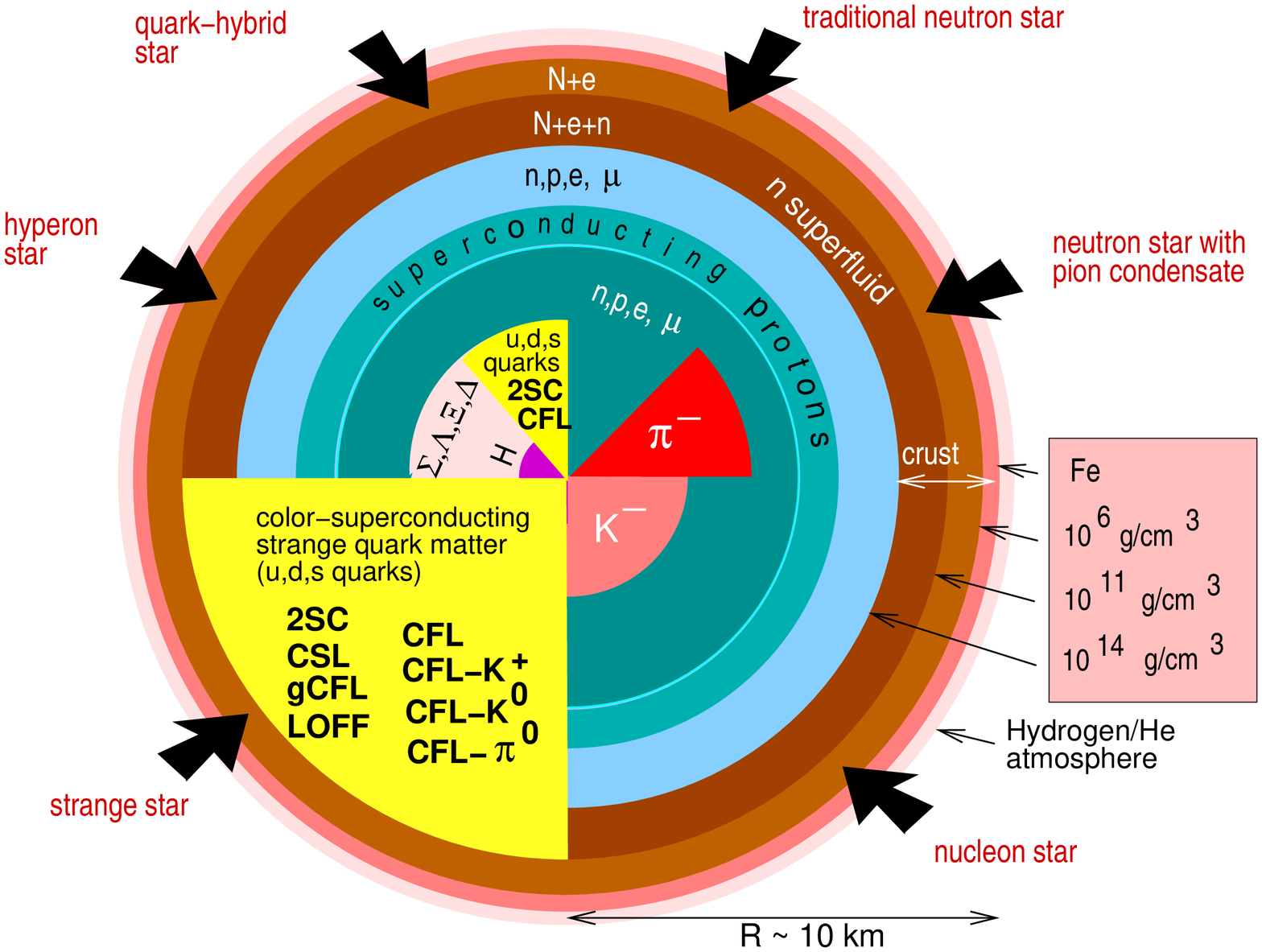,width=12cm,angle=00}
\begin{minipage}[t]{16.5 cm}
\caption{Competing structures and novel phases of subatomic matter
predicted by theory to make their appearance in the cores ($R\ls
8$~km) of neutron stars \cite{weber99:book}.}
\label{fig:cross}
\end{minipage}
\end{center}
\end{figure}

Neutron stars are dense, neutron-packed remnants of massive stars that
blew apart in supernova explosions. They are typically about 20
kilometers across and spin rapidly, often making several hundred
rotations per second.  Many neutron stars form radio pulsars, emitting
radio waves that appear from the Earth to pulse on and off like a
lighthouse beacon as the star rotates at very high speeds. Neutron
stars in X-ray binaries accrete material from a companion star and
flare to life with a burst of X-rays. Measurements of radio pulsars
and neutron stars in X-ray binaries comprise most of the neutron star
observations. Improved data on isolated neutron stars (e.g. \rxj1856,
\psrotwoofive) are now becoming available, and future investigations
at gravitational wave observatories like LIGO and VIRGO will focus on
neutron stars as major potential sources of gravitational
waves. Depending on star mass and rotational frequency, gravity
compresses the matter in the core regions of pulsars up to more than
ten times the density of ordinary atomic nuclei, thus providing a
high-pressure environment in which numerous subatomic particle
processes compete with each other. The most spectacular ones stretch
from the generation of hyperons and baryon resonances ($\Sigma,
\Lambda, \Xi, \Delta$) to quark ($u, d, s$) deconfinement to the
formation of boson condensates ($\pi^-$, $K^-$,
H-matter) \cite{glen97:book,weber99:book,heiselberg00:a,lattimer01:a}. There
are theoretical suggestions of even more exotic processes inside
neutron stars, such as the formation of absolutely stable strange
quark matter \cite{bodmer71:a,witten84:a,terazawa89:a}, a configuration
of matter more stable than the most stable atomic nucleus,
$^{62}$Ni.\footnote{It is common practice to compare the energy of
strange quark matter to $^{56}$Fe.  The energy per particle of
$^{56}$Fe, however, comes in only third after $^{62}$Ni and
$^{58}$Fe.}  In the latter event, neutron stars would be largely
composed of strange quark
matter \cite{alcock86:a,alcock88:a,madsen98:b} possibly enveloped in
thin nuclear crusts \cite{glen92:crust} whose density is less than
neutron drip. Another striking implication of the hypothesis is the
possible existence of a new class of white dwarfs
 \cite{weber93:b,glen94:a}. An overview of the conjectured composition
of neutron stars is shown in Fig.~\ref{fig:cross}.
Because of their complex interior structures, the very name neutron
star is almost certainly a misnomer. Instead these objects should be
coined nucleon stars, since relatively isospin symmetric nuclear
matter--in equilibrium with condensed $K^-$ mesons--may prevail in
their interiors \cite{brown95:a}, hyperon stars if hyperons ($\Sigma,
\Lambda, \Xi$, possibly in equilibrium with the $\Delta$ resonance)
become populated in addition to the nucleons \cite{glen85:b},
quark-hybrid stars if the highly compressed matter in the centers of
neutron stars should transform into $u, d, s$ quark matter
\cite{glen91:b}, or strange stars if strange quark matter should be
more stable than nuclear matter.  The idea that quark matter may exist
in the cores of neutron stars is not new but has already been
suggested by several authors
\cite{ivanenko65:a,itoh70:a,fritzsch73:a,baym76:a,keister76:a,chap77:a+b,fech78:a}.
For many years it has been thought that the deconfined phase of quarks
and hadrons is strictly excluded from neutron stars. Theoretical
studies, however, have shown that this was due to seemingly innocuous
idealizations \cite{glen91:pt,glen01:b}. Thus neutron stars may very
well contain quark matter in their cores, which ought to be in a color
superconducting state
\cite{rajagopal01:a,alford01:a,alford98:a,rapp98+99:a}. This
fascinating possibility has renewed tremendous interest in the physics
and astrophysics of quark matter.

Of course, at present one does not know from experiment at what
density the expected phase transition to quark matter occurs. Neither
do lattice Quantum Chromodynamical simulations provide a conclusive
guide yet.  From simple geometrical considerations it follows that,
for a characteristic nucleon radius of $r_N\sim 1$~fm, nuclei begin to
touch each other at densities of $\sim (4\pi r^3_N/3)^{-1} \simeq
0.24~\fmmt$, which is less than twice the baryon number density of
ordinary nuclear matter, $\rho_0 = 0.16~\fmmt$ (energy density
$\epsilon_0 = 140~\mevt$). Depending on rotational frequency and
stellar mass, such densities are easily surpassed in the cores of
neutron stars so that gravity may have broken up the neutrons ($n$)
and protons ($p$) in the centers of neutron stars into their
constituents. Moreover, since the mass of the strange quark ($s$) is
rather small, probably less than 100~MeV as indicated by the latest
lattice results  \cite{aubin04:a}, high-energetic up ($u$) and down
($d$) quarks may readily transform to strange quarks at about the same
density at which unconfined up and down quarks appear.

The phase diagram of quark matter, expected to be in a color
superconducting phase, is very complex
\cite{rajagopal01:a,alford01:a}. At asymptotic densities the ground
state of QCD with a vanishing strange quark mass is the color-flavor
locked (CFL) phase. This phase is electrically charge neutral without
any need for electrons for a significant range of chemical potentials
and strange quark masses \cite{rajagopal01:b}.  (Technically, there
are no electrons only at zero temperature. At finite temperature the
electron population is exponentially ($\exp (-\Delta/T)$) suppressed,
where $\Delta$ denotes the superconducting gap.)  If the strange quark
mass is heavy enough to be ignored, then up and down quarks may pair
in the two-flavor superconducting (2SC) phase.  Other possible
condensation patterns include the CFL-$K^0$ phase \cite{bedaque01:a}
and the color-spin locked (CSL) phase \cite{schaefer00:a}. The
magnitude of the gap energy lies between $\sim 50$ and
$100$~MeV. Color superconductivity, which modifies the \eos at the
order $(\Delta / \mu)^2$ level \cite{alford03:b,alford04:a}, thus
changes the volume energy by just a few percent.  Such a small effect
can be safely neglected in present determinations of models for the
\eos of neutron star matter and strange star matter. This is different
for phenomena involving the cooling by neutrino emission, the pattern
of the arrival times of supernova neutrinos, the evolution of neutron
star magnetic fields, rotational ($r$-mode) instabilities, and
glitches in rotation frequencies of pulsars (see Refs.\
\cite{rajagopal01:a,alford01:a,rajagopal00:a,alford00:a,alford00:b,%
blaschke99:a,blaschke01:a} and references therein).  Aside from
neutron star properties, an additional test of color superconductivity
may be provided by upcoming cosmic ray space experiments such as AMS
\cite{ams01:homepage} and ECCO \cite{ecco01:homepage}. As shown in
Ref.\ \cite{madsen01:a}, finite lumps of color-flavor locked strange
quark matter, which should be present in cosmic rays if strange matter
is the ground state of the strong interaction, turn out to be
significantly more stable than strangelets
\cite{berger87:a,gilson93:a} without color-flavor locking for wide
ranges of parameters. In addition, strangelets made of CFL strange
matter obey a charge-mass relation that differs significantly from the
charge-mass relation of strangelets made of ordinary strange quark
matter \cite{madsen98:b,madsen01:a}. This difference may allow an
experimental test of CFL locking in strange quark matter
\cite{madsen01:a}.

In this review I will describe the current status of our understanding
of the phases of superdense nuclear matter inside compact stars,
putting special emphasis on the role of strange quark matter in
astrophysics.  This is accompanied by a discussion of possible
observable signatures of the competing states of superdense matter in
the cores of compact stars. These signatures will provide most
valuable information about the phase diagram of superdense nuclear
matter at high baryon number density but low temperature, which is not
accessible to relativistic heavy ion collision experiments.

The article is organized as follows. Section \ref{sec:confinedhm}
discusses the properties and representative models for the \eos of
confined hadronic matter. This is followed by a brief primer on quark
matter presented in section \ref{sec:primerqm}. Relativistic stellar
models are discussed in section \ref{sec:constr}. The possible role of
strange quarks for compact stars and astrophysical phenomena
associated with such stars are reviewed in section
\ref{sec:scc}. Neutrino emission from compact stars and their cooling
behavior are discussed in section \ref{sec:cooling}. Finally, possible
astrophysical signals of quark matter in compact stars are reviewed in
section \ref{sec:mqdec}, followed by general concluding remarks provided in
section \ref{sec:summary}.

\goodbreak
\section{Confined Hadronic Matter}\label{sec:confinedhm}

The \eos of neutron star matter below neutron drip, which occurs at
densities around $4\times 10^{11}\,\gcmt$, and at densities above
neutron drip but below the saturation density of nuclear matter is
relatively well known.  This is to a less extent the case for the \eos
in the vicinity of the saturation density of normal nuclear
matter. Finally the physical properties of matter at still higher
densities are highly uncertain and the models derived for the \eos of
such matter differ considerably with respect to the functional
dependence of pressure on density. This has its origin in various
sources, which concern the many-body technique used to determination
the \eosp, the model of the nucleon-nucleon interaction, the
alteration of hadrons properties by immersion in dense matter, the
fundamental constituents of neutron star matter (including phase
transitions to meson condensates and quark matter), and the
theoretical possibility that strange quark matter may be the true
ground state of the strong interaction rather than nuclear matter. In
the subsection below we introduce a collection of relativistic models
for the \eos which account for these uncertainties. Non-relativistic
models for the \eos will be studied in section \ref{sec:primerqm}.

\goodbreak
\subsection{\it Effective Nuclear Field Theories}

Up until the mid 1970s nearly all dense nuclear matter studies were
based on non-relativistic potential models to describe the
nucleon-nucleon interaction. The relativistic, field-theoretical
approach to nuclear matter was pioneered primarily by Walecka and
collaborators  \cite{walecka74:a,serot86:a}. The generalization of
Walecka's model lagrangian to superdense neutron star matter has the
following form  \cite{glen97:book,weber99:book,glen85:b,weber89:e}
\begin{eqnarray}
\lefteqn{
{\cal L} = \sum_B\bar\psi_B \left( i \slash\partial - m_B
 \right) \psi_B + {1\over 2} \left( \partial^\mu \sigma \partial_\mu
 \sigma - m_\sigma^2 \sigma^2 \right) - {1 \over 4} F^{\mu\nu}
 F_{\mu\nu} + {1 \over 2} m_\omega^2 \omega^\nu \omega_\nu + {1\over
 2} \left( \partial^\mu {\bpi} \bcdot \partial_\mu {\bpi} - m^2_\pi
 {\bpi} \bcdot \bpi \right) }   \label{eq:nuclear_lagrangian} \\
&&- {1\over4} {\bfG}^{\mu\nu} \bcdot {\bfG}_{\mu\nu} + {1\over2}
m^2_\rho {\brho}^\mu \bcdot{\brho}_\mu - \sum_B \Bigl( g_{\sigma B}
\bar\psi_B \sigma \psi_B + g_{\omega B} \bar\psi_B \slash\omega \psi_B
+ { {f_{\omega B}}\over{4 m_B} } \bar \psi_B \sigma^{\mu\nu}
F_{\mu\nu} \psi_B + { {f_{\pi B}}\over{m_\pi} } \bar\psi_B \gamma^5
\slash\partial {\btau} \bcdot {\bpi} \psi_B \nonumber \\
&& + g_{\rho B} \bar\psi_B \gamma^\mu {\btau} \bcdot {\brho}_\mu
\psi_B + { {f_{\rho B}}\over{4 m_B} } \bar\psi_B \sigma^{\mu\nu}
{\btau} \bcdot {\bfG}_{\mu\nu} \psi_B \Bigr) 
- {1\over 3} m_N b_N \left(g_{\sigma N} \sigma \right)^3 - {1\over 4}
  c_N \left( g_{\sigma N} \sigma \right)^4 + \sum_L \bar\psi_L \left(
  i \slash\partial - m_L \right) \psi_L  \nonumber
\end{eqnarray}
where $B$ denotes baryons ($p,n,\Sigma, \Lambda,\Xi$), $L$ stands for
leptons ($e^-, \mu^-$), and $\slash\omega = \gamma^\mu\omega_\mu$,
$\slash\partial = \gamma^\mu\partial_\mu$, $\sigma^{\mu\nu} = i
[\gamma^\mu, \gamma^\nu]/2$. The baryons are described as Dirac
particles which interact via the exchange of $\sigma, \, \omega, \,
\pi$ and $\rho$ mesons. The $\sigma$ and $\omega$ mesons are
responsible for nuclear binding while the $\rho$ meson is required to
obtain the correct value for the empirical symmetry energy.  The cubic
and quartic $\sigma$ terms in Eq.\ (\ref{eq:nuclear_lagrangian}) are
necessary (at the relativistic mean-field level) to obtain the
empirical incompressibility of nuclear matter \cite{bodmer77:a}. The
field equations for the baryon fields follow from
(\ref{eq:nuclear_lagrangian}) as follows \cite{weber99:book,weber89:e},
\begin{eqnarray}
\left( i \gamma^\mu\partial_\mu-m_B \right) \psi_B &=& g_{\sigma B}
  \sigma \psi_B + \Bigl( g_{\omega B}\gamma^\mu\omega_\mu +{f_{\omega
  B}\over{4m_B}} \sigma^{\mu\nu} F_{\mu\nu}\Bigr) \psi_B \nonumber \\
  &&+ \Bigl( g_{\rho B} \gamma^\mu \ {\btau}\bcdot{\brho}_\mu +
  {f_{\rho B}\over{4\,m_B}} \sigma^{\mu\nu}
  {\btau}\bcdot{\bfG}_{\mu\nu} \Bigr) \psi_B + {f_{\pi B}\over{m_\pi}}
  \gamma^\mu \gamma^5 \bigl(\partial_\mu {\btau}\bcdot{\bpi}\bigr)
  \psi_B \, .
\label{eq:eompsi}
\end{eqnarray}
The meson fields in (\ref{eq:eompsi}) are obtained as solution of the
following field equations,
\begin{eqnarray}
\bigl( \partial^\mu\partial_\mu+m^2_\sigma) \sigma &=& - \sum_B
  g_{\sigma B} \bar\psi_B \psi_B - m_N b_N g_{\sigma N} \left(
  g_{\sigma N} \sigma \right)^2 - c_N\, g_{\sigma N} \left( g_{\sigma
  N} \sigma \right)^3 \, ,
\label{eq:eomsigma} \\
\partial^\mu F_{\mu\nu} + m_\omega^2 \, \omega_\nu &=& \sum_B
\Bigl( g_{\omega B} \bpsiB \gamma_\nu \psiB 
- {{f_{\omega B}}\over{2m_B}} \partial^\mu \left(
\bpsiB \sigma_{\mu\nu} \psiB \right) \Bigr) \, ,
\label{eq:eomom5} \\
  \left( \partial^\mu \partial_\mu + m^2_\pi \right)  {\bpi} &=&
  \sum_B {f_{\pi B}\over{m_\pi}}  \partial^\mu \left( \bar \psi_B
  \gamma_5 \gamma_\mu  {\btau}  \psi_B \right)\, ,
\label{eq:eompi4} \\ 
\partial^\mu {\bfG}_{\mu\nu} + m_\rho^2 \, \brho_\nu &=& \sum_B
\Bigl( g_{\rho B} \bpsiB \btau \gamma_\nu \psiB - {{f_{\rho
B}}\over{2m_B}}  \partial^\mu \left( \bpsiB \btau
\sigma_{\mu\nu} \psiB \right) \Bigr) \, ,
\label{eq:eomrho5}
\end{eqnarray} with the field tensors $F_{\mu\nu}$ and ${\bfG}_{\mu\nu}$ 
defined as $F_{\mu\nu} = \partial_\mu\omega_\nu -
\partial_\nu\omega_\mu$ and ${\bfG}_{\mu\nu} = \partial_\mu{\brho}_\nu
- \partial_\nu{\brho}_\mu$. For neutron star matter, Eqs.\
(\ref{eq:eompsi}) through (\ref{eq:eomrho5}) are to be solved subject
to the conditions of electric charge neutrality and chemical
equilibrium. The condition of electric charge neutrality reads
\begin{eqnarray}
  \sum_B q_B^{\rm el} (2J_B+1) {{\kFBt}\over{6\pi^2}} - \sum_L {
    {k_{F_L}^3}\over{3\pi^2}} - \rho_M \, \Theta(\mu^M - m_M) = 0 \, ,
\label{eq:chargeneut}
\end{eqnarray} 
where $J_B$ and $q^{\rm el}_B$ denote spin and electric charge number
of a baryon $B$, respectively.  The last term in (\ref{eq:chargeneut})
accounts for the electric charge carried by condensed bosons.  The
only mesons that may plausibly condense in neutron star matter are the
$\pi^-$ or, alternatively, the more favored $K^-$
 \cite{kaplan86:a,brown87:a}.  Equation~(\ref{eq:chargeneut})
constrains the Fermi momenta of baryons and leptons, $\kFB$ and $\kFL$
respectively. Leptons in neutron star matter are treated as free
relativistic particles,
\begin{equation}
\left( i \gamma^\mu\partial_\mu - m_L \right) \psi_L = 0 \, .
\label{eq:ffermiions}
\end{equation} The baryon and lepton Fermi momenta are further
constrained by the condition of chemical equilibrium.  Since neutron
star matter is characterized by the existence of two conserved
charges, electric charge and baryon charge, the chemical potential of
an arbitrary baryon, $B$, created in a neutron star can be expressed
in terms of two independent chemical potentials.  Choosing $\mu^n$ and
$\mu^e$ as the independent chemical potentials, one has
\begin{equation}
\mu^B = q_B \ \mu^n - q^{\rm el}_B \ \mu^e \, ,
\label{eq:chemeq}
\end{equation}
with $q_B$ the baryon number of particle $B$. Since $q_B \equiv q_f =
{1 \over 3}$ for quark flavors $f = u, d, s$, the chemical potentials
of quarks follow from (\ref{eq:chemeq}) as
\begin{equation}
\mu^f = {1 \over 3} \ \mu^n - q^{\rm el}_f \ \mu^e \, .
\label{eq:mq}
\end{equation}
Relativistic Greens functions constitute an elegant and powerful
technique which allows one to derive from (\ref{eq:eompsi}) to
(\ref{eq:eomrho5}) a set of three coupled equations which are
numerically tractable \cite{weber99:book}. The first one of these
equations is the Dyson equation which determines the two-point baryon
Green function $S^B$ in matter,
\begin{equation} 
 S^B = S_0^B - S_0^B \, \Sigma^B \, S^B \, .
\label{eq:dyson}
\end{equation} 
The second equation determines the effective baryon-baryon scattering
amplitude in matter, $T^{BB'}$, which is given by
\begin{eqnarray}
T^{BB'}= V^{BB'} - V^{BB'}_{\rm ex} + \int V^{B\bar B} \, \Lambda^{\bar
B \bar B'} \, T^{\bar B' B'} \, .
\label{eq:tmatrix}
\end{eqnarray} 
Here $V$ and $V_{\rm ex}$ denote the direct and exchange term of a
given one-boson-exchange potential, which serves as an input, and the
quantity $\Lambda^{\bar B \bar B'} \propto S^{\bar B} S^{\bar B'}$
describes the propagation of baryons $\bar B$ and $\bar B'$ in
intermediate scattering states. A popular and physically most
suggestive choice for $\Lambda$ is the so-called Brueckner
propagator \cite{weber99:book,huber95:a}. This propagator describes the
propagation of two baryons in intermediate scattering states in terms
of the full single-particle energy-momentum relation and, in addition,
guarantees that these particles obey the Pauli principle too.  The system
of equations is closed by the expression for the self energy of a
baryon in matter, $\Sigma^B$, given by
\begin{equation} 
\Sigma^B  =  i  \sum_{B'}
 \int  \left( \, {\rm Tr}\, \left(
T^{BB'} \, S^{B'} \right) - T^{BB'} \, S^{B'}\, \right) \, .
\label{eq:selfenergy}
\end{equation} The one-boson-exchange potentials, $V^{BB'}$, sum
the contributions arising from different types of mesons that
exchanged among the baryons,
\begin{equation}
  \langle 12|V^{BB'}|34 \rangle = \sum_{M=\sigma,\omega,\pi,\rho,
  \ldots} \delta^4(1,3) \delta^4(2,4) \Gamma^M(1,3) \Gamma^M(2,4)
\Delta^M(1,2) \, ,
\label{eq:V}
\end{equation}
where $\Gamma^M$ denote meson-nucleon vertices, and $\Delta^M$ free
meson propagators.  
\begin{figure}[tb] 
\begin{center}
\epsfig{figure=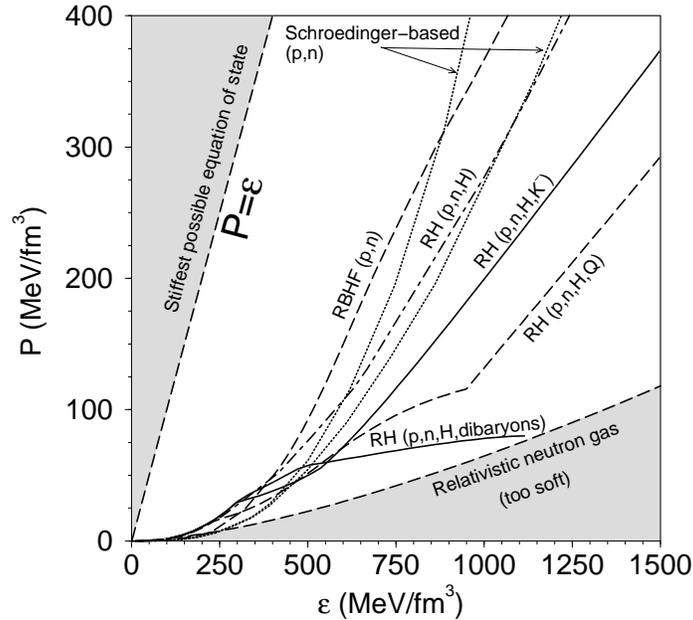,width=9.0cm,angle=0}
\begin{minipage}[t]{16.5 cm}
\caption{Models for the \eos of neutron star matter computed for
different compositions and many-body techniques described in the
text \cite{weber99:book}.  (p,n denotes protons and neutrons; H, K$^-$,
Q stand for hyperons, K$^-$ condensate, and quarks, respectively.)}
\label{fig:EOSs}
\end{minipage}
\end{center}
\end{figure}
Relativistic models for the equation of state of
neutron star matter are obtained by solving Eqs.\ (\ref{eq:dyson})
through (\ref{eq:selfenergy}) self-consistently in combination with
Eqs.\ (\ref{eq:chargeneut}) and (\ref{eq:chemeq}) for electric charge
neutrality and chemical equilibrium.  This has been accomplished for
several different approximation schemes.  With increasing level of
complexity, these are the relativistic Hartree (RH), relativistic
Hartree-Fock (RHF), and the relativistic Brueckner-Hartree-Fock (RBHF)
approximations \cite{weber99:book}.  The first two approximations are
obtained by keeping only the Born term in the $T$-matrix equation
(\ref{eq:tmatrix}), that is, by setting $T = V - V_{\rm ex}$ (and
readjusting the parameters on the boson-exchange potential). The mass
operator (\ref{eq:selfenergy}) is then given
\begin{eqnarray}
  \Sigma^B(p) = - i \sum_{B'} g_{\sigma B} \, g_{\sigma B'}
  \int\!\dfq\, e^{i \eta q^0} \left( \Delta\!^\sigma(0) S^{B'}(q) -
  \delta_{BB'} \Delta\!^\sigma(p-q) S^B(q) \right) \pm \ldots
\label{eq:SigmaHF}
\end{eqnarray}
where the $T$-matrix is replaced, according to Eq.\ (\ref{eq:V}), by
free meson propagators. This simplifies the solution of the many-body
equations considerably. Because of this approximation, however, the
coupling constants of the theory need to be adjusted to fit the
properties of infinite nuclear matter, hypernuclear data and neutron
star properties (see, for instance, Refs.\
 \cite{weber99:book,glen91:b,glen97:book,glen91:rec}). The five nuclear
matter properties are the binding energy $E/A$, effective nucleon mass
$m^*_{\rm N}/m_{\rm N}$, incompressibility $K$, and the symmetry
energy $a_{\rm s}$ at nuclear matter saturation density $\rho_0$
$(=0.16~{\rm fm}^{-3})$,
\begin{equation}
  E/A=-16.0~\mev, ~ m^*_N/m_N=0.79, ~K\simeq 225~\mev, ~a_s=32.5~\mev
  \, .
\label{eq:dcp.20}
\end{equation} Of the five, the value for the incompressibility 
of nuclear matter carries the biggest
uncertainty \cite{heiselberg00:a}. Its value is currently believed to
lie in the range between about 220 and 250~MeV. In contrast to RH and
RBHF, the RBHF approximation makes use of one-boson-exchange
\begin{table}[tb]
\begin{center}
\begin{minipage}[t]{16.5 cm}
\caption{Saturation properties of nuclear matter computed for numerous
different nuclear forces and many-body
techniques \cite{weber99:book,weber89:e,oyamatsu03:a}.}
\label{tab:saturation}
\end{minipage}
\begin{tabular}{l|c|c|c|c|r|r} \hline
 Method/Force &  $\rho_0$  & $E/A$  & $K$  & $a_{\rm s}$  & $L$ & $y$ \\
       &  (${\rm fm}^{-3})$  &(MeV)  &(MeV)  &(MeV)  &(MeV) 
       &(MeV~${\rm fm}^3$) \\
\hline
 SG-0  &   0.168  &$-16.7$  &  253 & 35.6 &  41.6 &$-430$\\
 SGI   &   0.154  &$-15.8$  &  261 & 28.3 &  64.1 &$-250$\\
 SGII  &   0.158  &$-15.6$  &  215 & 26.8 &  37.6 &$-322$\\
 SkM   &   0.160  &$-15.8$  &  217 & 30.7 &  49.3 &$-281$\\
 SkM$^*$ & 0.160  &$-15.8$  &  217 & 30.0 &  45.8 &$-296$\\
 $E_\sigma$ &0.163  &$-16.0$  &  249 & 26.4 & $-36.9$ &364\\
 $Z_\sigma$ &0.163  &$-15.9$  &  233 & 26.7 & $-29.4$ &432\\
 $Z_\sigma^*$ & 0.163  &$-16.0$  &  235 & 28.8 & $-4.58$ &3030\\
 $R_\sigma$   &0.158  &$-15.6$  &  238 & 30.6 &  85.7 &$-179$\\
 $G_\sigma$   &0.158  &$-15.6$  &  237 & 31.4 &  94.0 &$-167$\\
 SkT6  &   0.161  &$-16.0$  &  236 & 30.0 &  30.8 &$-475$\\
 SkP   &   0.163  &$-16.0$  &  201 & 30.0 &  19.5 &$-632$\\
 SkSC4 &   0.161  &$-15.9$  &  235 & 28.8 & $-2.17$ &6460\\
 SkX   &   0.155  &$-16.1$  &  271 & 31.1 &  33.2 &$-545$\\
 MSk7  &   0.158  &$-15.8$  &  231 & 27.9 &  9.36 &$-1460$\\
 BSk1  &   0.157  &$-15.8$  &  231 & 27.8 &  7.15 &$-1908$\\
 SLy4  &   0.160  &$-16.0$  &  230 & 32.0 &  45.9 &$-335$\\
 SLy7  &   0.158  &$-15.9$  &  230 & 32.0 &  47.2 &$-328$\\
 TM1   &   0.145  &$-16.3$  &  281 & 37.9 &  114 &$-215$\\
NL\,1  &0.152     &$-16.4$  &212   &43.5  &140     &$-145$    \\
FRDM   &0.152     &$-16.3$  &240   &32.7  &$-$     &$-$      \\
HV     &0.145     &$-15.98$ &285   &32.5  &        &        \\
Bro~$A$     &0.174 &$-16.5$  &280   &34.4  &81.9    &$-225$    \\
Bro~$B$     &0.172 &$-15.7$  &249   &32.8  &90.2    &$-175$    \\
Bro~$C$     &0.170 &$-14.4$  &258   &31.5  &76.1    &$-209$    \\
$\eut$      &0.157 &$-16.6$  &261   &30.8  &$-$     &$-$   \\
$\euu$      &0.175 &$-11.5$  &202   &29.3  &$-$     &$-$   \\
$\eau$      &0.194 &$-12.4$  &209   &27.6  &$-$     &$-$   \\
$\eaix$     &0.16  &$-16.00$ &$-$   &$-$   &$-$     &$-$   \\
TF96        &0.161 &$-16.04$ &234   &32.0  &$-$     &$-$   \\
\hline
\end{tabular}
\end{center}
\vskip -0.8cm
\end{table}
potentials whose parameters are adjusted to the properties of the
deuteron and relativistic nucleon-nucleon scattering data. This
approximation is therefore referred to as a parameter free. In passing
we mention that in recent years a new class of effective field
theories was developed which treat the meson-nucleon couplings density
dependent. These field theories provide a very good description of the
properties of nuclear matter, atomic nuclei as well as neutron stars
 \cite{lenske95:a,fuchs95:a,typel99:a,niksic02:a,ban04:a}.  We conclude
this section with a brief discussion of the total energy density of
the system follows from the stress-energy density tensor,
${T}_{\mu\nu}$, as \cite{weber99:book}
\begin{eqnarray} 
\epsilon = \langle {T}_{00} \rangle  = 
 \sum_{\chi=B, L} \partial_0 \psi_\chi
\, {{\partial {\cal L}} \over{\partial
\, (\partial^0\,\psi_\chi)}} - g_{00} \, {\cal L} \, .
\label{eq:epsilon}
\end{eqnarray} 
The pressure, and thus the \eosp, are obtained from Eq.\
(\ref{eq:epsilon}) as
\begin{eqnarray}
P = \rho^2 \partial/\partial\rho (\epsilon/\rho) \, .
\label{eq:pressure}
\end{eqnarray}
Finally, the energy per particle is given in terms of the energy
density and baryon number density as
\begin{eqnarray}
\epsilon = (E/A + m_{\rm N}) \rho \, .
\label{eq:epsilon2}
\end{eqnarray}
Figure~\ref{fig:EOSs} shows several model equations of state based on
RH and RBHF assuming different particle compositions of neutron star
matter \cite{weber99:book}.  Of particular interest later (section
\ref{sec:mqdec}) will be the \eos in Fig.\ \ref{fig:EOSs} accounting
for quark deconfinement  \cite{glen97:book} which, for this model, sets
in at $230~,\mevt$, which is less than two times the energy density of
nuclear matter, $\epsilon_0= 140~\mevt$. Pure quark matter exists for
densities greater than $950~ \mevt$, which is around seven times
$\epsilon_0$.  Of key importance for the possible occurrence of a
quark-hadron phase in neutron stars is that pressure in the mixed
phase of quarks and hadrons varies with density \cite{glen91:pt}. If
this is not the case, hydrostatic equilibrium would strictly excluded
the mixed phase from neutron stars.  Table~\ref{tab:saturation} shows
the properties of nuclear matter computed for different many-body
techniques as well as nuclear forces.
In addition to the saturation properties listed in ({\ref{eq:dcp.20})
this table also shows the symmetry energy density derivative, $L$, and
the slope, $y$, of the saturation curve of isospin asymmetric nuclear
matter, defined as $L = 3 \rho_0 (\partial a_{\rm s} /
\partial\rho)_{\rho_0}$ and $y = - K a_{\rm s} (3 \rho_0 L)^{-1}$,
for several models. 
which vary considerably from one model to another.

\goodbreak
\subsection{\it Non-relativistic Treatments}

The most frequently used non-relativistic treatments of dense nuclear
matter studies are the hole-line expansion (Brueckner theory)
 \cite{baldo01:b,baldo03:b}, the coupled cluster method, self-consistent Green
functions, the variational approach (Monte Carlo techniques), the
semiclassical Thomas-Fermi method \cite{myers95:a,strobel97:a}, and the
density functional approach based on Skyrme effective interactions
(for an overview of these methods and additional references, see Refs.\
 \cite{heiselberg00:a,muether00:a}). Aside from the density functional
approach, the starting point in each case is a phenomenological
nucleon-nucleon interaction, which, in some cases, is supplemented
with a three-nucleon interaction $V_{ijk}$ introduced to achieve the
correct binding energy of nuclear matter at the empirical saturation
density, $\rho_0=0.15$~nucleons/fm$^3$.  Figure \ref{fig:EOSs} shows
two Schroedinger-based models for the \eos which are obtained for
variational calculations based on the Urbana $V_{14}$ two-nucleon
interaction supplemented with the three-body interaction UVII (left
curve) and TNI (right curve) three-nucleon
interaction \cite{heiselberg00:a,wiringa88:a,akmal97:a}. The
Hamiltonian is thus of the form
\begin{equation}
  {\cal H} = \sum_i \, {{-\,\hbar^2}\over{2\, m}} \, \nabla_i^2 +
  \sum_{i<j} \, {V}_{ij} + \sum_{i<j<k} \, {V}_{ijk} \, .
\label{eq:hamil}
\end{equation}
In the variational approach the Schroedinger equation ${\cal H} |\Psi
\rangle = E |\Psi\rangle$ is solved using a variational trial function
$|\Psi_v \rangle$, which is constructed from a symmetrized product of
two-body correlation operators ($F_{ij}$) acting on an unperturbed
ground state (Fermi gas wave function) $|\Phi \rangle$,
\begin{equation}
  |\Psi_v \rangle = \Bigl( S \, \prod_{i<j} \, F_{ij} \Bigr)
   |\Phi\rangle \, .
\label{eq:psiv}
\end{equation}
The antisymmetrized Fermi-gas wave function is given by $|\Phi\rangle
= A (\Pi_{i<{\rm F}} |i \rangle)$.  The correlation operator contains
variational parameters (14 for the $\rm{UV}_{14}$ and ${\rm AV}_{14}$
models \cite{wiringa88:a} and 18 for the more recent ${\rm A}_{18}$
model \cite{akmal98:a}) which are varied to minimize the energy per
baryon for a given density $\rho$,
\begin{equation}
  E(\rho) = {\rm min} ~ { {\langle \Psi_v | {\cal H} | \Psi_v \rangle}
\over {\langle \Psi_v | \Psi_v \rangle} }  \, ,
\label{eq:evar}
\end{equation} 
which constitutes an upper bound to the ground-state energy of the
system.  Pressure and energy density of the stellar matter are
obtained from relations (\ref{eq:pressure}) and (\ref{eq:epsilon2}),
respectively. The new Thomas-Fermi approach of Myers and Swiatecki,
TF96, is based on a Seyler-Blanchard potential generalized by the
addition of one momentum dependent and one density dependent term
 \cite{myers95:a},
\begin{eqnarray}
  V_{12} = - \frac{2\, T_{0}}{\rho_0} \, Y\bigl(r_{12}\bigr) \left(
 \frac{1}{2} (1 \mp \xi) \alpha - \frac{1}{2}(1 \mp \zeta) \Bigl(\beta
 \Bigl( \frac{p_{12}}{\kFo} \Bigr)^2 - \gamma \frac{\kFo}{p_{12}} +
 \sigma \Bigl( \frac{2 \bar \rho}{\rho_0} \Bigr)^{\frac{2}{3}} \Bigr)
 \right) \, .
\label{eq:v.tf96}
\end{eqnarray} The upper (lower) sign in (\ref{eq:v.tf96}) corresponds
to nucleons with equal (unequal) isospin.  The quantities $\kFo$,
$T_{0}$ $(= \kFosq / 2m)$, and $\rho_0$ are the Fermi momentum, the
Fermi energy and the saturation density of symmetric nuclear
matter. The potential's radial dependence is described by a
Yukawa-type interaction of the form
\begin{equation}
  Y\bigl(r_{12}\bigr) = {{1}\over{4 \pi a^3}} { {e^{-
  r_{12}/a}}\over{r_{12}/a} } \, .
\label{eq:2.ms96}
\end{equation} 
Its strength depends both on the magnitude of the particles' relative
momentum, $p_{12}$, and on an average of the densities at the
locations of the particles.  The parameters $\xi$ and $\zeta$ were
introduced in order to achieve better agreement with asymmetric
nuclear systems. The behavior of the optical potential is improved by
the term $\sigma (2\bar\rho/\rho_0)^{2/3}$ with the average density
defined as $\overline{\rho}^{2/3} = (\rho_{1}^{2/3} +
\rho_{2}^{2/3})/2$, and $\rho_1$ and $\rho_2$ the densities of
interacting neutron or protons at points 1 and 2. The seven free
parameters of the theory are adjusted to the properties of finite
nuclei, the parameters of the mass formula, and the behavior of the
optical potential \cite{myers95:a}. The nuclear matter properties at
saturation density obtained for TF96 are listed in
table~\ref{tab:saturation}. 

\goodbreak
\section{Primer on Quark Matter}\label{sec:primerqm}

\subsection{\it Models for the Equation of State}

The field theory of quarks and gluons, Quantum Chromodynamics, is a
non-Abelian gauge theory with ${\rm {SU(3)}}_{\rm c}$ as a gauge
group. The QCD Lagrangian has the form
\begin{equation}
{\cal L} = {\bar\psi}^a_f \left( i \gamma_\mu D^\mu_{ab} - m_f
\right) {\psi}^b_f - {{1}\over{4}} F^i_{\mu\nu}
F_i^{\mu\nu} \, ,
\label{eq:LQEC}
\end{equation} where ${\psi}^a_f$ are the quark fields for each
flavor $f$ and $m_f$ the current quark masses (see
table~\ref{tab:quarks}). In color space, the fields $\psi^a_f$ are
three-component columns with $a=1,2,3$. The color gauge-covariant
derivative $D^\mu$ given by
\begin{equation}
D^\mu_{a b} = \delta_{a b} \, \partial^\mu - i \, {{g_s}\over{2}} \,
[\lambda_i]_{a b} \, G_i^\mu \, ,
\label{eq:cgcd}
\end{equation} where $g_s$ is the strong interaction coupling constant.
The quantities $G^\mu_i$ are the gluon fields with color indices
$i=1,\ldots,8$, and $\lambda_i$ the Gell-Mann ${\rm SU(3)}_{\rm c}$
matrices. The quantity $F^i_{\mu\nu}$ is the gluon field tensor
defined as
\begin{equation}
F^i_{\mu\nu} = \partial_\mu G^i_\nu - \partial_\nu G^i_\mu +
g_s \, f_{ijk} \, G^j_\mu \, G^k_\nu \, ,
\label{eq:Fi}
\end{equation}
where $f_{ijk}$ are the ${\rm SU(3)}_{\rm c}$ structure constants.
The equations of motion of the coupled quark and gluon fields derived
from Eq.~(\ref{eq:LQEC}) are then as follows,
\begin{eqnarray}
\left( i \gamma_\mu \partial^\mu - m_f \right) \, \psi^a_f &=& -
g_s \, \gamma_\mu \left( {{\lambda_i}\over{2}} \right)_{ab}
\psi^b_f \, G^\mu_i \, ,
\label{eq:eom.quarks} \\
\partial^\mu F^i_{\mu \nu} + g_s \, f_{ijk} \, G^{j \mu} \,
F^k_{\mu \nu} &=& - g_s \, {\bar\psi}^a_f \, \gamma_\nu \left(
{{\lambda_i}\over{2}} \right)_{ab} \psi^b_f \, .
\label{eq:eom.gluons}
\end{eqnarray}
Considerable efforts are made to solve the QCD equations of motion on
\begin{table}
\caption{Approximate masses, $m_f$, and electric charge numbers,
$q_f^{\rm el}$ of quarks.}
\label{tab:quarks}
\begin{center}
\begin{tabular}{l|c|c|c|c|c|c}\hline
Quark flavor ($f$)  &$u$    &$d$    &$c$     &$s$    &$t$      &$b$  \\ \hline 
$m_f$ (GeV)             &0.005   &0.01   &1.5     &0.1    &180      &4.7  \\ 
$q_f^\el$               &$+{2\over 3}$ &$-{1\over 3}$ &$+{2\over 3}$ 
 &$-{1\over 3}$ &$+{2\over 3}$ &$-{1\over 3}$\\  \hline
\end{tabular}
\end{center}
\end{table}
the lattice. At present, however, such simulations do not provide a
guide at finite baryon number density yet, and it is necessary to rely
on non-perturbative QCD models for quark matter which incorporate the
basic properties expected for QCD. Three different categories of
models have emerged. These are (1) phenomenological models (MIT bag
model) where quark masses are fixed and confinement is described in
terms of a bag constant, and more advanced (2) dynamical models and
(3) Dyson-Schwinger equation models where the properties of quark
(matter) are determined self-consistently.  The most widely used of
these models is the MIT bag
model \cite{chodos74:a,chodos74:b,farhi84:a}. For this model, the
pressure $P^i$ of the individual quarks and leptons contained in the
bag is counterbalanced by the total external bag pressure $P+B$
according to
\begin{eqnarray}
  P + B = \sum_f  P^f \, ,
\label{eq:pbag} 
\end{eqnarray} while the total energy density of the quark flavors 
confined in the bag is given by
\begin{eqnarray}
\epsilon = \sum_f \epsilon^f + B \, .
\label{eq:ebag}
\end{eqnarray} 
The quantity $B$ denotes the bag constant, and $\epsilon^f$ are the
contributions of the individual quark to the total energy density. The
condition of electric charge neutrality among the quarks reads
\begin{eqnarray}
3 \sum_f q_f^{\rm el} \ \kFft - \sum_L \ {k_{F_L}^3} = 0 \, ,
\label{eq:ladn}
\end{eqnarray} where $q_f^{\rm el}$ denotes the electric charge number
of a quark of flavor $f$, listed in table~\ref{tab:quarks}.  The 
contributions of each quark flavor to pressure, energy density and
baryon number density are determined by the thermodynamic potentials
$d\Omega^f = -S^f d T - P^f d V - A^f d\mu^f$, from which one obtains
\begin{eqnarray} 
P^f = \frac{\nu_{i}}{6 \pi^{2}} \int_{0}^{\kFf}\! d k \, {{k^4} \over
{\sqrt{k^2 + m_f^2}} } \, , \ \ \epsilon^{f} = \frac{\nu_{f}}{2
\pi^{2}} \int_{0}^{\kFf}\! d k \, k^2 \, \sqrt{k^2 + m_f^2} \, , \ \
\rho^f = \frac{{\nu_f}}{6 \pi^2} \, \kFft \, .
\label{eq:pint}
\end{eqnarray}  The quantity $\mu^f$ denotes the chemical 
potential of quark flavor $f$, and $m_f$ stands for its mass.  The
phase space factor $\nu_f$ is equal to $2~ ({\rm spin})\times 3 ~({\rm
color}) = 6$.  Chemical equilibrium among the quark flavors and the
leptons is maintained by the following weak reactions,
\begin{eqnarray} 
  d \leftrightarrow u + e^- + \bar\nu^{e} \, , \ \ s \leftrightarrow u + e^-
  + \bar\nu^{e} \, , \ \ s \leftrightarrow c + e^- + \bar\nu^{e} \, , \ \
  s + u \leftrightarrow d + u \, , \ \  c + d \leftrightarrow u + d \, .
  \label{eq:5}
\end{eqnarray} 
Since neutron stars loose the neutrinos within the first few seconds
after birth, the chemical potentials of neutrinos and anti-neutrinos
obey $\mu^\nu = \mu^{\bar\nu} = 0$ and one obtains from the weak
reactions~(\ref{eq:5})
\begin{eqnarray}
  \mu^d = \mu^u + \mu^{e} \, , \ \ \mu^c = \mu^u \, , \ \  \mu
  \equiv \mu^d = \mu^s \, .
\label{eq:ch1}
\end{eqnarray} 
The \eos of relativistic quark matter at zero temperature made up of
massless, non-interacting particles is readily calculated from
(\ref{eq:pint}). One obtains
\begin{equation} 
  P^{f} = \frac{\nu_{f}}{24 \pi^{2}} \ (\mu^f)^{4} = {1\over 3} \
 \epsilon^f \, , \ \ \rho^{f} = \frac{\nu_{f}}{6 \pi^{2}} \
 (\mu^f)^{3} \, ,
\label{eq:eosqm.0}
\end{equation} with $\nu_f=6$. The \eos  of such matter is obtained 
from equations~(\ref{eq:pbag}) and (\ref{eq:ebag}) as
\begin{eqnarray} 
  P = ( \epsilon - 4  B) / 3 \, .
\label{eq:bag1}
\end{eqnarray} 
From Eq.\ (\ref{eq:bag1}) one sees that the external pressure acting
on a bag filled with quarks vanishes for $\epsilon = 4 B$. The mass
contained inside the bag is given by $M = \int_0^R \epsilon dV = (4
\pi / 3) \epsilon R^3$, which is the generic mass-radius relation of
self-bound matter. The consequences of this relation for strange quark
matter systems are illustrated in Fig.\ \ref{fig:nucl_sizes}.
\begin{figure}[tb]
\begin{center}
\epsfig{figure=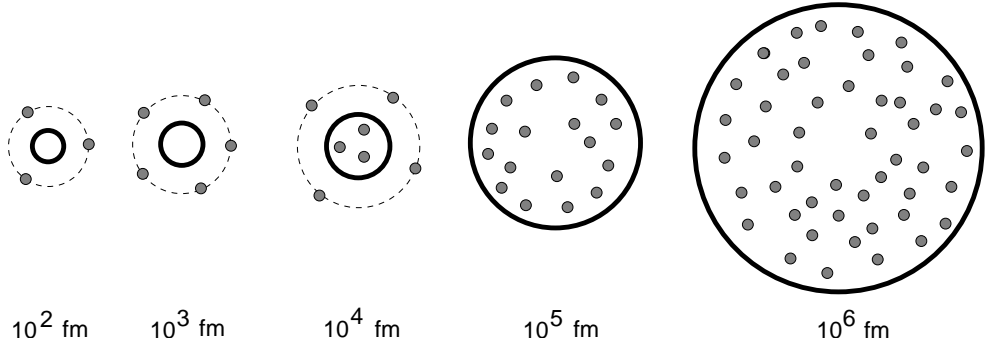,width=13.0cm,angle=0}
\begin{minipage}[t]{16.5 cm}
\caption{Radii of quark bags  \cite{giacomelli03:b}.  For masses less
than $10^9$~GeV the electrons (gray dots) are outside the quark bags
(indicated by thick solid circles) and the core+electron system has a
size of $\sim 10^5$~fm. For masses between $10^9$ and $10^{15}$~GeV
the electrons are partially inside the core. For masses greater than
$10^{15}$~GeV all electrons are inside the core.}
\label{fig:nucl_sizes}
\end{minipage}
\end{center}
\end{figure} 
The condition of electric charge neutrality of stellar 
quark matter, Eq.~(\ref{eq:ladn}), leads to
\begin{eqnarray} 
 2  \rho^{u} - \rho^{d} - \rho^{s} = 0 \, .
\label{eq:chn0}
\end{eqnarray} Since $\mu^u = \mu^d = \mu^s$ for massless quarks,
one finds from Eq.~(\ref{eq:pbag}) that for zero external pressure,
$P$, the bag constant is related to the quark chemical potential as $B
= 3 \mu^4 / 4 \pi^2$. The energy per baryon number of quark
matter follow as
\begin{eqnarray}
  {E\over A} \equiv \frac{\epsilon}{\rho} = 4 \frac{B} {(\rho^{u} +
    \rho^{d} + \rho^{s}) /3} = 4 \frac{B}{\rho^{u}} = 4 \pi^2 {{B}
    \over {\mu^3}} \, ,
\label{eq:ebary}
\end{eqnarray}
with $\rho$ the total baryon number density defined as
\begin{equation}
\rho = \sum_f \rho^f/3 \, .
\label{eq:rho_quarks}
\end{equation}
In the next step we turn to the determination of the \eos if the
charm and strange quarks are given their finite mass values listed in
\ref{tab:quarks}. In this case Eq.\ (\ref{eq:pint}) leads for
pressure, mass density, and baryon number density of quark matter to
the following expressions,
\begin{eqnarray}
  P^f & = & \frac{\nu_f (\mu^f)^4}{24 \pi^{2}} \Biggl(
  \sqrt{1-z_f^{2}} \ \Bigl( 1 -\frac{5}{2} z_f^{2} \Bigr) +
  \frac{3}{2} z_f^{4} \ \ln \frac{ 1 +\sqrt{1-z_f^{2}} } {z_f} \, 
  \Biggr) \, ,
\label{eq:pm1} \\ %
\rho^f & = & \frac{\nu_f (\mu^f)^3}{6 \pi^{2}} \,(1 -
z_f^{2})^{\frac{3}{2}} \, ,
\label{eq:nm1} \\ %
\epsilon^f & = & \frac{\nu_f (\mu^f)^4}{8 \pi^{2}} \Biggl(
\sqrt{1 - z_f^{2}} \ \Bigr( 1 - \frac{1}{2} z_f^{2} \Bigr) -
\frac{z_f^{4}}{2} \ \ln \frac{1 + \sqrt{ 1 - z_f^{2}}}{z_f} \, 
\Biggr) \, ,
\label{eq:em1a}
\end{eqnarray} with $z_i$ defined as $z_f = m_f/\mu^f$.
Ignoring the two most massive quark flavors, $t$ and $b$, which are
\begin{figure}[tb]
\begin{center}
\epsfig{figure=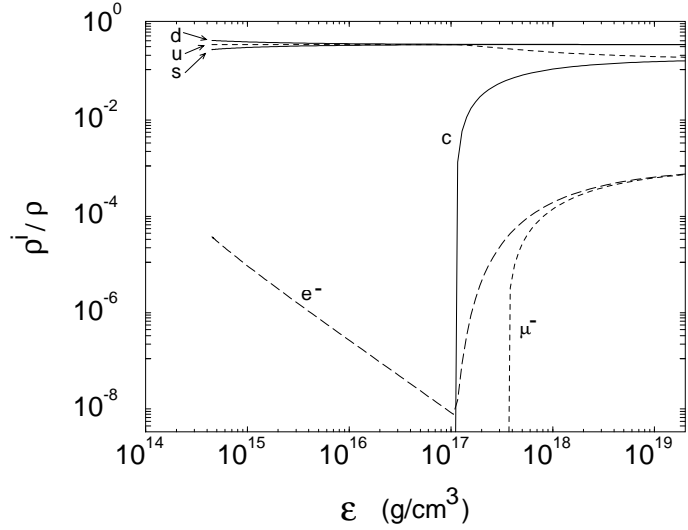,width=7.0cm,angle=-90}
\begin{minipage}[t]{16.5 cm}
\caption{Relative quark and lepton densities in quark-star matter as a
  function of mass density (from Ref.\  \cite{kettner94:b}).}
\label{fig:qlcomp1}
\end{minipage}
\end{center}
\end{figure} 
far too heavy to become populated in compact stars (see Fig.\
\ref{fig:qlcomp1}), the condition of electric charge neutrality,
expressed in Eq.\ (\ref{eq:ladn}), reads
\begin{eqnarray}
  2 ( \rho^{u} + \rho^{c} ) - ( \rho^{d} + \rho^{s} ) - 3 (\rho^e +
  \rho^\mu) = 0 \, . 
\label{eq:lm1.add}
\end{eqnarray} Upon substituting (\ref{eq:nm1}) into  (\ref{eq:lm1.add}),  
this relation can be written as
\begin{eqnarray}
2 (1 - (\mu^e/\mu)^{3}) \left( { 1 + ( 1 - z_{c}^{2} )^{\frac{3}{2}}}
\right) - \left( { 1 + ( 1 - z_{s}^{2} )^{\frac{3}{2}}} \right) -
(\mu^e/\mu)^{3} \left( { 1 + ( 1 - z_\mu^{2} )^{\frac{3}{2}}} \right)
= 0 \, . 
\label{eq:lm1}
\end{eqnarray} An expression for the pressure of the system is obtained by 
substituting (\ref{eq:pm1}) into (\ref{eq:pbag}). This leads to
\begin{eqnarray}
P + B = {{\mu^4}\over{4 \pi^2}} \left( (1-(\mu^e/\mu)^4) \Phi(z_c) +
\Phi(z_s) \right) \, ,
\label{eq:gg1}
\end{eqnarray}
with $\Phi$ defined as
\begin{eqnarray}
\Phi(z_f) = 1 + \sqrt{1 - z_f^2} \, \left( 1 - \frac{5}{2} z_f^2
 \right) + \frac{3}{2} z_f^2 \ \ln \frac{1 + \sqrt{1 - z_f^2}} {z_f}
 \, .
\label{eq:Phi}
\end{eqnarray}
The total energy density follows from (\ref{eq:ebag}) as
\begin{eqnarray}
\epsilon = ( 3 P + 4 B ) + {1\over{4 \pi^2}} \sum_{f=s,c} \nu_f
(\mu^f)^4 z_f^{-2} \Biggl( \sqrt{1-z_f^{2}} - z_f^{2} \ln \frac{ 1 +
\sqrt{1-z_f^{2}}}{z_f} \, \Biggr) .
\label{eq:eos1}  
\end{eqnarray} 
The first term on the right-hand side of Eq.~(\ref{eq:eos1})
represents the \eos of a relativistic gas of massless quarks derived
in Eq.\ (\ref{eq:bag1}) while the second term accounts for finite
strange and charm quark masses.  Finally, the total baryon number
density of such quark matter is given by
\begin{eqnarray}
\rho  = \frac{\mu^{3}}{3\pi^{2}} \left( (1-(\mu^e/\mu)^{3}) \left( 1 + ( 1 -
z_{c}^{2})^{\frac{3}{2}} \right) + \left( 1 + ( 1 - z_{s}^{2}
)^{\frac{3}{2}} \right) \right) \, .
\label{eq:nb1}
\end{eqnarray} 
The relative quark-lepton composition of absolutely stable
($\bag=145$~MeV) quark-star matter at zero temperature is shown in
Fig.~\ref{fig:qlcomp1}.  All quark flavor states that become populated
in such matter up to densities of $10^{19}~\gcmt$ are taken into
account. Since the Coulomb interaction is so much stronger than
gravity, quark star matter in the lowest energy state must be charge
neutral to very high precision \cite{alcock88:a}.  Therefore, any net
positive quark charge must be balanced by a sufficiently large number
of negatively charged quarks and leptons, which determines the lepton
concentration shown in Fig.\ \ref{fig:qlcomp1}. Because of their
relatively large masses, the presence of charm quarks requires
densities greater than $10^{17}\, \gcmt$ in order
\begin{figure}[tb]
\begin{center}
\parbox[t]{7.5cm} {\epsfig{figure=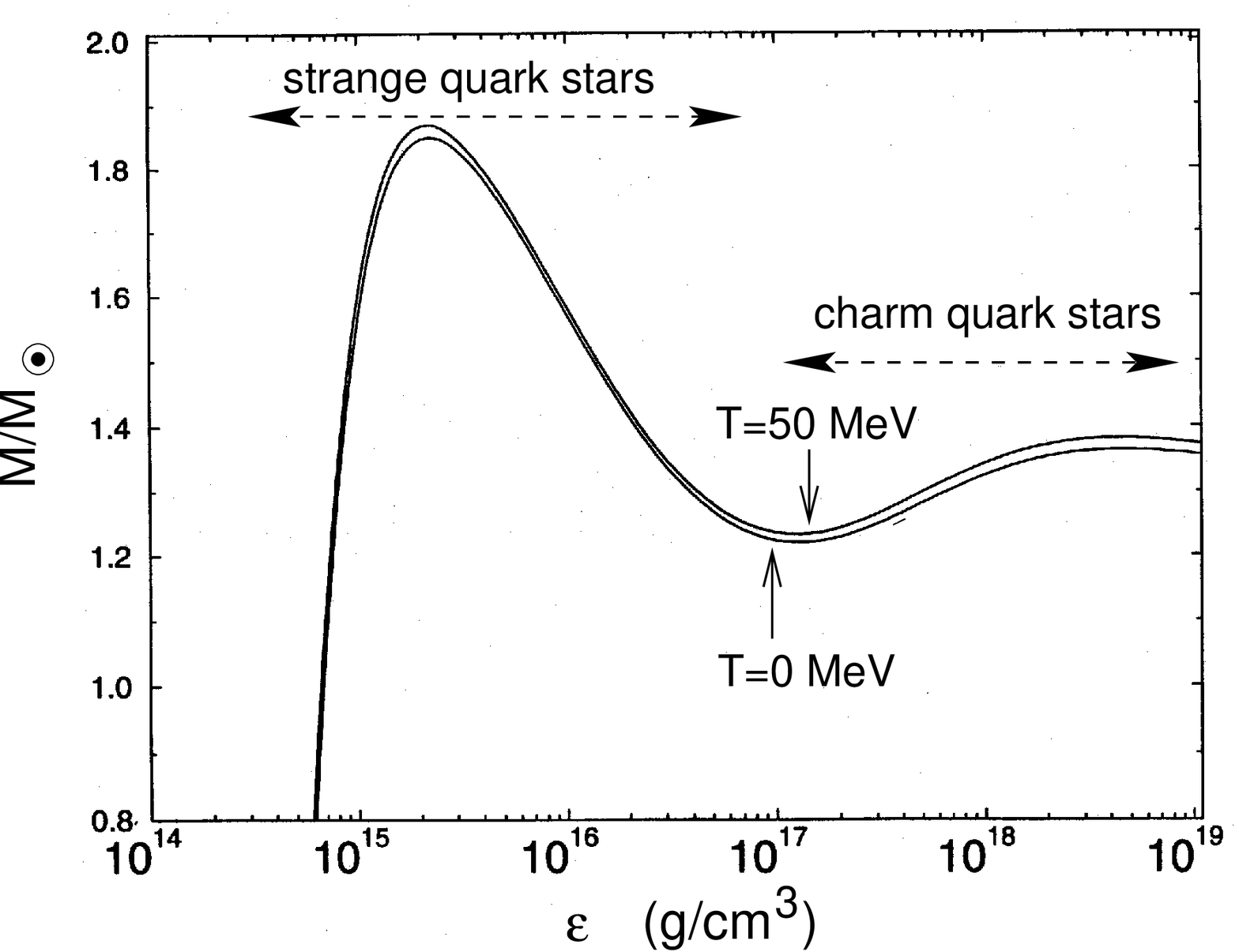,width=7.0cm,angle=0}
{\caption{Mass of quark stars at zero and finite temperature versus
central star density \cite{kettner94:b}.}
\label{fig:Mvsec_charm}}}
\ \hskip 1.0cm   \
\parbox[t]{7.5cm} {\epsfig{figure=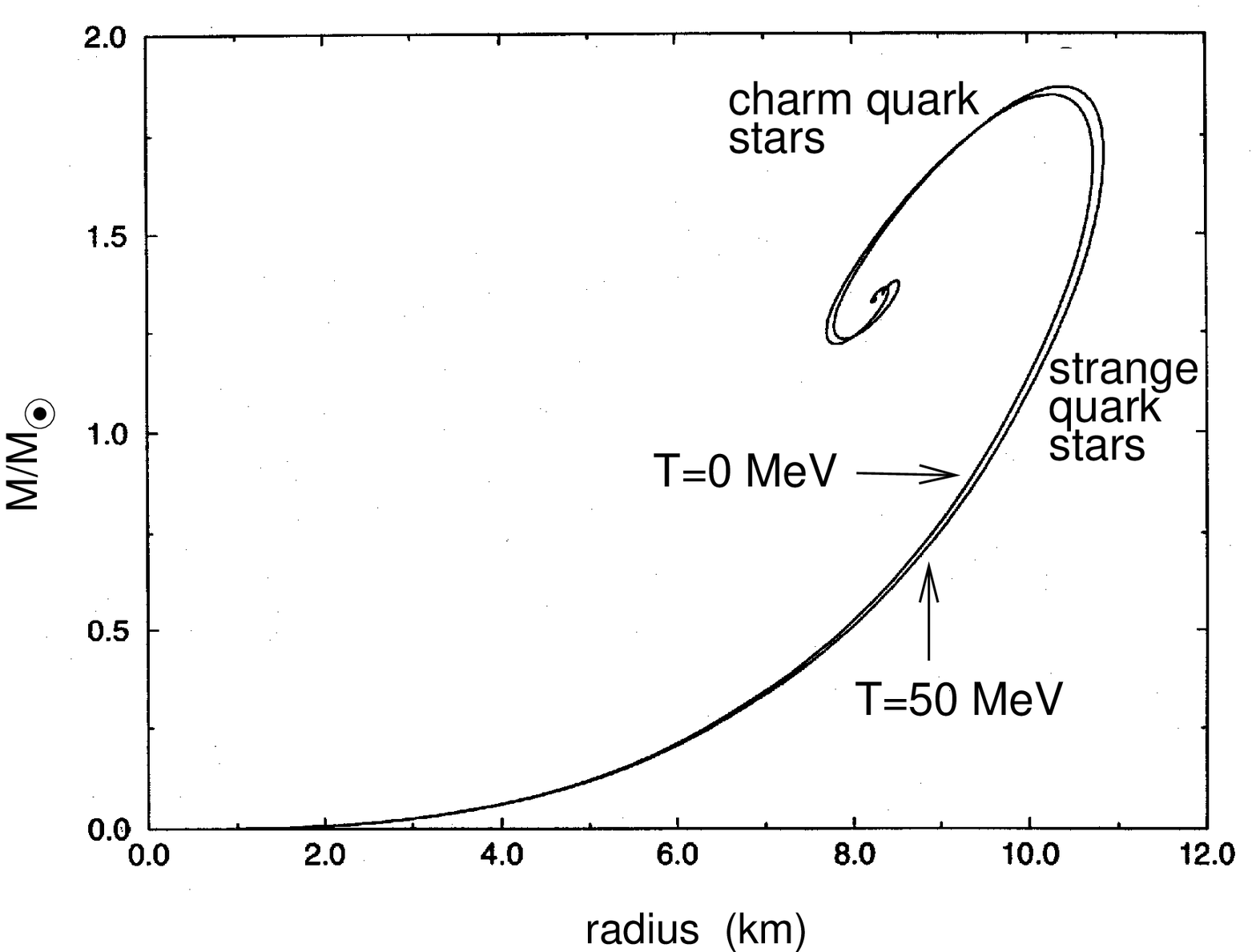,width=7.0cm,angle=0}
{\caption{Mass versus radius of the quark-star sequences shown in
Fig.~\ref{fig:Mvsec_charm} \cite{kettner94:b}.}
\label{fig:MvsR_charm}}}
\end{center}
\end{figure} 
to be present. Stellar sequences containing strange and charm quarks
are shown in Figs.\ \ref{fig:Mvsec_charm} and \ref{fig:MvsR_charm}.  A
stability analysis of these sequences against radial oscillations (see
section \ref{sec:sdwarfs}), however, shows that only the strange star
sequence is stable but not the charm star sequence. Finally, we
mention that a value for the bag constant of $\bag=145$~MeV places the
energy per baryon number of (non-interacting) strange quark matter at
$E/A = 829$~MeV  \cite{farhi84:a}, which corresponds to strange quark
matter strongly bound with respect to $^{56}$Fe whose energy per
baryon is $M(^{56}{\rm Fe})c^2/56=930.4$~MeV, with $M(^{56}{\rm Fe})$
the mass of the $^{56}$Fe atom.

\goodbreak
\subsection{\it Color-Superconductivity}\label{sec:color}

There has been much recent progress in our understanding of quark
matter, culminating in the discovery that if quark matter exists it
ought to be in a color superconducting state
 \cite{rajagopal01:a,alford01:a,alford98:a,rapp98+99:a}. This is made
possible by the strong interaction among the quarks which is very
attractive in some channels. Pairs of quarks are thus expected to form
Cooper pairs very readily. Since pairs of quarks cannot be
color-neutral, the resulting condensate will break the local color
symmetry and form what is called a color superconductor.  The phase
diagram of such matter is expected to be very
complex \cite{rajagopal01:a,alford01:a}, as can be seen from Figs.\
\ref{fig:phase} and \ref{fig:rtwo}.
\begin{figure}[t]
\begin{center}
\parbox{0.45\hsize}{
\begin{center} Light strange quark\\ or large pairing gap \end{center}
 \includegraphics[width=0.8\hsize]{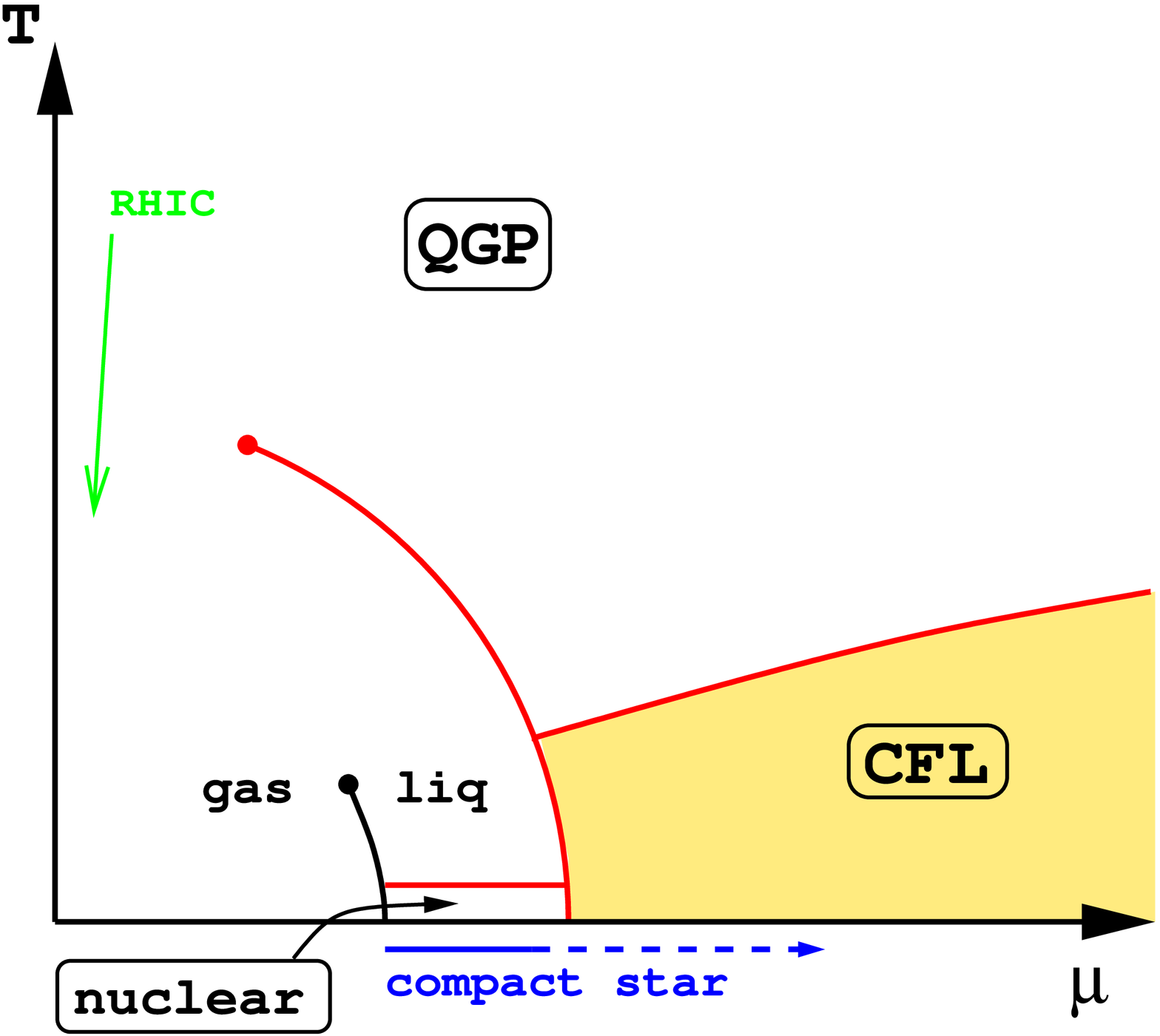}}
\parbox{0.45\hsize}{
\begin{center} Heavy strange quark\\ or small pairing gap \end{center}
 \includegraphics[width=0.8\hsize]{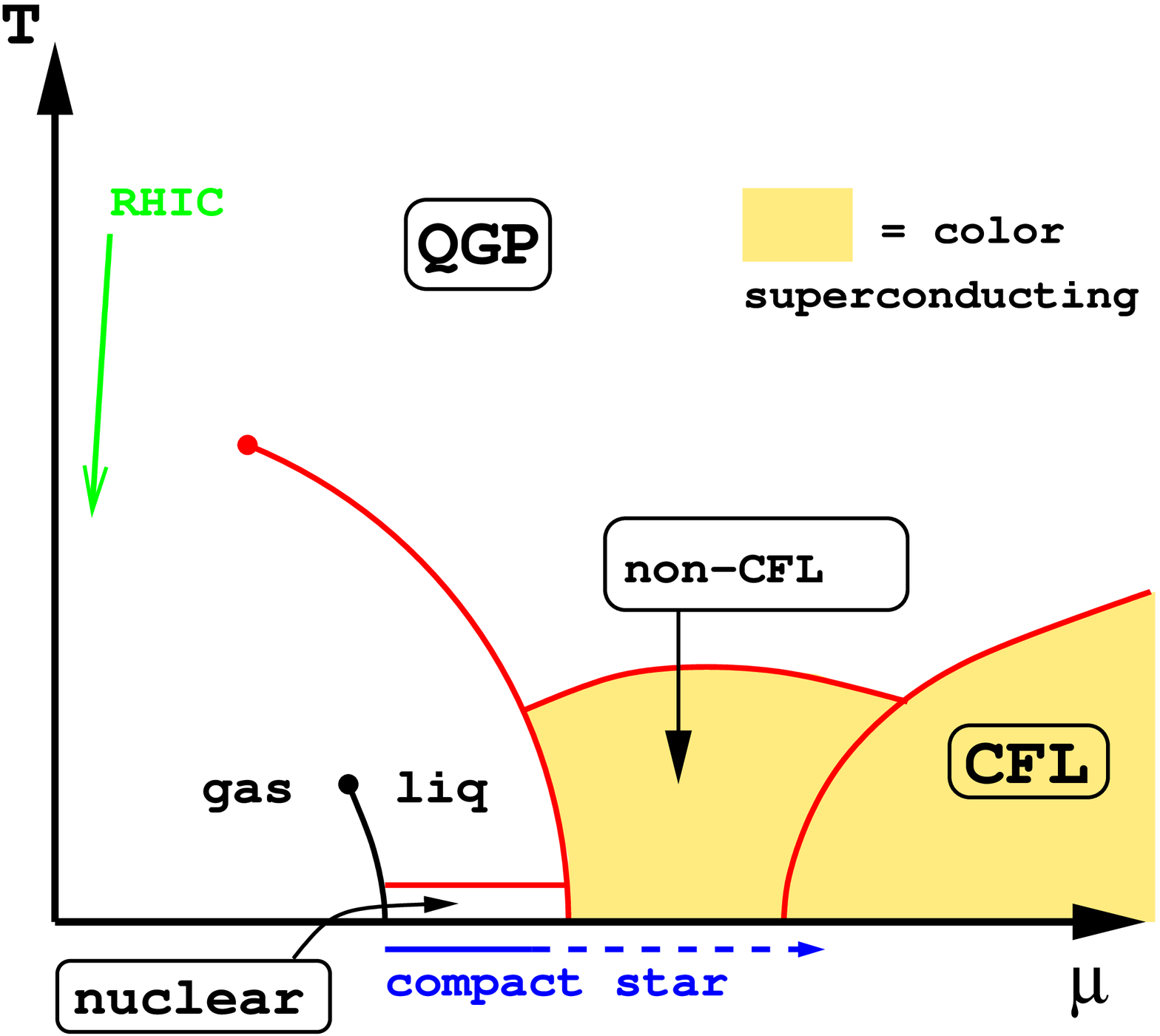}}
\begin{minipage}[t]{16.5 cm}
\caption{Conjectured phase diagram for QCD \cite{alford04:a}.  For
small $m_s^2/\Delta$ there is a direct transition from nuclear matter
to CFL color superconducting quark matter. For large $m_s^2/\Delta$
there is an intermediate phase where condensation patterns like the
CFL-$K^0$, CFL-$K^+$, CFL-$\pi^0$, gCFL, 2SC, 1SC, CSL, and LOFF phase
(see text) may exist. Reprinted figure with permission from M. Alford,
J.~Phys.~G~30 (2004) S441. Copyright 2004 by the Institute of Physics
Publishing.}
\label{fig:phase}
\end{minipage}
\end{center}
\end{figure}
This is caused by the fact that quarks come in three different colors,
different flavors, and different masses.  Moreover, bulk matter is
neutral with respect to both electric and color charge, and is in
\begin{figure}[tb]
\begin{center}
\epsfig{file=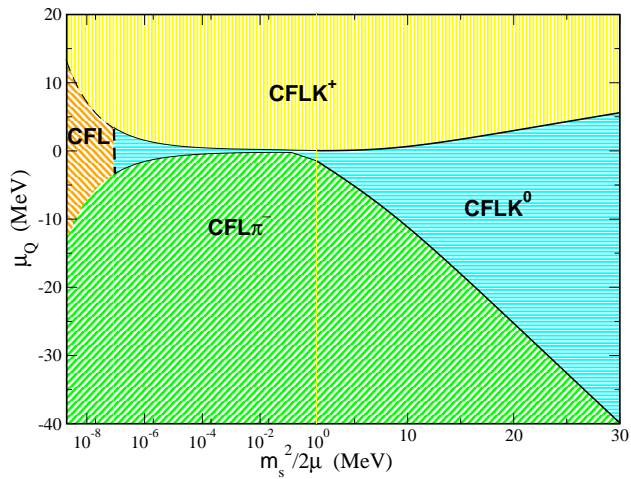,width=7.0cm,angle=-90}
\begin{minipage}[t]{16.5 cm}
\caption{Possible meson condensed phases in the neighborhood of the
symmetric CFL state for a strange quark mass of $m_s = 150$~MeV.
$\mu$ is the quark chemical potential and $\mu_Q$ the chemical
potential for positive electric charge. Solid and dashed lines
indicate first- and second-order transitions
respectively \cite{kaplan02:a}.  Reprinted figure with permission from
D. B. Kaplan and S. Reddy, Phys.~Rev.~D~65 (2002) 054042.  Copyright
2002 by the American Physical Society.}
\label{fig:rtwo}
\end{minipage}
\end{center}
\end{figure}
chemical equilibrium under the weak interaction processes that turn
one quark flavor into another. To illustrate the condensation pattern
briefly, we note the following pairing ansatz for the quark
condensate \cite{alford03:a},
\begin{eqnarray}
\langle \psi^\alpha_{f_a} C\gamma_5 \psi^\beta_{f_b} \rangle \sim
\Delta_1 \epsilon^{\alpha\beta 1}\epsilon_{{f_a}{f_b}1} + \Delta_2
\epsilon^{\alpha\beta 2}\epsilon_{{f_a}{f_b}2} + \Delta_3
\epsilon^{\alpha\beta 3}\epsilon_{{f_a}{f_b}3} \, ,
\label{eq:pairing_ansatz}
\end{eqnarray}
where $\psi^\alpha_{f_a}$ is a quark of color $\alpha=(r,g,b)$ and
flavor ${f_a}=(u,d,s)$. The condensate is a Lorentz scalar,
antisymmetric in Dirac indices, antisymmetric in color, and thus
antisymmetric in flavor. The gap parameters $\Delta_1$, $\Delta_2$ and
$\Delta_3$ describe $d$-$s$, $u$-$s$ and $u$-$d$ quark Cooper pairs,
respectively. The following pairing schemes have emerged. At
asymptotic densities ($m_s \rightarrow 0$ or $\mu \rightarrow \infty$)
the ground state of QCD with a vanishing strange quark mass is the
color-flavor locked (CFL) phase (color-flavor locked quark pairing),
in which all three quark flavors participate symmetrically.  The
gaps associated with this phase  are
\begin{equation}
\Delta_3 \simeq \Delta_2 = \Delta_1 = \Delta \, ,
\label{eq:delta_CFL}
\end{equation}
and the quark condensates of the CFL phase are approximately of the form
\begin{equation}
 \langle \psi^{\alpha}_{f_a} C\gamma_5 \psi^{\beta}_{f_b} \rangle
\sim \Delta \, \epsilon^{\alpha \beta X} \epsilon_{{f_a} {f_b} X}
\, ,
\label{eq:CFL1}
\end{equation}
with color and flavor indices all running from 1 to 3. Since
$\epsilon^{\alpha\beta X} \epsilon_{{f_a} {f_b} X} =
\delta^\alpha_{f_a}\delta^\beta_{f_b} -
\delta^\alpha_{f_b}\delta^\beta_{f_a}$ one sees that the condensate
(\ref{eq:CFL1}) involves Kronecker delta functions that link color and
flavor indices. Hence the notion color-flavor locking. The CFL phase
has been shown to be electrically neutral without any need for
electrons for a significant range of chemical potentials and strange
quark masses \cite{rajagopal01:b}. If the strange quark mass is heavy
enough to be ignored, then up and down quarks may pair in the
two-flavor superconducting (2SC) phase.  Other possible condensation
patterns are CFL-$K^0$ \cite{bedaque01:a}, CFL-$K^+$ and
CFL-$\pi^{0,-}$ \cite{kaplan02:a}, gCFL (gapless CFL phase)
 \cite{alford03:a}, 1SC (single-flavor-pairing)
 \cite{alford03:a,buballa02:a,schmitt04:a}, CSL (color-spin locked
phase)  \cite{schaefer00:a}, and the LOFF (crystalline pairing)
 \cite{alford00:a,bowers02:a,casalbuoni04:a} phase, depending on $m_s$,
$\mu$, and electric charge density.  Calculations performed for
massless up and down quarks and a very heavy strange quark mass ($m_s
\rightarrow \infty$) agree that the quarks prefer to pair in the
two-flavor superconducting (2SC) phase where
\begin{equation}
\Delta_3 > 0\, , \quad {\rm and} \quad  \Delta_2 = \Delta_1 = 0 \, .
\label{eq:delta_2SC}
\end{equation}
In this case the pairing ansatz (\ref{eq:pairing_ansatz}) reduces to
\begin{equation}
 \langle \psi^{\alpha}_{f_a} C \gamma_5 \psi^{\beta}_{f_b} \rangle
\propto \Delta \, \epsilon_{ab} \epsilon^{\alpha \beta 3} \, .
\label{eq:2SC}
\end{equation}
Here the resulting condensate picks a color direction (3 or blue in
the example (\ref{eq:2SC}) above), and creates a gap $\Delta$ at the
\begin{figure}[tb]
\begin{center}
\includegraphics[width=0.4\textwidth]{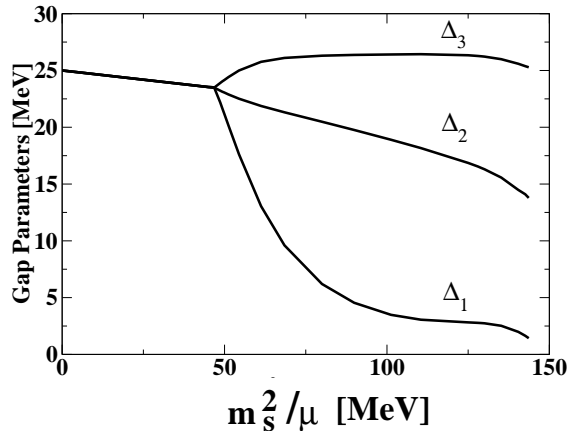}
\begin{minipage}[t]{16.5 cm}
\caption{Gap parameters $\Delta_1$, $\Delta_2$, and $\Delta_3$ as a
  function of $m_s^2/\mu$ for $\mu=500$~MeV, in an NJL
  model \cite{alford03:a}.  There is a second order phase transition
  between the CFL phase and the gapless CFL (gCFL) phase at $m_s^2/\mu
  = 2 \Delta$.  Reprinted figure with permission from M. Alford,
  C. Kouvaris, and K. Rajagopal, Phys.~Rev.~Lett.~92 (2004) 222001.
  Copyright 2004 by the American Physical Society.}
\label{fig:gaps}
\end{minipage}
\end{center}
\end{figure}
Fermi surfaces of quarks with the other two out of three colors (red
and green). The gapless CFL phase (gCFL) may prevail over the CFL and
2SC phases at intermediate values of $m^2_s/\mu$ with gaps given
obeying the relation (see Fig.\ \ref{fig:gaps})
\begin{equation}
\Delta_3 > \Delta_2 > \Delta_1 > 0 \, .
\label{eq:gCFL}
\end{equation}
As shown in Fig.\ \ref{fig:Delta_Kaplan}, for chemical potentials that
are of astrophysical interest(see sections \ref{sec:scc} through
\ref{sec:mqdec}), $\mu < 1000$~MeV, the gap is between 50 and
\begin{figure}[tb]
\begin{center}
\includegraphics[width=0.4\textwidth,angle=-90]{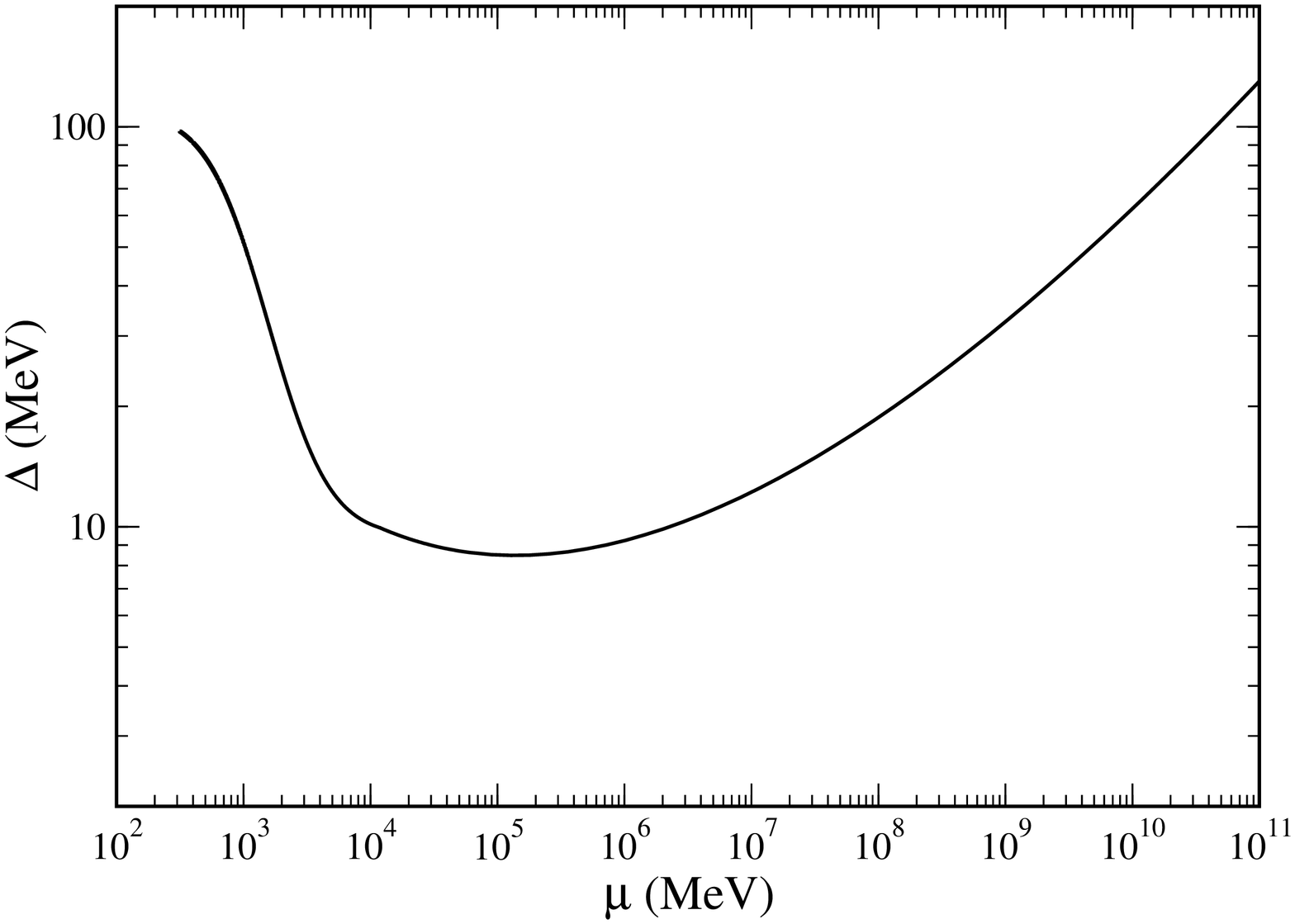}
\begin{minipage}[t]{16.5 cm}
\caption{Superconducting gap $\Delta$ versus quark chemical potential
$\mu$  \cite{kaplan02:a}.  Reprinted figure with permission from
D. B. Kaplan and S. Reddy, Phys.~Rev.~D~65 (2002) 054042.  Copyright
2002 by the American Physical Society.}
\label{fig:Delta_Kaplan}
\end{minipage}
\end{center}
\end{figure}
100~MeV. The order of magnitude of this result agrees with
calculations based on phenomenological effective
interactions \cite{rapp98+99:a,alford99:b} as well as with perturbative
calculations for $\mu > 10$~GeV \cite{son99:a}. We also note that
superconductivity modifies the equation of state at the order of
$(\Delta / \mu)^2$  \cite{alford03:b,alford04:a}, which is even for
such large gaps only a few percent of the bulk energy. Such small
effects may be safely neglected in present determinations of models
for the equation of state of quark-hybrid stars. There has been much
recent work on how color superconductivity in neutron stars could
affect their properties. (See Refs.\
 \cite{rajagopal01:a,alford01:a,rajagopal00:a,alford00:a,alford00:b,%
blaschke99:a} and references therein.)  These studies reveal that
possible signatures include the cooling by neutrino emission, the
pattern of the arrival times of supernova neutrinos, the evolution of
neutron star magnetic fields, rotational stellar instabilities, and
glitches in rotation frequencies. Several of these issues will be
discussed in sections \ref{sec:scc} to \ref{sec:mqdec}.

\goodbreak
\subsection{\it The Strange Quark Matter Hypothesis}\label{ssec:shyp}

The theoretical possibility that strange quark matter may constitute
the true ground state of the strong interaction rather than $^{56}$Fe
was proposed by Bodmer \cite{bodmer71:a}, Witten \cite{witten84:a},
and Terazawa \cite{terazawa89:a}.  A schematic illustration of this
so-called strange matter hypothesis is given in Fig.~\ref{fig:uds},
which compares the energy per baryon of $^{56}{\rm Fe}$ and infinite
nuclear matter with the energy per baryon of two and three-flavor
quark matter.  Three-flavor quark matter is always lower in energy
than two-flavor quark matter due to the extra Fermi-well accessible to
the strange quarks. Theoretical arguments indicate that the energy of
three-flavor quark matter may be even smaller than 930~MeV in which
case strange matter would be more stable than nuclear matter and
atomic nuclei. This peculiar feature makes the strange matter
hypothesis to one of the most startling speculations of modern
physics, which, if correct, would have implications of fundamental
importance for our understanding of the early universe, its evolution
in time to the present day, astrophysical compact objects and
laboratory physics, as summarized in table \ref{tab:over}
\cite{farhi84:a,alcock86:a,madsen98:b,aarhus91:proc}.  In the
following we describe the possible absolute stability of strange quark
matter for a gas of massless $u, d, s$ quarks inside a confining bag
at zero temperature \cite{farhi84:a}.  For a massless quark-flavor
$f$, the Fermi momentum, $\pFf$, equals the chemical potential,
$\mu^f$. The number densities, energy densities and corresponding
pressures, therefore, follow from Eq.\ (\ref{eq:eosqm.0}) as
\begin{eqnarray}
\rho^f = (\mu^f)^3/\pi^2 \, , \quad \epsilon^f = 3 (\mu^f)^4/4\pi^2 \,
, \quad P^f = (\mu^f)^4/4\pi^2 = \epsilon^f/3 \, .
\label{eq:nep_sqm}
\end{eqnarray}
For a gas of massless $u$ and $d$ quarks the condition of electric
charge neutrality $2\rho^u - \rho^d = 0$, which follows from Eq.\
(\ref{eq:ladn}) as, requires that $\rho^d = 2 \rho^u$. Hence the
chemical potential of two-flavor quark matter is given by $\mu_2
\equiv \mu^u = \mu^d / 2^{1/3}$.  The corresponding two-flavor quark
pressure then follows as $P_2 \equiv P^u + P^d = (1 + 2^{4/3})
\mu_2^4/4\pi^2 = B$. From the expressions for the total energy density,
$\epsilon_2 = 3P_2 + B = 4B$, and baryon number density, $\rho_2 =
(\rho^u + \rho^d)/3 = \mu_2^3/\pi^2$, one then obtains for the energy
per baryon of two-flavor quark matter,
\begin{eqnarray}
  {E\over{A}}\biggr|_2 \equiv {\epsilon_2\over{\rho_2}} = {{4
      B}\over{\rho_2}} = 934~{\rm MeV} \times \bag_{145} \, ,
\label{eq:EA.2flavor}
\end{eqnarray} where $\bag_{145} \equiv \bag/145~{\rm MeV}$.  Values for $\bag$
smaller than 145~MeV are excluded. Otherwise two-flavor quark matter
would have a lower energy than $^{56}$Fe, which then would be made up
of up and down quarks, in contradiction to what is observed. In the
next step we add massless strange quarks to the system. The
three-flavor quark gas is electrically neutral for $\rho^u = \rho^d =
\rho^s$, i.e.\ $\mu_3 \equiv \mu^u = \mu^d = \mu^s$.  For a fixed bag
constant, the three-flavor quark gas should exert the same pressure as
the two-flavor gas, that is $P_3 = P_2$.  This implies for the
chemical potentials $\mu_3 = ((1 + 2^{4/3})/3)^{1/4} \mu_2$.  Hence,
the total baryon number in this case can be written as $\rho_3 =
\mu_3^3 / \pi^2 = ((1 + 2^{4/3})/3)^{3/4} \rho_2$.  The energy per
baryon is therefore given by
\begin{equation}
  {E\over{A}}\biggr|_3 \equiv {\epsilon_3\over{\rho_3}} = {{4B}\over{1.127
      \, \rho_2}} = 829~{\rm MeV} \times B^{1/4}_{145} \, ,
\label{eq:EA.3flavor}
\end{equation} since $\epsilon_3 = 3P_3 + B = 4B = \epsilon_2$.  It thus
follows that the energy per baryon of a massless non-interacting
three-flavor quark gas is of order 100~MeV per baryon lower than for
two-flavor quark matter. The difference arises from the fact that
baryon number can be packed denser in three-flavor quark matter,
$\rho_2/\rho_3 = (3/(1 + 2^{4/3}))^{3/4} \simeq 0.89$, due to the
extra Fermi-well accessible to the strange quarks.  The energy per
\begin{figure}[tb]
\begin{center}
\epsfig{figure=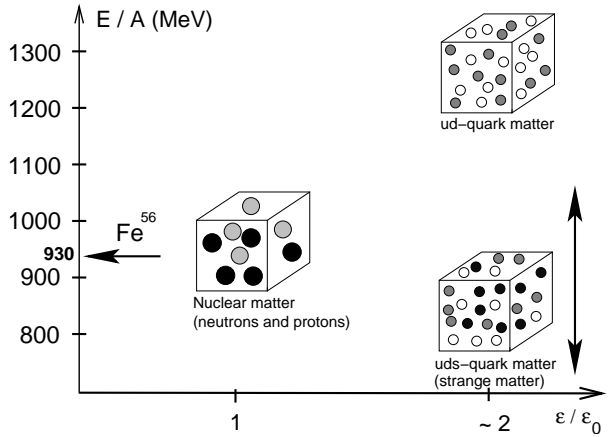,width=8.0cm,angle=0}
\begin{minipage}[t]{16.5 cm}
\caption{Comparison of the energy per baryon of $^{56}{\rm Fe}$ and
  nuclear matter with the energy per baryon of 2-flavor ($u,d$ quarks)
  and 3-flavor ($u, d, s$ quarks) strange quark matter. Theoretically
  the energy per baryon of strange quark matter may be below 930~MeV,
  which would render such matter more stable than nuclear matter.}
\label{fig:uds}
\end{minipage}
\end{center}
\end{figure}
baryon in a free gas of neutrons is equal to the neutron mass, $E/A =
939.6$~MeV.  For an $^{56}{\rm Fe}$ nucleus the energy per baryon is
$E/A = (56 m_N - 56 \times 8.8~{\rm MeV})/56 = 930~{\rm MeV}$, where
$m_N=938.9$ denotes the nucleon mass.  Stability of two-flavor quark
matter relative to neutrons thus corresponds to $(E/A)_2 < m_n$, or
$\bag < 145.9$~MeV ($\bag < 144.4$~MeV for stability relative to
$^{56}{\rm Fe}$). The stability argument is often turned around
\begin{table}[tb]
\caption{Strange matter phenomenology.}\label{tab:over}
\begin{center}
\begin{tabular}{@{}l|l@{}} \hline
Phenomenon                                             &References$^{\rm a}$ \\ \hline
Centauro cosmic ray events                             & \cite{bjorken79:a,chin79:a,chinellato90:a,wilk96:a}                 \\
High-energy gamma ray sources:                         &                           \\
~~~~Cyg~X-3 and Her~X-3                                & \cite{jaffe77:a,baym85:a} \\  
Strange matter hitting the Earth:                      &                   \\
~~~~Strange meteors                                    & \cite{rujula84:a}  \\
~~~~Nuclearite-induced earthquakes                     & \cite{rujula84:a,larousserie02:a}  \\
~~~~Unusual seismic events                             & \cite{anderson03:a} \\
~~~~Strange nuggets in cosmic rays                     & \cite{terazawa91:a,terazawa93:a,gladysz97:a,rybczynski00:a,rybczynski01:a,madsen03:paper+reply}  \\
Strange matter in supernovae                           & \cite{michel88:a,benvenuto89:a,horvath92:a} \\
Strange star phenomenology     & \cite{weber99:book,alcock86:a,alcock88:a,glen92:crust,weber93:b,kettner94:b,aarhus91:proc,haensel86:a,usov01:a,haensel89:a} \\
Strange dwarfs:                                        &                                  \\
~~~~Static properties and stability                    & \cite{weber93:b,weber95:a,romanelli86:a}                              \\
~~~~Thermal evolution                                  & \cite{althaus96:a}              \\
Strange planets                                        & \cite{weber93:b,glen94:a,weber95:a}                   \\
Strange MACHOS                                         & \cite{weber97:vulcano}                 \\
Strangeness production in dense stars                  & \cite{ghosh96:a}                \\
Burning of neutron stars to strange stars              & \cite{olinto87:a,olinto89:a,horvath88:a}                           \\
Gamma-ray bursts, Soft Gamma Repeaters                 & \cite{alcock86:a,usov01:a,horvath93:a,usov01:c,zhang00:a,usov01:b,ouyed04:a,usov98:a,cheng03:a}                  \\
Cosmological aspects of strange matter                 & \cite{witten84:a,madsen88:a,madsen86:a,madsen91:a,cho94:a}    \\
Strange matter as compact energy source                & \cite{shaw89:a} \\
Strangelets in nuclear collisions                      & \cite{shaw84:a,greiner87:a,greiner88:a} \\
\hline
\end{tabular}
\begin{minipage}[t]{16.5 cm}
\vskip 0.1cm
\noindent
$^{\rm a}$ Numerous additional references are provided in the text. 
\end{minipage}
\vskip -0.4cm
\end{center}
\end{table}  
because one observes neutrons and $^{56}{\rm Fe}$ in nature rather
than two-flavor quark matter. Hence the bag constant must be larger
than about 145~MeV.  Bulk three-flavor quark matter is absolutely
stable relative to $^{56}{\rm Fe}$ for $\bag < 162.8$~MeV, metastable
relative to a neutron gas for $\bag < 164.4$~MeV, and metastable
relative to a gas of $\Lambda$ particles for $\bag < 195.2$~MeV. These
numbers are upper limits. A finite strange-quark mass as well as a
non-zero strong coupling constant decrease the limits on $\bag$
\cite{farhi84:a,madsen98:b}.  The presence of ordinary nuclei in
nature is not in contradiction to the possible absolute stability of
strange matter. The reason being that conversion of an atomic nucleus
of baryon number $A$ into a lump of strange quark matter requires the
simultaneous transformation of roughly $A$ up and down quarks into
strange quarks.  The probability for this to happen involves a weak
transition $\propto G_F^{2 A}$ which makes nuclei with $A \gs 6$
stable for more than $10^{60}$~years.
\begin{figure}[tb]
\begin{center}
\epsfig{file=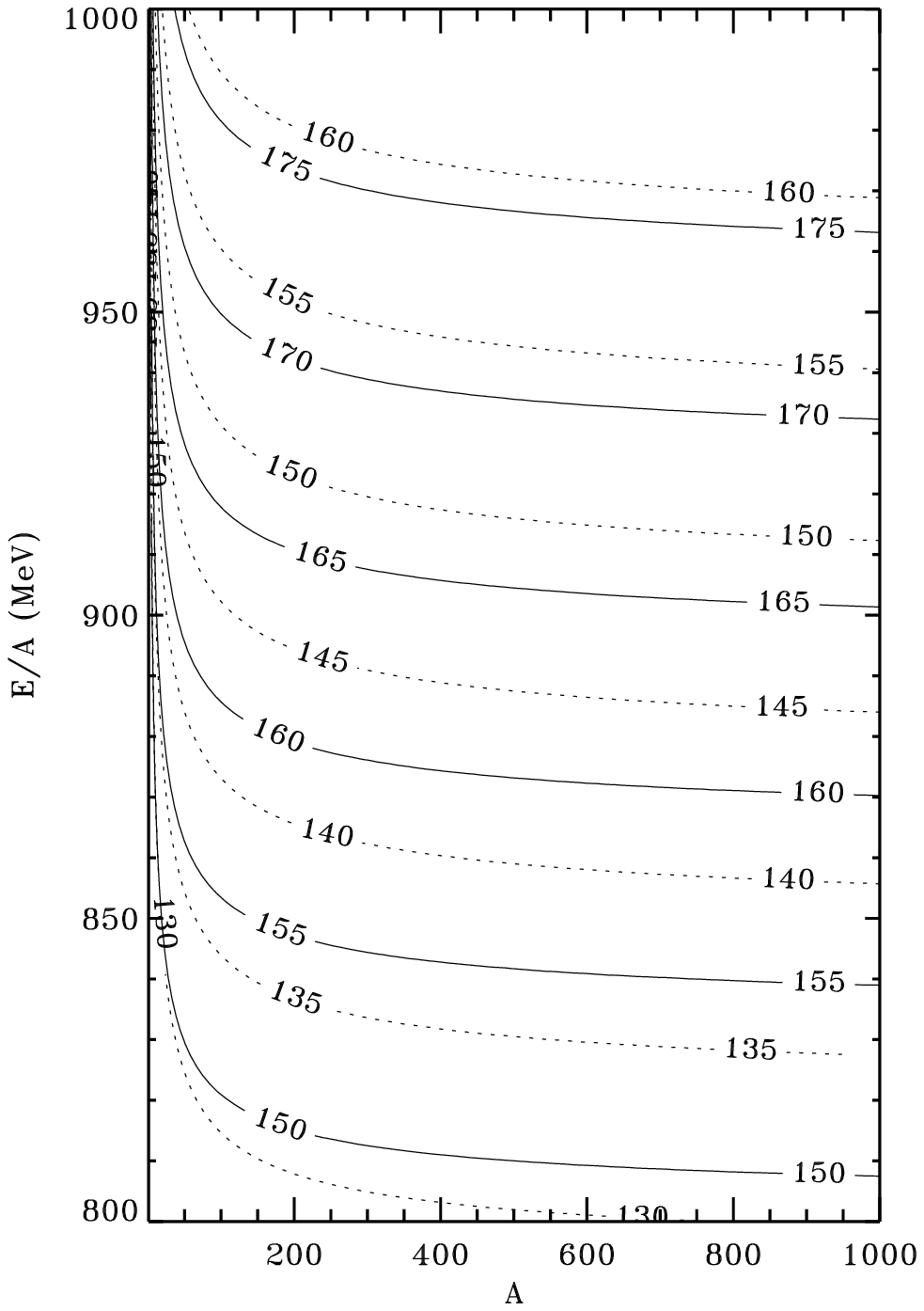,width=8.0cm}
\begin{minipage}[t]{16.5 cm}
\caption{Energy per baryon in MeV as a function of $A$ for ordinary
strangelets (dashed curves) and CFL strangelets (solid curves) for
$B^{1/4}$ in MeV as indicated, $m_s=150$~MeV, and
$\Delta=100$~MeV \cite{madsen01:a}.  Reprinted figure with permission
from J. Madsen, Phys.~Rev.~Lett.~87 (2001) 172003.  Copyright 2001 by
the American Physical Society.}
\label{fig:eovera}
\end{minipage}
\vskip -0.4cm
\end{center}
\end{figure}
The conversion of very light nuclei into strange matter is determined
by finite-size and shell effects which dominate over the volume energy
of strange matter at small $A$ values.  An example for the mass
formula of strange matter is \cite{madsen98:b,berger87:a,gilson93:a}
\begin{equation}
  {E\over{A}} \simeq \left( 829~{\rm MeV} + 351~{\rm MeV} ~ A^{-2/3}
\right) ~ \bag_{145} \, ,
\label{eq:18.Mad93}
\end{equation}
in which case strange matter becomes absolutely stable for $A > 6$.
If quark matter is in the CFL phase, metastability or even absolute
stability of strange quark matter may become more likely than hitherto
thought since the binding energy from pairing of the quarks should
reduce the energy of the system by a contribution proportional to
$\Delta^2$  \cite{madsen01:a}. Figure \ref{fig:eovera} shows the energy
per baryon for ordinary quark matter and CFL quark matter. For high
$A$ values a bulk value is approached, but for low $A$ the finite-size
contributions (surface tension and curvature) increase the energy per
baryon significantly. The pairing contribution is on the order of
100~MeV per baryon for $\Delta \approx 100$~MeV for fixed values of
the strange quark mass and bag constant.  Another crucial difference
between non-CFL and CFL quark matter is the equality of all quark
Fermi momenta in CFL quark matter which leads to charge neutrality in
bulk without any need for electrons \cite{rajagopal01:b}. This has most
important consequences for the charge-to-mass ratios of strangelets.
For non-CFL strangelets one has
\begin{equation}
Z\approx 0.1 \left( m_s\over {150~{\rm MeV}} \right)^2 A ~~{\rm for}
~~A\ll 10^3 \, , \quad {\rm and} \quad Z\approx 8 \left( m_s\over
{150~{\rm MeV}} \right)^2 A^{1/3} ~~ {\rm for} ~~A\gg 10^3 \, ,
\end{equation}
while, in contrast to this, CFL strangelets have a charge-to-mass
ratio of \cite{madsen01:a}
\begin{equation}
Z\approx 0.3 \left( m_s\over {150~{\rm MeV}}\right) A^{2/3} \, .
\end{equation}
This difference may provide a test of color superconductivity by
upcoming cosmic ray space experiments such as AMS \cite{ams01:homepage}
and ECCO \cite{ecco01:homepage} (see section \ref{sec:ams+ecco} and
table \ref{tab:labexp}).

\goodbreak
\subsection{\it Searches for Strange Quark Matter}\label{sec:searchexp}

Experimental physicists have searched unsuccessfully for stable or
quasistable strange matter systems over the past two decades. These
searches fall in three main categories: (a) searches for strange
matter (strange nuggets or strangelets) in cosmic rays, (b) searches
for strange matter in samples of ordinary matter, and (c) attempts to
produce strange matter at accelerators.  An overview of these search
experiments is given in table~\ref{tab:labexp}.
\begin{table}[tb]
\begin{center}
\begin{minipage}[t]{16.5 cm}
\caption{Past, present, and future search experiments for strange quark
matter.}\label{tab:labexp}
\end{minipage}
\begin{tabular}{l|l} \hline
Experiment                                 &References                                                \\ \hline
Cosmic ray searches for strange nuggets:   &                                                          \\
~~~~~ AMS-02$^{\rm a}$                     & \cite{ams01:homepage,sandweiss04:a}                       \\
~~~~~ CRASH$^{\rm b}$                      & \cite{saito90:a,saito94:a,ichimura93:a}                   \\
~~~~~ ECCO$^{\rm c}$                       & \cite{ecco01:homepage}                                    \\
~~~~~ HADRON                               & \cite{shaulov96:a}                                        \\
~~~~~ IMB$^{\rm d}$                        & \cite{rujula83:a}                                         \\
~~~~~ JACEE$^{\rm e}$                      & \cite{miyamura95:a,lord95:a}                              \\
~~~~~ MACRO$^{\rm f}$                      & \cite{macro92:a,ambrosio00:a,ambrosio02:a,giacomelli02:a} \\
Search for strangelets in terrestrial matter: & \cite{lu04:a}                                          \\
~~~~~ Tracks in ancient mica                     & \cite{rujula84:a,price84:a}                         \\ 
~~~~~ Rutherford backscattering                  & \cite{bruegger89:a,isaac98:a}                       \\
Search for strangelets at accelerators:    &                                                          \\
~~~~~ Strangelet searches E858, E864, E878, E882-B, E896-A, E886  & \cite{thomas95:a,rusek96:a,buren99:a}           \\
~~~~~ H-dibaryon search                    & \cite{belz96:a,belz96:b}                                               \\
~~~~~ Pb+Pb collisions                     & \cite{dittus95:a,appelquist96:a,ambrosini96:a,klingenberg99:topr}      \\
\hline
\end{tabular}
\begin{minipage}[t]{16.5 cm}
\vskip 0.1cm
\noindent
$^{\rm a}$ AMS: Alpha Magnetic Spectrometer (scheduled for 2005-2008).  \\
$^{\rm b}$ CRASH: Cosmic Ray And Strange Hadronic matter.                \\
$^{\rm c}$ ECCO: Extremely-heavy Cosmic-ray Composition Observer.        \\
$^{\rm d}$ IMB: Irvine Michigan Brookhaven proton-decay detector (1980-1991). \\
$^{\rm e}$ JACEE: Japanese-American Cooperative Emulsion Chamber Experiment.  \\
$^{\rm f}$ MACRO: Monopole, Astrophysics and Cosmic Ray Observatory (1989-2000). 
\end{minipage} 
\vskip -0.4cm
\end{center}
\end{table}
The experiments searching for nuggets of strange matter which got
stuck in terrestrial matter focus on objects whose masses can range
from those of atomic nuclei to the upper limit of about $0.3\times
10^{-9}$~g. The latter carry a baryon number of $A\sim 2\times
10^{14}$ and have a radius of $R\sim 10^{-8}~\cm$. Strange nuggets
heavier than $0.3\times 10^{-9}$~g will not be slowed down and stopped
in the crust of the Earth. Finally nuggets of more than $\sim 10^{22}$
quarks (i.e. $A \sim 10^{21}$) would have too much momentum to be
stopped by the encounter and thus would pass through the Earth.  Such
encounters could take the form of unusual meteorite events, Earth
quakes, and peculiar particle tracks in ancient mica, in meteorites
and in cosmic ray detectors \cite{rujula84:a}. One distinguishing
feature of unusual meteorite events caused by strange nuggets could be
the apparent magnitude of $-1.4$ associated with a 20-gram nugget at a
distance of 10~km, which would rival with that of the brightest star,
Sirius. Another distinguishing feature could be the meteorite's
velocity which is smaller than about 70~km/s for an ordinary meteorite
bound to the solar system, but amounts to about 300~km/s for a strange
meteorite of galactic origin. An upper limit on the flux of cosmic
strange nuggets can be derived by assuming that the galactic dark
matter halo consists entirely of strange nuggets.  The expected flux
at the Earth is then on the order of $10^{6} \, A^{-1}\, v_{\rm 250}
\, \rho_{24} \; {\rm cm}^{-2}\, {\rm s}^{-1} \, {\rm sterad}^{-1}$,
where $\rho_{24} = \rho / (10^{-24}~ \gcmt)$ and $v_{\rm 250}$ is the
speed in units of 250~km/s \cite{madsen98:b}.  Experiments sensitive at
this flux level or better have been able to rule out quark nuggets as
being the dark matter for baryon numbers $10^8 < A < 10^{25}$
 \cite{aarhus91:proc,price88:a}.  This however does not rule out a low
flux level either left over from the Big Bang or arising from
collisions of strange stars. If the strange matter hypothesis is
valid, one should indeed expect a significant background flux of
nuggets from collisions of strange stars in binary systems, which are
ultimately colliding because of the loss of angular momentum emitted
from the binary system as gravitational radiation. If such collisions
spread as little as $0.1\,\msun$ of strangelets with baryon number $A
~ (\sim 10^3)$, a single collision will lead to a flux of $10^{-6} \,
A^{-1} \, v_{\rm 250} \; {\rm cm}^{-2} \, {\rm s}^{-1} \, {\rm
sterad}^{-1}$ \cite{madsen98:b}, assuming that the nuggets are spread
homogeneously in a galactic halo of radius 10~kpc.  This would lead to
a concentration of nuggets in our galaxy of less than
$10^{-8}$~nuggets/cm$^3$, translating to a nugget concentration in
terrestrial crust matter of at most $10^{9}~\cmmt$ (mass density $\sim
10^{-29}~\gcmt$). Such a nugget density would corresponds to a
concentration of nuggets per nucleon, $N_{\rm strange}/N_{\rm
nucleons}$, that is much less than $10^{-14}$ \cite{glen91:a}.  The
upper limit on the concentration of strange nuggets per nucleon in
terrestrial matter established experimentally by Br{\"u}gger \etal\
 \cite{bruegger_etal89:a} and Perillo Isaac \etal\  \cite{isaac98:a} is
$10^{-14}$, which falls short of the upper limits that follow from
strange star-strange star collisions as well as he flux of strange
nuggets in cosmic rays.  Consequently the results of Br{\"u}gger
\etal\  \cite{bruegger_etal89:a} and Perillo Isaac \etal\
 \cite{isaac98:a} shown in Fig.\ \ref{fig:smfig2}
\begin{figure}[tb]
\begin{center}
\epsfig{file=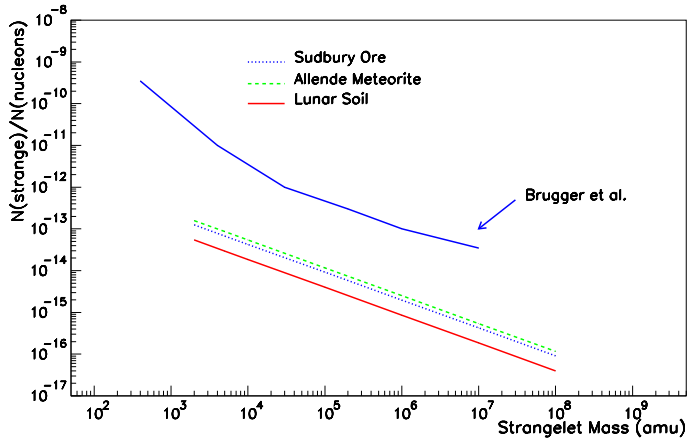,width=10.0cm}
\begin{minipage}[t]{16.5 cm}
\caption{Experimental limits on the concentration of strangelets per
nucleons, $N_{\rm strange} / N_{\rm nucleons}$, contained in three
studied samples \cite{isaac98:a}. These are a meteorite, terrestrial
nickel ore, and lunar soil. The results from Br{\"u}gger \etal\
 \cite{bruegger_etal89:a} obtained with an iron meteorite are shown for
comparison.  Reprinted figure with permission from M. C. Perillo Isaac
{\it et al.}, Phys.~Rev.~Lett. 81 (1998) 2416. Copyright 1998 by the
American Physical Society.}
\label{fig:smfig2}
\end{minipage}
\end{center}
\end{figure}
do not rule out the existence of strange matter.  This figure shows
the results for three samples, a heavy ion activation experiment
mostly sensitive to light strangelets ($A < 10^9$) which, if present
as cosmic rays, would have been absorbed into the Earth's
atmosphere. This would be different for the Moon, which has no
atmosphere. Since its surface has been exposed to cosmic rays for
millions of years, the upper limit of the concentration of strange
matter in the lunar soil can be used to deduce a limit for the flux of
impinging strangelets. The lunar sample was collected from the top 0.5
to 1~cm surface, at the base of the Sculptured Hills, Station 8. The
presence of high cosmic ray track densities in the sample suggests
that the integrated lunar surface exposure age is about 100~Myr. Using
the range
\begin{figure}[tb]
\begin{center}
\epsfig{file=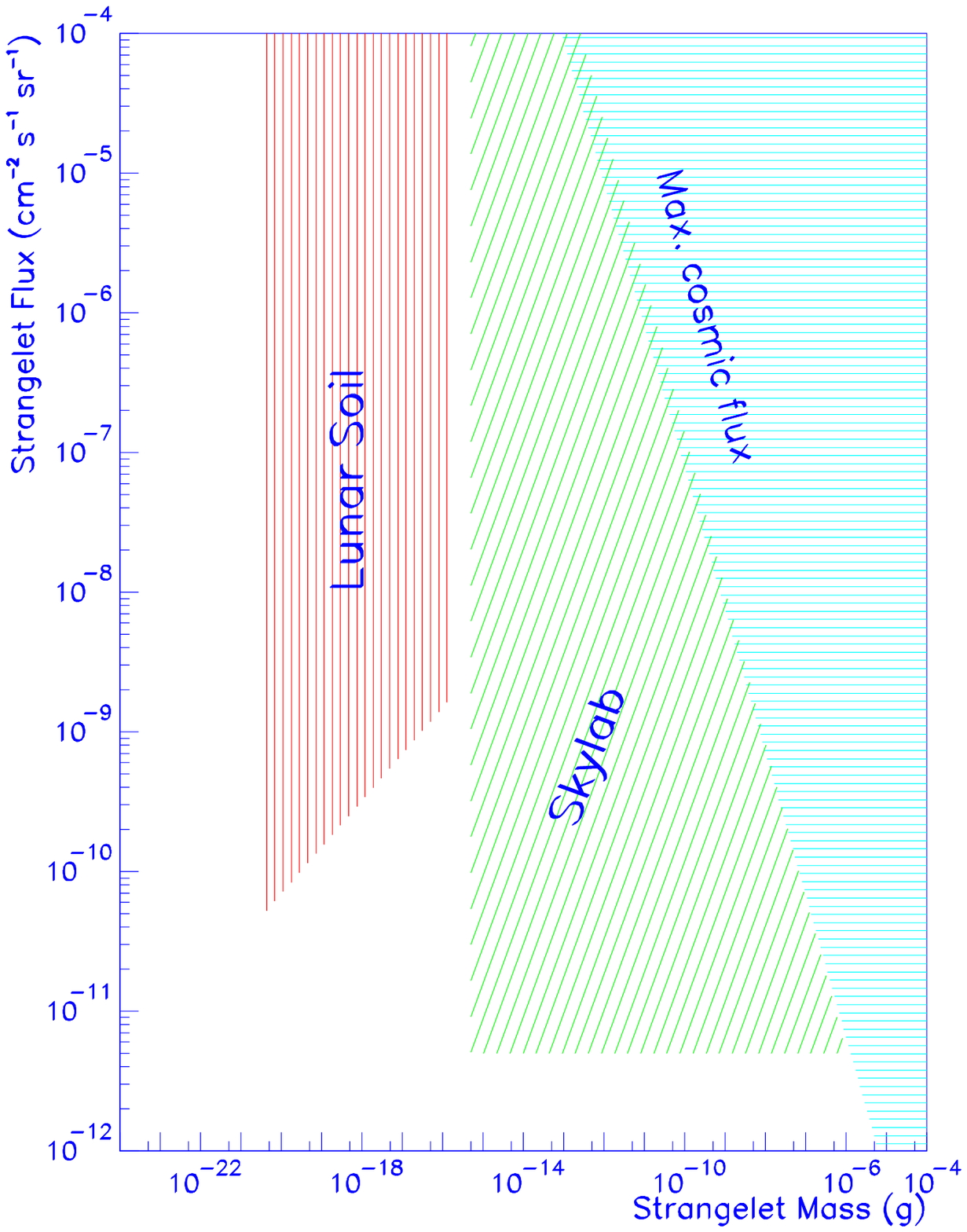,width=10.0cm}
\begin{minipage}[t]{16.5 cm}
\caption{Limits on the flux of strangelets impinging on the lunar
surface \cite{isaac98:a}. Maximum cosmic flux refers to the cosmic flux
of strangelets, assuming that all of the dark matter in the Universe
is composed of strangelets.  Reprinted figure with permission from
M. C. Perillo Isaac {\it et al.}, Phys.~Rev.~Lett. 81 (1998)
2416. Copyright 1998 by the American Physical Society.}
\label{fig:smfig3}
\end{minipage}
\end{center}
\end{figure}
of strange matter in normal matter suggested in \cite{rujula84:a}, 
a limit on the flux of strangelets on the surface of the Moon was deduced
in \cite{isaac98:a} and is shown in Fig.\ \ref{fig:smfig3}.

A limit on the total amount of strange matter in the Universe follows
from the observed abundance of light isotopes. This is so because the
strange nuggets formed in the Big Bang would have absorbed free
neutrons which reduces the neutron to proton ratio, $N_n/N_p$. This
effect in turn would lower the rate of production of the isotope
$^4{\rm He}$, whose abundance is well known from observation.  For a
given mass of the strange nuggets, this constrains their total surface
area.  To be consistent with the missing dark matter, assumed to be
strange quark matter, and the observed abundance of light isotopes,
the primordial quark nuggets had to be made of more than $\sim
10^{23}$ quarks. According to what has been said above, quark nuggets
that massive are not stopped by the Earth.

As summarized in table \ref{tab:labexp}, during the past years several
experiments have been using high energy heavy-ion collisions to
attempt to create strange quark matter (strangelets) in the
laboratory.  The detection of strangelets in relativistic heavy-ion
collisions is conceptually rather simple, because of the strangelet's
low $Z/A$ ratio.  Technically, however, such an attempt is very
difficult because of several reasons. First of all one has to defeat
the finite-number destabilizing effect of strange nuggets, that is,
the number of free quarks produced in a nuclear collision must be
sufficiently large enough such that surface and shell effects do not
dominate over the volume energy of strange matter, destroying its
possible absolute stability. Moreover, since the dense and hot matter
is produced for $\sim 10^{-22}~\rms$, there is no time to develop a
net strangeness.  However it is believed that strange nuggets will
result from two types of simultaneous fluctuations that separate
strange and antistrange quarks, and that also cool the nuggets so that
they do not evaporate.  Finally, we mention the huge multiplicity of
particles produced in such collisions, which makes the particle
identification rather cumbersome \cite{kumar94:crete}.  The NA52
(Newmass) experiment has been searching for long-lived strangelets as
well as for antinuclei in Pb-Pb collisions at CERN SPS.  No evidence
for the production of long-lived charged strangelets has been
observed.  One very intriguing candidate for a strangelet of mass
$m=(7.4 \pm 0.3)$~GeV, electric charge $Z=-1$, and laboratory lifetime
$\tau > 0.85\times 10^{-6}~\sec$ was detected in the data of the 1996
run \cite{pretzl97:a}. This object, which could not be confirmed
however, could have been made of $6u+6d+9s$ quarks, carrying a baryon
number of $A=7$, or $7u+7d+10s$ quarks ($A=8$).

Evidence for the possible existence of strange matter in cosmic rays
may come from Centauro cosmic ray events \cite{bjorken79:a,chin79:a,%
chinellato90:a,wilk96:a,gladysz95:a}. Such events are seen in mountain
top emulsion chamber experiments.  The typical energy of such an event
is of order $\sim 10^3~\tev$, and the typical particle multiplicity is
50 to 100 particles.  Several intriguing Centauro events have been
reported from a Brazilian-Japanese collaboration first several decades
ago, where an interaction in the air 50 to about 500 meters above the
detector gave rise to a large number of charged hadrons and zero or
very few photons or electrons \cite{lattes80:a}. In one particularly
striking Centauro event, 49 hadrons were observed to interact in the
detector but only one photon or electron.  The typical transverse
momentum (poorly determined) of about $\sim 1~\gev$ seems to be larger
than that for a typical event of the same energy. The striking feature
is that there seem to be no photons produced in the primary
interaction which makes the Centauro. This is unusual because in
high-energy collisions, $\pi^0$ mesons are always produced, and they
decay into photons.  A Centauro event is much like a nuclear
fragmentation. If a nucleus were to fragment, then there would be many
nucleons, and if the interaction which produced the fragmentation was
sufficiently peripheral, there would be few pions. This possibility is
ruled out because the typical transverse momentum is so large, and
more important because a nucleus would not survive to such a great
depth in the atmosphere.  Being much more tightly bound together than
an ordinary nucleus, a strangelet with a baryon number around $A\sim
10^3$ explains many of these unusual features.  So it is conceivable
that strangelets incident upon the top of the atmosphere or produced
at the top of the atmosphere could survive to mountain altitude. It
may have lost a significant amount of baryon number before getting to
this depth however. A peripheral interaction might be sufficient to
unbind it, since it certainly will not be so tightly bound with
reduced baryon number. The problem with this explanation is that it
does not explain the high transverse momenta.  At transverse momenta
of $\sim 1~\gev$ one would expect final-state interactions to generate
some pions, and therefore an electromagnetic component, which as
mentioned above, is not observed \cite{mclerran90:peniscola}.  Strange
matter constitutes not the only explanation of Centauros.  One
alternative explanation would be that Centauro (and Anti-Centauro)
events are manifestations of disoriented chiral
condensates \cite{bjorken92:a}. Another interpretation (of the
Chacaltaya Centauro events) suggests that they are due to fragments of
heavy primary cosmic rays. However the survival probability of heavy
primary nuclei to this depth in the atmosphere appears much too low to
account for the number of Centauros reported.

Besides the peculiar Centauro events which may act as agents for
strange matter, the high-energy $\gamma$-ray sources Cygnus~X-3 and
Hercules~X-1 may give additional independent support for the existence
of strange matter.  The reason is that air shower experiments on
Her~X-1 and Cyg~X-3 indicate that these air showers have a muon
content typical of cosmic rays. This muon content is a surprising
result.  Typical cosmic rays at energies between 10 and $10^5~\tev$
are protons. To arrive from the direction of Cyg~X-1 or Her~X-1, the
particle must be electrically neutral. To survive the trip, it must be
long-lived. The only known particle which can quantitatively satisfy
this constraint is the photon. Photon air showers however have only a
small muon component. Photons shower by producing electron pairs.
When only underground data was known, it was proposed that the most
likely candidate for the initiating particle is a hadron, and in order
for interstellar magnetic fields not to alter its direction relative
to that of the source, the hadron--known in the literature as the
cygnet--must be neutral. A natural candidate for the cygnet appears to
be the H particle (see section \ref{sec:hdibaryons}), the electrically
neutral strangeness-2 di-baryons with the quantum numbers of two
lambdas ($Z=0$, $A=2$) proposed by Jaffe \cite{jaffe77:a}. In the
theory of hadrons composed of colored quarks and antiquarks,
combinations other than the usual $qqq$ and $q\bar q$ are allowed as
long as they are color singlets.  Jaffe found that a six-quark
$uuddss$ color-singlet state (H) might have strong enough
color-magnetic binding to be stable against strong decay.  That is,
$\mH$ could be less than the strong-decay threshold, twice the
$\Lambda^0$ ($uds$) mass, $\mH<2 m_{\Lambda^0}$. Estimated lifetimes
for H range from $\sim 10^{-10}~\rms$ for $\mH$ near the
$\Lambda^0\Lambda^0$ threshold to $>10^7$ for light H particles near
the $nn$ threshold.  The potentially long lifetimes raise the
possibility that H particles may be present as components of existing
neutral particle beams (e.g.\ E888 experiment listed in
\ref{tab:labexp}). To make the H long-lived enough, it is necessary to
make the H have a mass below single weak decay stability. To generate
a large enough flux of H particles, the source is assumed to be a
strange star.  Studies of Her~X-1 however seem to rule out this
hypothesis, since studies of the correlation in arrival time with the
known period of Her~X-1 give an upper limit of the particle mass of
about 100~MeV.  The source of radiation must be either due to
anomalous interactions of photons or neutrinos, or from some exotic as
yet undiscovered light-mass, almost stable particles.  The problem
with Cyg~X-3 may be that it is accreting mass and thus has a crust,
such that there is no exposed strange matter surface where small
strangelets could be produced and subsequently accelerated
electrodynamically to high energies into the atmosphere of the
companion star where H particles were created via spallation
reactions.

Anomalously massive particles, which could be interpreted as
strangelets, have been observed in a number of independent cosmic ray
experiments \cite{wilk96:a}.  Two such anomalous events, which are
consistent with electric charge values $Z\simeq 14$ and baryon numbers
$A\simeq 350$, have been observed by a balloon-borne counter
experiment devoted to the study of primary cosmic rays by Saito \etal
\  \cite{saito90:a}. A balloon-borne experiment carried out by the
Italian/Japanese CRASH (table \ref{tab:labexp}) collaboration,
however, could not confirm the existence of such objects in cosmic
rays \cite{saito94:a}. Evidence for the presence of strangelets in
cosmic rays has also been pointed out by
Shaulov \cite{shaulov96:a,shaulov04:a}. This experiment, known as
HADRON, was carried out at Tien-Shan Station between 1985 and 1993. Is
is based on a combination of extensive air shower arrays and large
emulsion chambers. The strangelet component in this experiment was
estimated to be about $1~{\rm m}^{-2}\, {\rm yr}^{-1}$. The data taken
by HADRON indicate that some primary cosmic rays may contain
non-nucleus components which generate extended air showers that
contain both a large number of muons as well as very high energetic
photons \cite{shaulov04:a}. Another group of data, associated with the
absorption of high energy photons in the atmosphere, suggests that
cosmic rays may contain an unusual component with an absorption length
a few times greater than for ordinary nuclei \cite{shaulov04:a}.  These
features are nicely explained if one assumes that they are caused by
stable or metastable strangelets \cite{shaulov04:a}.  Besides that, the
so-called Price event \cite{price78:a} with $Z\simeq 46$ and $A>1000$,
regarded previously as a possible candidate for the magnetic monopole,
turned out to be fully consistent with the $Z/A$ ratio for strange
quark matter \cite{saito95:a}. Finally we mention an exotic track event
with $Z\simeq 20$ and $A\simeq 460$ observed in an emulsion chamber
exposed to cosmic rays on a balloon has reported by
Miyamura \cite{ichimura93:a}. This exotic track event motivated the
balloon-borne emulsion chamber experiment JACEE \cite{miyamura95:a} and
Concorde aircraft \cite{capdevielle95:a} experiments.  JACEE was flown
near the top of the atmosphere. At least two events have been observed
which have been referred to as Anti-Centauros \cite{lord95:a}.

\goodbreak
\subsection{\it Unusual Seismographic Events}

As already described in section \ref{sec:searchexp}, De R{\'u}jula and
Glashow speculated about the presence of lumps of stable strange
matter, also referred to as strange nuggets or nuclearites, in the
cosmic radiation \cite{rujula83:a}.  The seismic signals caused by
these nuclearites passing through the Earth would be very different
from the seismic signals caused by an earthquake or a nuclear
explosion \cite{larousserie02:a,rujula83:a}.  This follows from the
rate of seismic energy produced by strange nuggets given by $dE/dt = f
\sigma \rho v^3$, where $\sigma$ is the nugget cross section, $\rho$
the nominal Earth density, $v$ the nugget speed, and $f$ the fraction
of nugget energy loss that results in seismic waves rather than other
dissipation such as heat or breaking rock
 \cite{anderson03:a}. Underground nuclear explosions have $f \simeq
0.01$, chemical explosions $f \simeq 0.02$.  In contrast to this,
strange nuggets with a mass of several tons (size of $~10^{-3}$~cm)
passing through the Earth would imply that $f\simeq 0.05$.  Anderson
\etal\  \cite{anderson03:a} looked at more than a million records
collected by the US Geological Survey between 1990 and 1993 that were
not associated with traditional seismic disturbances such as
earthquakes. The seismic signature would be caused by the large ratio
of the nuclearites speed, estimated to be around 400~km/s. Strange
nuggets might thus pass through the Earth at 40 times the speed of
seismic waves. Most interestingly, Anderson \etal\ were able to single
out two seismic events exhibiting this behavior. One event occurred on
22 October 1993, the other event occurred on 24 November 1993.  In the
first case, an unknown object seem to have entered the Earth off
Antarctica and left it south of India. It was recorded at seven
monitoring stations in India, Australia, Bolivia and Turkey.  In the
second case, an object seem to have entered in the South Pacific,
south of Australia, and left the Earth 16.8 seconds later in the Ross
Ice Shelf near the South Pole.  This event was recorded at nine
monitoring stations in Australia and Bolivia. The chord length between
the entry and exit points of the 24 November 1993 event is 4204~km so
that the duration measured for this event, if caused by the passage of
an object through the Earth, would imply a velocity for the
hypothetical object of 250~km/s. The interpretation of the data as
being caused by strange nuggets penetrating the Earth is backed by a
Monte Carlo study that was used to identify the extent to which
nuclearites could be detected by seismographic
stations \cite{anderson03:a,herrin96:a}.  The study showed that one
would expect to detect as many as 25 4-ton nuclearite events per year
if 4~ton strange nugget were to saturate the halo dark matter
density. If 10\% of the dark matter density were distributed in
strange nuggets over the mass range from 0.25 to 100 tons one would
expect about an event per year.  Detection of a nuclearite would
require at least six station sites to fix its impact time, location
and velocity, and seismic detection of signals by at least seven
stations is required in order to separate strange nuggets events from
random spurious coincidences.

\goodbreak
\subsection{\it AMS and ECCO}\label{sec:ams+ecco}

As shown in Ref.\  \cite{madsen01:a}, finite lumps of color-flavor
locked strange quark matter, which should be present in cosmic rays if
strange matter is the ground state of the strong interaction, turn out
to be significantly more stable than strangelets without color-flavor
locking for a wide range of parameters. In addition, strangelets made
of CFL strange matter obey a charge-mass relation of $Z/A \propto
A^{-1/3}$, which differs significantly from the charge-mass relation
of strangelets made of ordinary strange quark matter, as discussed in
section \ref{ssec:shyp}. In the latter case, $Z/A$ would be constant
for small baryon numbers $A$, and $Z/A \propto A^{-2/3}$ for large $A$
 \cite{madsen98:b,madsen01:a,aarhus91:proc}. This difference may allow
an astrophysical test of CFL locking in strange quark
matter \cite{madsen01:a}. The test in question is the upcoming cosmic
ray experiment AMS-02 on the International Space Station
scheduled \cite{ams01:homepage}. AMS-02 is a roughly a $1~{\rm m}^2$
sterad detector which will provide data from October 2005 for at least
3 years. AMS-02 will probe the dark matter content in various channels
(anti-protons, anti-deuteron, $e^+$, $\gamma$), cosmic rays and
$\gamma$ astrophysics, and, as already mentioned, strangelets.  The
expected flux of strangelets of baryon number $A < 6\times 10^6$ at
AMS-02 is \cite{madsen01:a}
\begin{eqnarray}
F \simeq 5\times 10^5 \; ({\rm m^2~y~sterad})^{-1}\times R_{-4} \times
M_{-2} \times V_{100}^{-1} \times t_7 ,
\label{eq:sFlux}
\end{eqnarray}
where $R_{-4}$ is the number of strange star collisions in our galaxy
per $10^4$ years, $M_{-2}$ is the mass of strangelets ejected per
collision in units of $10^{-2}\, \msun$, $V_{100}$ is the effective
galactic volume in units of $100~{\rm kpc}^3$ over which strangelets
are distributed, and $t_7$ is the average confinement time in units of
$10^7$ years.  All these factors are of order unity if strange matter
is absolutely stable, though each with significant uncertainties
 \cite{madsen01:a}.

Another intriguing instrument is ECCO (table \ref{tab:labexp}), whose
primary goal will be the measurement of the abundances of the
individual long-lived actinides (Th, U, Pu, Cm) in the galactic cosmic
rays with excellent resolution and statistics. ECCO is a large array
of passive glass-track-etch detectors to be exposed on orbit for at
least three years \cite{ecco01:homepage}.  The detectors will
passively record the tracks of relativistic ultra-heavy galactic cosmic
rays during exposure on orbit. After recovery, the detectors are
calibrated, etched, and analyzed.  ECCO is one of two instruments on
the HNX (Heavy Nuclei eXplorer) spacecraft, which is under
consideration as a Small Class Explorer Mission.  The HNX mission is
planned for launch in October 2005 into a 475-km circular orbit.
Recovery is planned nominally for 3 years following launch.

\goodbreak
\section{Relativistic Stellar Models}\label{sec:constr}

\subsection{\it Particles in Curved Space-Time}\label{sec:curved}

Neutron stars are objects of highly compressed matter so that the
geometry of spacetime is changed considerably from flat space.  Thus
models of such stars are to be constructed in the framework of
Einstein's general theory of relativity combined with theories of
superdense matter. The effects of curved space-time are included by
coupling the energy-momentum density tensor for matter fields to
Einstein's field equations. The generally covariant Lagrangian
density is
\begin{equation}
{\cal L} = {\cal L}_{\rm E} + {\cal L}_{\rm G}  \, ,
\label{eq:L+G}
\end{equation}
where the dynamics of particles are introduced through ${\cal L}_{\rm E}$
and the gravitational Lagrangian density ${\cal L}_{\rm G}$ is given by
\begin{equation}
{\cal L}_{\rm G} = g^{1/2} \,  R = g^{1/2} \, g^{\mu\nu} \, R_{\mu\nu} \, ,
\label{eq:L_G2}
\end{equation}
where $g^{\mu\nu}$ and $R_{\mu\nu}$ denote the metric tensor and the
Ricci tensor, respectively. The latter is given by
\begin{eqnarray}
  R_{\mu\nu} = \Gamma_{\mu\sigma, \, \nu}^\sigma - \Gamma_{\mu\nu, \,
  \sigma }^\sigma + \Gamma_{\kappa\nu}^\sigma \,
  \Gamma_{\mu\sigma}^\kappa - \Gamma_{\kappa\sigma}^\sigma \,
  \Gamma_{\mu\nu}^\kappa \, ,
\label{eq:14.27} 
\end{eqnarray} 
with the commas denoting derivatives with respect to space-time
coordinates, e.g. ${,\nu} = {{\partial}/{\partial x^\nu}}$ etc.  The
Christoffel symbols $\Gamma$ in (\ref{eq:14.27}) are defined as
\begin{eqnarray}
  \Gamma_{\mu\nu}^{\sigma} = {{1}\over{2}}\, g^{\sigma\lambda}\,
  \left( g_{\mu\lambda,\, \nu} + g_{\nu\lambda,\, \mu } - g_{\mu\nu,
  \, \lambda} \right)\, .
\label{eq:14.17}
\end{eqnarray} 
The connection between both branches of physics is provided by
Einstein's field equations
\begin{eqnarray}
  G^{\mu\nu} \equiv R^{\mu\nu} - {{1}\over{2}} g^{\mu\nu} R = 8 \pi
T^{\mu\nu}(\epsilon,P(\epsilon)) \, ,
\label{eq:intro.1}
\end{eqnarray}
($\mu, \nu= 0, 1, 2, 3$) which couples the Einstein curvature tensor,
$G^{\mu\nu}$, to the energy-momentum density tensor, $T^{\mu\nu}$, of
the stellar matter. The quantities $g^{\mu\nu}$ and $R$ in
(\ref{eq:intro.1}) denote the metric tensor and the Ricci scalar
(scalar curvature) \cite{weber99:book}.  The tensor $T^{\mu\nu}$
contains the equation of state, $P(\epsilon)$, on the stellar matter,
discussed in sections \ref{sec:confinedhm} and \ref{sec:primerqm}.  In
general, Einstein's field equations and the many-body equations were
to be solved simultaneously since the baryons and quarks move in
curved spacetime whose geometry, determined by Einstein's field
equations, is coupled to the total mass energy density of the matter.
In the case of neutron stars, as for all astrophysical situations for
which the long-range gravitational forces can be cleanly separated
from the short-range forces, the deviation from flat spacetime over
the length scale of the strong interaction, $\sim 1~\fm$, is however
practically zero up to the highest densities reached in the cores of
such stars (some $10^{15}~\gcmt$).  This is not to be confused with
the global length scale of neutron stars, $\sim 10$~km, for which $M /
R \sim 0.3$, depending on the star's mass.  That is to say, gravity
curves spacetime only on a macroscopic scale but leaves it flat to a
very good approximation on a microscopic scale. To achieve an
appreciable curvature on a microscopic scale set by the strong
interaction, mass densities greater than $\sim 10^{40}~\gcmt$ would be
necessary \cite{thorne66:a}!  This circumstance divides the
construction of models of compact stars into two distinct
problems. Firstly, the effects of the short-range nuclear forces on
the properties of matter are described in a comoving proper reference
frame (local inertial frame), where space-time is flat, by the
parameters and laws of (special relativistic) many-body physics.
Secondly, the coupling between the long-range gravitational field and
the matter is then taken into account by solving Einstein's field
equations for the gravitational field described by the general
relativistic curvature of space-time, leading to the global structure
of the stellar configuration.

\goodbreak
\subsection{\it Stellar Structure Equations of Non-rotating Stars}

For many studies of neutron star properties it is sufficient to treat
neutron star matter as a perfect fluid. The energy-momentum tensor of
such a fluid is given by
\begin{eqnarray}
  T^{\mu\nu} = u^\mu \, u^\nu \, \bigl(\, \epsilon + P \, \bigr) \, +
  \, g^{\mu\nu} \, P \, ,
\label{eq:85.7}
\end{eqnarray} where $u^\mu$ and $u^\nu$ are four-velocities defined as
\begin{eqnarray}
  u^\mu \equiv {{dx^\mu}\over{d\tau}} \, , \qquad u^\nu \equiv
  {{dx^\nu}\over{d\tau}} \, .
\label{eq:4vel}
\end{eqnarray} They are the components of the macroscopic velocity of
the stellar matter with respect to the actual coordinate system that
is being used to derive the stellar equilibrium equations.  The
production of curvature by the star's mass is specified by Einstein's
field equations,
\begin{eqnarray}
  G_{\mu\nu} = 8\, \pi\, T_{\,\mu\nu}\, , \quad \makebox{where} \quad
  G_{\mu\nu} \equiv R_{\,\mu\nu} - {{1}\over{2}}\, g_{\mu\nu}\, R
\label{eq:14.26}
\end{eqnarray} is the Einstein tensor.
The scalar curvature of spacetime $R$ in Eq.\ (\ref{eq:14.26}), also
known as Ricci scalar, follows from Eq.\ (\ref{eq:14.27}) as
\begin{equation}
  R = R_{\,\mu\nu}\, g^{\mu\nu} \, .
\label{eq:14.27a}
\end{equation}
Finally, we need to specify the metric of a non-rotating body in
general relativity theory. Assuming spherical symmetry, the metric
has the form
\begin{eqnarray}
  ds^2 = - e^{2\,\Phi(r)} \, dt^2 + e^{2\,\Lambda(r)} \, dr^2 +
  r^2 \, d\theta^2 + r^2 \, {\rm sin}^2\theta \, d\phi^2\, ,
\label{eq:15.20}
\end{eqnarray} where $\Phi(r)$ and $\Lambda(r)$ are radially
varying metric functions. Introducing the covariant components of the
metric tensor, 
\begin{equation}
g_{t t} = - \, {e^{2\,\Phi(r)}} \, , ~
g_{r r} = {e^{2\,\Lambda(r)}} \, , ~
g_{\theta \theta} =  {r}^{2} \, , ~
g_{\phi \phi} =  {r}^{2} \sin^2\!\theta \, ,
\label{eq:15.34}
\end{equation} 
the non-vanishing Christoffel symbols of a spherically symmetric body
are
\begin{eqnarray}
&&\Gamma_{t t}^r = 
 {e^{2\,\Phi(r)-2\,\Lambda(r)}} \, \Phi'(r) \, , ~
\Gamma_{t r}^t =  \Phi'(r) \, , ~
\Gamma_{r r}^r = \Lambda'(r)  \, , ~ 
\Gamma_{r \theta}^\theta = {r}^{-1}  \, , ~
\Gamma_{r \phi}^\phi =  {r}^{-1}  \, , ~
\Gamma_{\theta \theta}^r = - \, r \; e^{-2\,\Lambda(r)}  \, , 
\nonumber \\
&& \Gamma_{\theta \phi}^\phi =  {\frac {\cos\theta}{\sin\theta}} \, , ~
\Gamma_{\phi \phi}^r = - \, r \, \sin^2\!\theta \;
e^{-2\,\Lambda(r)}  \, , ~
\Gamma_{\phi \phi}^\theta = -\sin\theta \, \cos\theta \, ,
\label{eq:15.54c}
\end{eqnarray} 
where primes denote differentiation with respect to the radial
coordinate.  From Eqs.\ (\ref{eq:85.7}), (\ref{eq:14.26}) and
(\ref{eq:15.54c}) one derives the structure equations of spherically
symmetric neutron stars known as Tolman-Oppenheimer-Volkoff
equations \cite{oppenheimer39,tolman39:a}
\begin{eqnarray}
{{dP}\over{dr}} = - \, \frac{\epsilon(r)\, m(r)}{r^2} \; \frac{\left( 1
+ P(r)/\epsilon(r) \right) \, \left( 1 + 4 \pi r^3 P(r)/m(r) \right)}
{1 - 2 m(r)/r} \, .
\label{eq:f28}
\end{eqnarray}
Note that we use units for which the gravitational constant and
velocity of light are $G=c=1$ so that $\msun = 1.475$~km.  The
boundary condition to (\ref{eq:f28}) is $P(r=0) \equiv P_c =
P(\epsilon_c)$, where $\epsilon_c$ denotes the energy density at the
star's center, which constitutes an input parameter.  The pressure is
to be computed out to that radial distance where $P(r=R)=0$ which
determines the star's radius $R$.  The mass contained in a sphere of
radius $r~(\leq R)$, denoted by $m(r)$, follows as $m(r) = 4 \pi
\int^r_0 dr'\; r'^2 \; \epsilon(r') \, .$ The star's total
gravitational mass is thus given by $M\equiv m(R)$.

Figure \ref{fig:mrad2} shows the gravitational mass of non-rotating
neutron stars as a function of stellar radius for several sample
equations of state discussed in sections \ref{sec:confinedhm} and
\ref{sec:primerqm}.  Each star sequence is shown up to densities that
are slightly larger than those of the maximum-mass star (indicated by
tick marks) of each sequence. Stars beyond the mass peak are unstable
against radial oscillations and thus cannot exist stably in nature.
\begin{figure}[tb]
\begin{center}
\epsfig{file=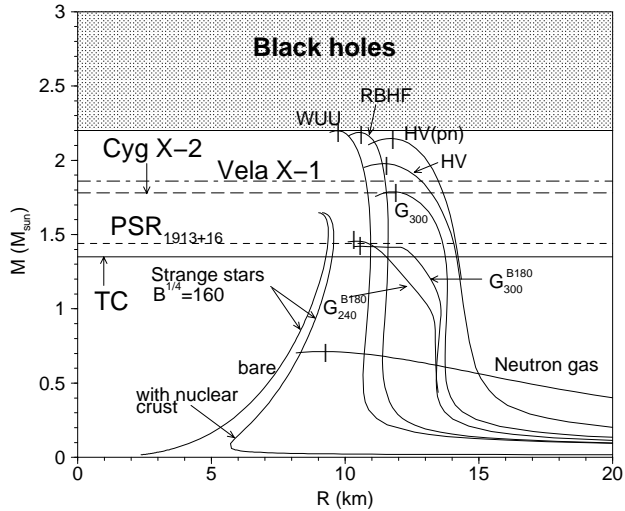,scale=0.5,angle=0}
\begin{minipage}[t]{16.5 cm}
\caption{Neutron star mass versus radius for different \eossp. The
  broken horizontal lines refer to the masses of Vela X-1 ($1.86\pm
  0.16\, \msun$) \cite{barziv01:a}, Cyg X-2 ($1.78\pm 0.23 \,
  \msun$) \cite{orosz99:a}, and PSR 1913+16 ($1.442\pm 0.003\, \msun$)
   \cite{taylor89:a}. The line labeled `TC' denotes the average neutron
  star mass ($1.350 \pm 0.004\, \msun$) derived by Thorsett and
  Chakrabarty \cite{thorsett99:a}.
\label{fig:mrad2}}
\end{minipage}
\end{center}
\vskip -0.5cm
\end{figure} One sees that all equations of state are able to support 
neutron stars of canonical mass, $M \sim 1.4 \, \msun$.  Neutron
stars more massive than about $2\, \msun$ on the other hand, are only
supported by \eoss that exhibit a very stiff behavior at supernuclear
densities and disfavor exotic (e.g., $K^-$ mesons, quark matter)
degrees of freedom. Knowledge of the maximum possible mass on neutron
stars is of great importance for two reasons. Firstly, because the
largest known neutron star mass imposes a lower bound on the maximum
mass of a theoretical model. The current lower bound is about
$1.55\,\msun$ for neutron star Cyg X-2 \cite{orosz99:a}, which does
not exclude the existence of exotic phases of matter in the core of
Cyg X-2. The situation could easily change if a future determination
of the mass of this neutron star should result in a value that is
close to its present upper limit of $\sim 2\,\msun$.  The second
reason is that the maximum mass is essential in order to identify
black hole candidates \cite{brown94:a,bethe95:a}.  For example, if the
mass of a compact companion of an optical star is determined to exceed
the maximum mass of a neutron star it must be a black hole. Since the
maximum mass of stable neutron stars studied here is $\sim
2.2\,\msun$, compact companions being more massive than that value
would be black holes.

\goodbreak
\subsection{\it Rotating Star Models}\label{sec:rotst}

The structure equations of rotating compact stars are considerably
more complicated that those of non-rotating compact stars
 \cite{weber99:book}.  These complications have their cause in the
rotational deformation, that is, a flattening at the pole accompanied
with a radial blowup in the equatorial direction, which leads to a
dependence of the star's metric on the polar coordinate, $\theta$, in
addition to the mere dependence on the radial coordinate, $r$.
Secondly, rotation stabilizes a star against gravitational
collapse. It can therefore carry more mass than it would be the case
if the star would be non-rotating. Being more massive, however, means
that the geometry of spacetime is changed too. This makes the metric
functions associated with a rotating star depend on the star's
rotational frequency.  Finally, the general relativistic effect of the
dragging of local inertial frames implies the occurrence of an
additional non-diagonal term, $g^{t\phi}$, in the metric tensor
$g^{\mu\nu}$. This term imposes a self-consistency condition
\begin{figure}[tb] 
\begin{center} 
\includegraphics[width=0.6\textwidth]{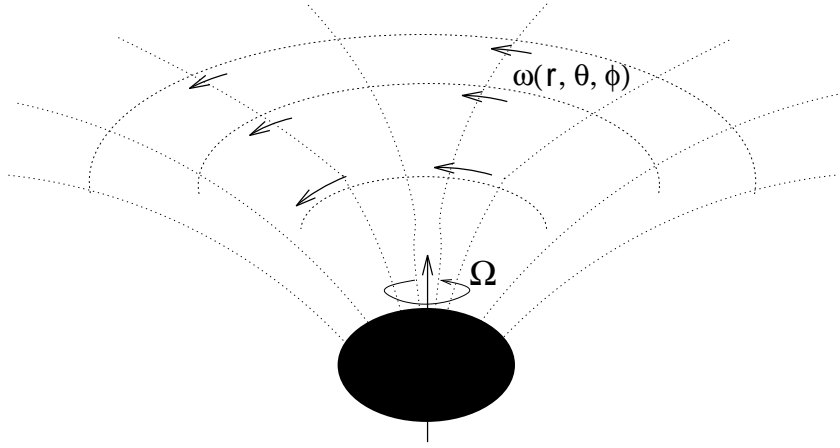}
\begin{minipage}[t]{16.5 cm}
\caption{Features of a rotating compact star in general
relativity. Indicated is the deformation of the geometry of spacetime
and the dragging of the local inertial frames. The latter rotate at a
position dependent angular frequency $\omega(r,\theta,\phi)$, which is
to be calculated self-consistently from Einstein's field
equations. The dragging frequencies inside three stellar
configurations are shown in Fig.~\protect{\ref{fig:fdragg}}.}
\label{fig:RStar}
\end{minipage}
\end{center}
\end{figure}
on the stellar structure equations, since the extent to which the
local inertial frames are dragged along in the direction of the star's
rotation, indicated schematically in Fig.~\ref{fig:RStar}, is
determined by the initially unknown stellar properties like mass and
rotational frequency. The covariant components of the metric tensor of
a rotating compact star are thus given
by \cite{weber99:book,friedman86:a}
\begin{eqnarray}
g_{t t} = - {e^{2\,{\nu}}} + {e^{2\,\psi}} \omega^2 \, , ~ g_{t \phi}
 = - {e^{2\,\psi}} \omega \, , ~ g_{r r} = {e^{2\,\lambda}} \, , ~
 g_{\theta \theta} = {e^{2\,{\mu}}} \, , ~ g_{\phi \phi} =
 {e^{2\,\psi}} \, ,
\label{eq:11.4bk} 
\end{eqnarray}
which leads for the line element to
\begin{eqnarray}
  d s^2 = g_{\mu\nu} dx^\mu dx^\nu = - \, e^{2\,\nu} \, dt^2 +
e^{2\,\psi} \, \bigl( d\phi - \omega \, dt \bigr)^2 + e^{2\,\mu} \,
d\theta^2 + e^{2\,\lambda} \, dr^2 \, .
\label{eq:f220.exact} 
\end{eqnarray} Here each metric function, i.e.\ $\nu$, $\psi$,  $\mu$ and
$\lambda$, as well as the angular velocities of the local inertial
frames, $\omega$, depend on the radial coordinate $r$ and polar angle
$\theta$, and implicitly on the star's angular velocity $\Omega$.  Of
particular interest is the relative angular frame dragging frequency,
$\bar\omega$, defined as
\begin{equation}
  \bar\omega(r,\theta,\Omega) \equiv \Omega - \omega(r,\theta,\Omega)
  \, ,
\label{eq:bar.omega}
\end{equation} which is the angular velocity of the star, $\Omega$,
relative to the angular velocity of a local inertial frame,
$\omega$. It is this frequency that is of relevance when discussing
the rotational flow of the fluid inside the star, since the magnitude
of the centrifugal force acting on a fluid element is governed--in
general relativity as well as in Newtonian gravitational theory--by
the rate of rotation of the fluid element relative to a local inertial
frame \cite{hartle67:a}. In contrast to Newtonian theory, however, the
inertial frames inside (and outside) a general relativistic fluid are
not at rest with respect to the distant stars, as pointed out just
above.  Rather, the local inertial frames are dragged along by the
rotating fluid. This effect can be quite strong, as
\begin{figure}[tb]
\begin{center}
\epsfig{figure=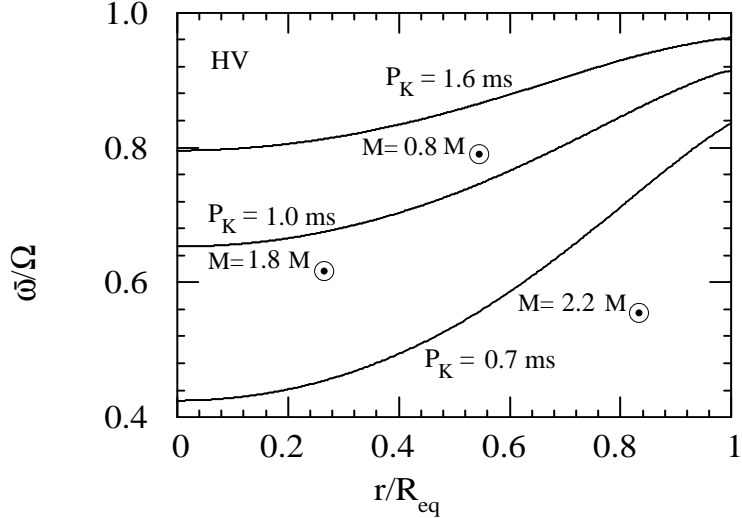,width=7.0cm,angle=-90}
\begin{minipage}[t]{16.5 cm}
\caption{Dragging of the local inertial frames inside rotating neutron
stars in equatorial direction. $\pkgr$ and $M$ denote Kepler period,
defined in Eq.\ (\ref{eq:okgr}), and gravitational mass.
(Fig.\ from Ref.\  \cite{weber99:book}.)}
\label{fig:fdragg}
\end{minipage}
\end{center}
\end{figure} shown in Fig.~\ref{fig:fdragg}. For a heavy neutron star 
rotating at its Kepler frequency, one sees that $\bar\omega/\Omega$
varies typically between about 15\% at the surface and 60\% at the
center, where the mass density is highest.

\goodbreak
\subsection{\it Kepler Frequency}\label{sec:kepler}

No simple stability criteria are known for rapidly rotating stellar
configurations in general relativity. However an absolute upper limit
on stable neutron star rotation is set by the Kepler frequency
$\okgr$, which is the maximum frequency a star can have before mass
loss (mass shedding) at the equator sets in.  In classical mechanics,
the expression for the Kepler frequency, determined by the equality
between the centrifugal force and gravity, is readily obtained as
$\okgr = \sqrt{M/R^3}$. In order to derive its general relativistic
counterpart, one applies the extremal principle to the circular orbit
of a point mass rotating at the star's equator.  Since
$r=\theta=\const$ for a point mass there one has $d r=d\theta=0$.  The
line element (\ref{eq:f220.exact}) then reduces to
\begin{eqnarray}
  d s^2 = \left( e^{2\,\nu} - e^{2\,\psi} \, (\Omega -
\omega)^2 \right) \, d t^2 \, .
\label{eq:12.1bk}
\end{eqnarray} Substituting this expression into $J \equiv
\int^{s_2}_{s_1} d s$, where $s_1$ and $s_2$ refer to points located at
that particular orbit for which $J$ becomes extremal, gives
\begin{eqnarray}
  J = \int_{s_1}^{s_2}\! d t \, \sqrt{e^{2\,\nu} - e^{2\,\psi}
    \, (\Omega - \omega)^2} \, .
\label{eq:12.2bk}
\end{eqnarray} Application of the extremal condition $\delta J=0$ to
Eq.~(\ref{eq:12.2bk}) and noticing that 
\begin{equation}
  V = e^{\psi-\nu} \, (\Omega - \omega) \, ,
\label{eq:Veq}
\end{equation} one obtains from Eq.~(\ref{eq:12.2bk})
\begin{eqnarray}
  {{\partial\psi}\over{\partial r}}\, e^{2\,\nu} \, V^2 -
  {{\partial\omega}\over{\partial r}}\, e^{\nu+\psi}\, V -
  {{\partial\nu}\over{\partial r}}\, e^{2\,\nu} = 0 \, . 
\label{eq:12.6bk}
\end{eqnarray} It constitutes a simple quadratic equation for the 
orbital velocity $V$ of a particle at the star's equator. Solving
(\ref{eq:Veq}) for $\Omega=\okgr$ gives the fluid's general
relativistic Kepler frequency in terms of $V$, the metric functions
$\nu$ and $\psi$, and the frame dragging frequency $\omega$, each
quantity being a complicated function of all the other quantities.  In
this manner $\okgr$ is given by \cite{weber99:book}
\begin{eqnarray}
  \okgr = \omega +\frac{\omega^\prime}{2\psi^\prime} +e^{\nu -\psi} \sqrt{
    \frac{\nu^\prime}{\psi^\prime} + \Bigl(\frac{\omega^\prime}{2
      \psi^\prime}e^{\psi-\nu}\Bigr)^2 } \, , \qquad \pkgr \equiv {{2 \pi}
    \over {\okgr}} \, ,
\label{eq:okgr}  
\end{eqnarray} which is to be evaluated at the star's equator (primes denote
radial derivatives).
Figure~\ref{fig:PK} shows $\pkgr$ as a function of rotating star mass. The
rectangle indicates both the approximate range of observed neutron star masses
as well as the observed rotational periods which, currently, are $P \geq 1.6$
ms.  One sees that
\begin{figure}[tb]
\begin{center}
\epsfig{figure=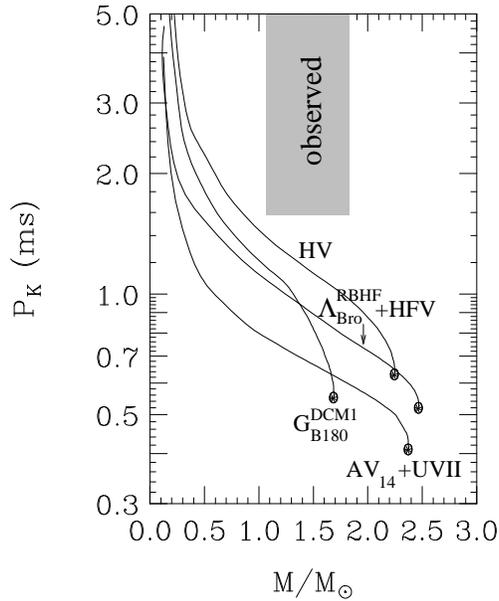,width=8.0cm,angle=90}
\begin{minipage}[t]{16.5 cm}
\caption{Onset of mass shedding from rapidly spinning neutron stars,
computed for a collection of equations of state (Fig.\ from Ref.\
 \cite{weber99:book}). The Kepler period is defined in Eq.\
(\ref{eq:okgr}).}
\label{fig:PK}
\end{minipage}
\end{center}
\end{figure} all pulsars so far observed rotate below the
mass shedding frequency and so can be interpreted as rotating neutron
stars. Half-millisecond periods or even smaller ones are excluded for
neutron stars of mass $1.4\,
\msun$ \cite{weber99:book,friedman86:a,friedman89:a}. The situation
appears to be very different for stars made up of self-bound strange
quark matter, the so-called strange stars which will be introduced in
section \ref{sec:sstars}. Such stars can withstand stable rotation
against mass shedding down to rotational periods in the
half-millisecond regime or even below \cite{glen91:a}. Consequently,
the possible future discovery of a single sub-millisecond pulsar
spinning at say 0.5~ms could give a strong hint that
strange stars actually exist, and that the deconfined self-bound
phase of three-flavor strange quark matter is in fact the true ground
state of the strong interaction rather than nuclear matter.
Strange stars of a canonical pulsar mass around $1.4\,\msun$ have
Kepler periods in the range of $0.55 ~{\rm ms} \ls \pkgr \ls 0.8 ~
{\rm ms}$, depending on the thickness of the nuclear curst and the bag
constant \cite{glen92:crust,weber93:b}. This range is to be compared
with $\pkgr\sim 1~ {\rm ms}$ obtained for standard (i.e., no phase
transition) neutron stars of the same mass. Phase transitions in
neutron stars, however, may lower this value down to Kepler periods
typical of strange stars \cite{burgio03:a}.

\goodbreak
\subsection{\it Moment of Inertia of Rotating Compact Stars}\label{sec:MoI}

To derive the expression for the moment of inertia of a rotationally
deformed, axisymmetric star in hydrostatic equilibrium, we start from
the following expression,
\begin{eqnarray}
  I({\cal A}, \Omega) \equiv {{1}\over{\Omega}} \, \int_{\cal A} \! d r\,
  d\theta\, d\phi \; T_{\,\phi}{^t}(r,\theta,\phi,\Omega)\, \sqrt{-
    g(r,\theta,\phi,\Omega)} \, .
\label{eq:11.64bk}
\end{eqnarray} We assume stationary rotation, which
is well justified for our investigations.  The quantity ${\cal A}$
denotes an axially symmetric region in the interior of a stellar body
where all matter is rotating with the same angular velocity $\Omega$,
and $\sqrt{-g}=e^{\lambda+\mu+\nu+\psi}$.  The component
$T_{\,\phi}{^t}$ of the energy-momentum tensor is given by
\begin{equation}
  T_{\,\phi}{^t} = (\epsilon+P) \, u_\phi \, u^t \, .
\label{eq:11.TPt}
\end{equation} Let us focus next on the determination of the fluid's
four-velocity, $u^\kappa = (u^t,u^r,u^\theta,u^\phi)$.  From the
general normalization relation $u^\kappa u_\kappa=-1$ one readily derives
\begin{eqnarray}
  - 1 &=& (u^t)^2\, g_{tt} + 2\, u^t\, u^\phi\, g_{t\phi} + (u^\phi)^2\,
  g_{\phi\phi} \, .
\label{eq:11.65bk}
\end{eqnarray} This relation can be rewritten by noticing that
\begin{equation}
  u^\phi = \Omega u^t \, ,
\label{eq:51.HS67}
\end{equation} which extremizes the total mass energy of the
stationary stellar fluid subject to the constraint that the angular
momentum about the star's symmetry axis, $J_z$, and its baryon number,
$A$, remain fixed \cite{hartle67:b}. Substituting (\ref{eq:51.HS67})
into (\ref{eq:11.65bk}) leads to
\begin{eqnarray}
  -1 &=& (u^t)^2 \, (g_{tt} + 2\, g_{t\phi}\, \Omega + g_{\phi\phi}\,
  \Omega^2) \, ,
\label{eq:11.66bk} 
\end{eqnarray} which can be solved for $u^t$. This gives
\begin{eqnarray}
  u^t = \left( - (g_{tt} + 2\, g_{t\phi}\, \Omega + g_{\phi\phi}\,
        \Omega^2) \right)^{-1/2}\, .
\label{eq:11.67bk}
\end{eqnarray} Replacing $g_{tt}$, $g_{t\phi}$ and $g_{\phi\phi}$ 
with the expressions given in (\ref{eq:11.4bk}) and rearranging terms
leads for (\ref{eq:11.67bk}) to
\begin{eqnarray}
  u^t &=& {e^{-\nu}} \ \left( 1 - (\omega - \Omega)^2\, e^{2\psi
        -2\nu} \right)^{-1} \, .
\label{eq:11.68bk}
\end{eqnarray} Last but not least we need an expression for $u_\phi$ of Eq.\
(\ref{eq:11.TPt}) in terms of the star's metric functions. To
accomplish this we write $u_\phi$ as $u_\phi=g_{\phi\kappa} u^\kappa =
g_{\phi t} u^t + g_{\phi\phi} u^\phi$. Upon substituting the
expressions for $g_{\phi t}$ and $g_{\phi\phi}$ from
(\ref{eq:11.4bk}) into this relation, we arrive at
\begin{eqnarray}
  u_\phi = (\Omega - \omega) \; e^{2\, \psi} \; u^t \, .
\label{eq:11.69bk}
\end{eqnarray} 
Substituting the four-velocities (\ref{eq:11.68bk}) and
(\ref{eq:11.69bk}) into (\ref{eq:11.TPt}) gives the required
expression for the energy-momentum tensor,
\begin{eqnarray}
  T_{\,\phi}{^t} = (\epsilon + P)\, (\Omega - \omega)\, e^{2\,\psi}
      \left( e^{2\, \nu} - (\omega - \Omega)^2\, e^{2\, \psi}\right)^{-1}
      \, .
\label{eq:11.70bk}
\end{eqnarray} Finally, inserting this expression into (\ref{eq:11.64bk}) leads 
for the moment of inertia of a rotationally deformed star to
\begin{eqnarray}
  I(\Omega) = 2\, \pi \int_0^\pi \! d\theta \int_0^{R(\theta)} d r \;
  e^{\lambda+\mu+\nu+\psi} \, {{\epsilon + P(\epsilon)}\over{e^{2\nu -
  2\psi} - (\omega - \Omega)^2}} \, {{\Omega - \omega}\over{\Omega}}
  \, . \nonumber \\
\label{eq:11.71bk} 
\end{eqnarray} 
Relativistic corrections to the Newtonian expression for $I$, which
for a sphere of uniform mass density, $\epsilon(r)={\rm const}$, is
given by $I={2\over 5} M R^2$, come from the dragging of local
inertial frames ($(\Omega - \omega) / \Omega < 1$) and the curvature
of space.

\goodbreak
\section{Strangeness in Compact Stars}\label{sec:scc}

\subsection{\it Neutron Stars}

Physicists know of three types of compact stars.  These are black
holes, neutron stars, and white dwarfs.  Neutron stars and white
dwarfs are in hydrostatic equilibrium so that at each point inside the
star gravity is balanced by the degenerate particle pressure, as
described mathematically by the Tolman-Oppenheimer-Volkoff equation
(\ref{eq:f28}).  These stars, therefore, exhibit the generic
mass-radius relationship shown in Fig.\ \ref{fig:nswd}.
\begin{figure}[tb]
\begin{center}
\leavevmode
\psfig{figure=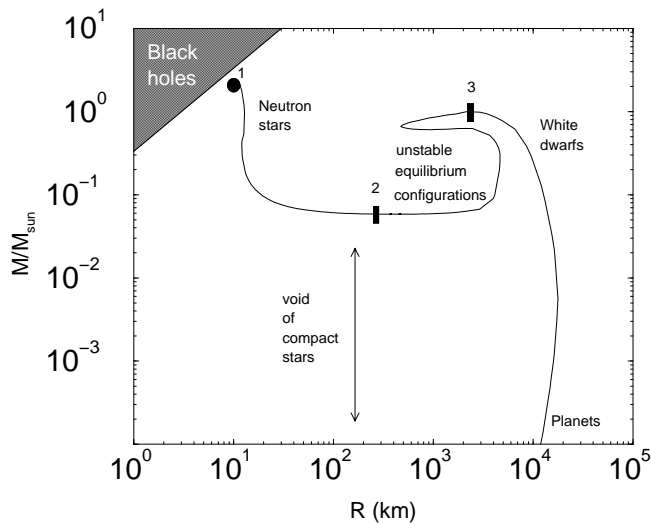,width=7.0cm,angle=-90}
\caption{Mass versus radius of neutron stars, white dwarfs, and
  planets. Stars located between the lightest neutron star (marked
  `2') and the heaviest white dwarf (marked `3') are unstable against
  radial oscillations and thus cannot exist stably in nature.  If
  strange matter would be more stable than nuclear matter, an enormous
  region in the $M$--$R$ plane void of compact objects could be
  populated with strange matter configurations (see Fig.\
  \ref{fig:mr145}).}
\label{fig:nswd}
\end{center}
\end{figure} 
Depending on composition the maximum neutron star mass (marked `1')
lies somewhere between $\sim 1.5$ and $2.5\, \msun$ (see also Fig.\
\ref{fig:mrad2}). The minimum mass of neutron stars (marked `2') is
around $0.1\, \msun$. These stars are considerably less dense ($\sim
0.1 \epsilon_0$) and thus much bigger ($R \ls 300$~km) than the
neutron stars of canonical mass, which is $\sim 1.4\, \msun$.  White
dwarfs at the Chandrasekhar mass limit (marked `3') have densities
around $10^9~\gcmt$, which is five orders of magnitude smaller than the
typical densities encountered in neutron stars.

\goodbreak
\subsubsection{\it Hyperons}\label{sec:hyperons}

Model calculations indicate that only in the most primitive
conception, a neutron star is made of only neutrons.  At a more
accurate representation, a neutron star may contain neutrons ($n$) and
protons ($p$) whose charge is balanced by electrons ($e^-$) and muons
($\mu^-$), strangeness-carrying hyperons ($\Sigma, \Lambda, \Xi$),
meson condensates ($K^-$ or $p^-$), $u,\, d,\, s$ quarks, or possibly
H-dibaryons.  The particle composition is determined by the conditions
of electric charge neutrality and chemical equilibrium as well as the
in-medium properties of the constituents calculated for a given
microscopic many-body theory (cf.\ sections \ref{sec:confinedhm} and
\ref{sec:primerqm}). In general, the population of negatively charged
hadronic states is favored over the population of positively charged
hadronic states since the negative charge carried by the hadrons can
replace high-energy electrons, by means of which a lower energy state
is reached. Aside from electric charge, the isospin orientation of the
possible constituents is of key importance for the population too. The
reason is that neutron star matter constitutes a highly excited state
of matter relative to isospin symmetric nuclear matter. Hence, as soon
as there are new hadronic degrees of freedom accessible to neutron
star matter which allow such matter to become more isospin symmetric,
it will make use of them.

\goodbreak
\subsubsection{\it $K^-$ Meson Condensate}\label{sec:kminus}

The condensation of $K^-$ mesons in neutron stars is initiated by the
reaction
\begin{equation}
  e^- \rightarrow K^- + \nu \, .
\label{eq:kaon.1}
\end{equation} If this reaction becomes possible in a neutron star, it is
energetically advantageous for the star to replace the fermionic
electrons with the bosonic $K^-$ mesons. Whether or not this happens
depends on the behavior of the $K^-$ mass in neutron star matter. 
Experiments which shed light on the properties of the $K^-$ in nuclear
matter have been performed with the Kaon Spectrometer (KaoS) and the
FOPI detector at the heavy-ion synchrotron SIS at
GSI \cite{barth97:a,senger01:a,sturm01:a,devismes02:a}.  An analysis of
the early $K^-$ kinetic energy spectra extracted from Ni+Ni collisions
 \cite{barth97:a}
\begin{figure}[tb] 
\begin{center}
\leavevmode
\epsfig{figure=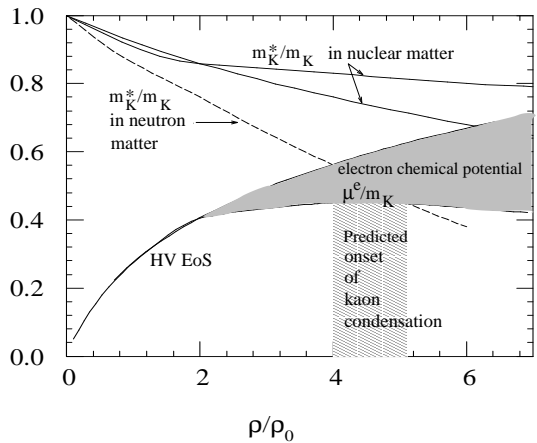,width=7.0cm,angle=0}
\caption{Effective kaon mass in nuclear \cite{mao99:a} and neutron
  star \cite{waas97:a} matter. (Fig.\ from Ref.\
  \cite{weber99:book}.)}
\label{fig:Kmass}
\end{center}
\end{figure}  showed that the attraction from nuclear matter would bring 
the $K^-$ mass down to $m^*_{K^-}\simeq 200~\mev$ at $\rho\sim 3\,
\rho_0$. For neutron-rich matter, the relation
\begin{equation}
  m^*_{K^-}(\rho) \simeq m_{K^-}  \left( 1 - 0.2 \, {{\rho}\over{\rho_0}} \right)
\label{eq:meff02}
\end{equation} was established \cite{li97:a,li97:b,brown97:a}, with
$m_K = 495$~MeV the $K^-$ vacuum mass.  Values around $m^*_{K^-}\simeq
200~\mev$ lie in the vicinity of the electron chemical potential,
$\mu^e$, in neutron star matter \cite{weber99:book,glen85:b} so that
the threshold condition for the onset of $K^-$ condensation, $\mu^e =
m^*_K$, which follows from Eq.\ (\ref{eq:kaon.1}), could be fulfilled
in the centers of neutron stars.  The situation is illustrated
graphically in Fig.\ \ref{fig:Kmass}. Equation (\ref{eq:kaon.1}) is
followed by
\begin{equation}
  n + e^- \rightarrow p + K^- + \nu \, ,
\label{eq:npK.1}
\end{equation} with the neutrinos leaving the star. By this conversion the
nucleons in the cores of newly formed neutron stars can become half
neutrons and half protons, which lowers the energy per baryon of the
matter \cite{brown96:a}. The relative isospin symmetric composition
achieved in this way resembles the one of atomic nuclei, which are
made up of roughly equal numbers of neutrons and protons.  Neutron
stars are therefore referred to, in this picture, as nucleon stars.
The maximal possible mass of this type of star, where Eq.\
(\ref{eq:npK.1}) has gone to completion, has been calculated to be
around $1.5\, \msun$  \cite{thorsson94:a}. Consequently, the collapsing
core of a supernova (e.g.\ 1987A), if heavier than this value, should
go into a black hole rather than forming a neutron
star \cite{brown94:a,li97:a,li97:b}. Another striking implication,
pointed out by Brown and Bethe, would be the existence of a large
number of low-mass black holes in our galaxy  \cite{brown94:a}.

\goodbreak
\subsubsection{\it Strange Quarks}\label{sec:strangeness_quarks}

Already several decades ago it has been suggested that, because of the
extreme densities reached in the cores of neutron stars, neutrons
protons plus the heavier constituents may melt, creating quark matter
being sought at the most powerful terrestrial heavy-ion
colliders \cite{ivanenko65:a,fritzsch73:a,baym76:a,keister76:a,chap77:a+b,fech78:a}.
At present one does not know from experiment at what density the
expected phase transition to quark matter occurs, and one has no
conclusive guide yet from lattice QCD simulations.  From simple
geometrical considerations it follows that nuclei begin to touch each
other at densities of $\sim (4\pi r^3_N/3)^{-1} \simeq 0.24~\fmmt$,
which, for a characteristic nucleon radius of $r_N\sim 1$~fm, is less
than twice the baryon number density $\rho_0$ of ordinary nuclear
matter \cite{glen97:book}.  Above this density, therefore, is appears
plausible that the nuclear boundaries of hadrons dissolve and the
originally confined quarks begin to populate
\begin{figure}[tb] 
\begin{center}
\begin{minipage}[t]{8.0 cm}
\epsfig{file=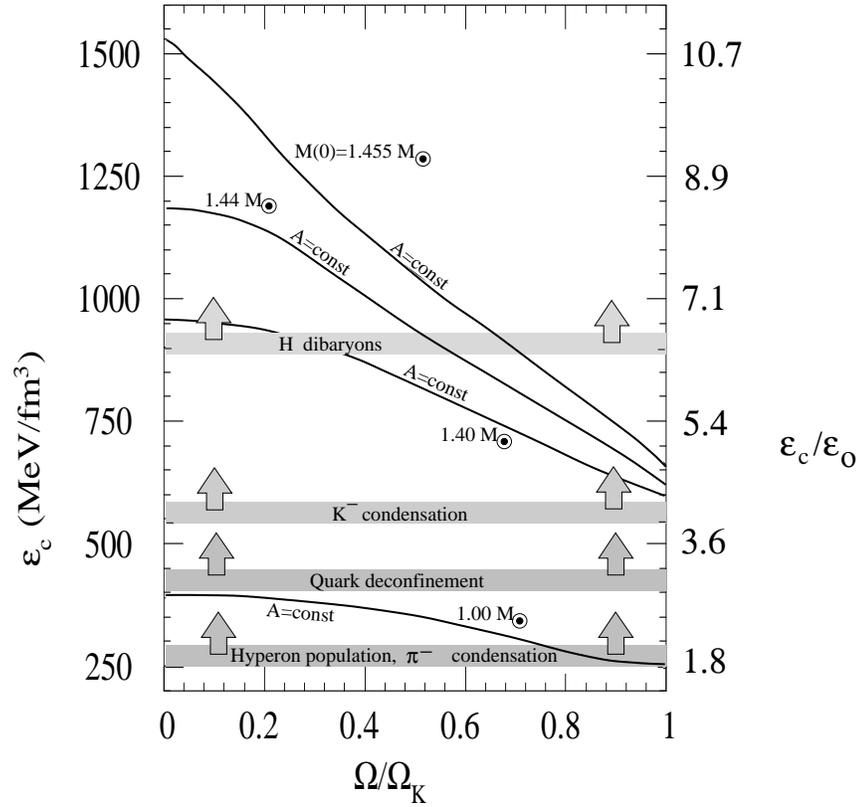,scale=0.4,angle=-90}
\end{minipage}
\begin{minipage}[t]{16.5 cm}
\caption{Central density versus rotational frequency for
several sample neutron stars. The stars' baryon number, $A$, is
constant in each case.  Theory predicts that the interior stellar
density could become so great that the threshold densities of various
novel phases of superdense matter are reached. $\epsilon_0 =
140~\mevt$ denotes the density of nuclear matter, $\okgr$ is the
Kepler frequency, and $M(0)$ is the stars' mass at zero
rotation. (From Ref.\  \cite{weber99:book}.)}
\label{fig:ec1445fig}
\end{minipage}
\end{center}
\end{figure} 
free states outside of the hadrons.  Depending on rotational frequency
and stellar mass, densities as large as two to three times $\rho_0$
are easily surpassed in the cores of neutron stars, as can be seen
from Figs.\ \ref{fig:ec1445fig} and \ref{fig:ecvsomega}, so that the
neutrons and protons in the centers of neutron stars may have been
broken up into their constituent quarks by
gravity \cite{weber99:book,weber99:topr}. More than that, since the
mass of the strange quark is so small, high-energetic up and down
quarks are expected to readily
\begin{figure}[tb]
\begin{center}
{\rotatebox{-90}
{\includegraphics[scale=0.40]{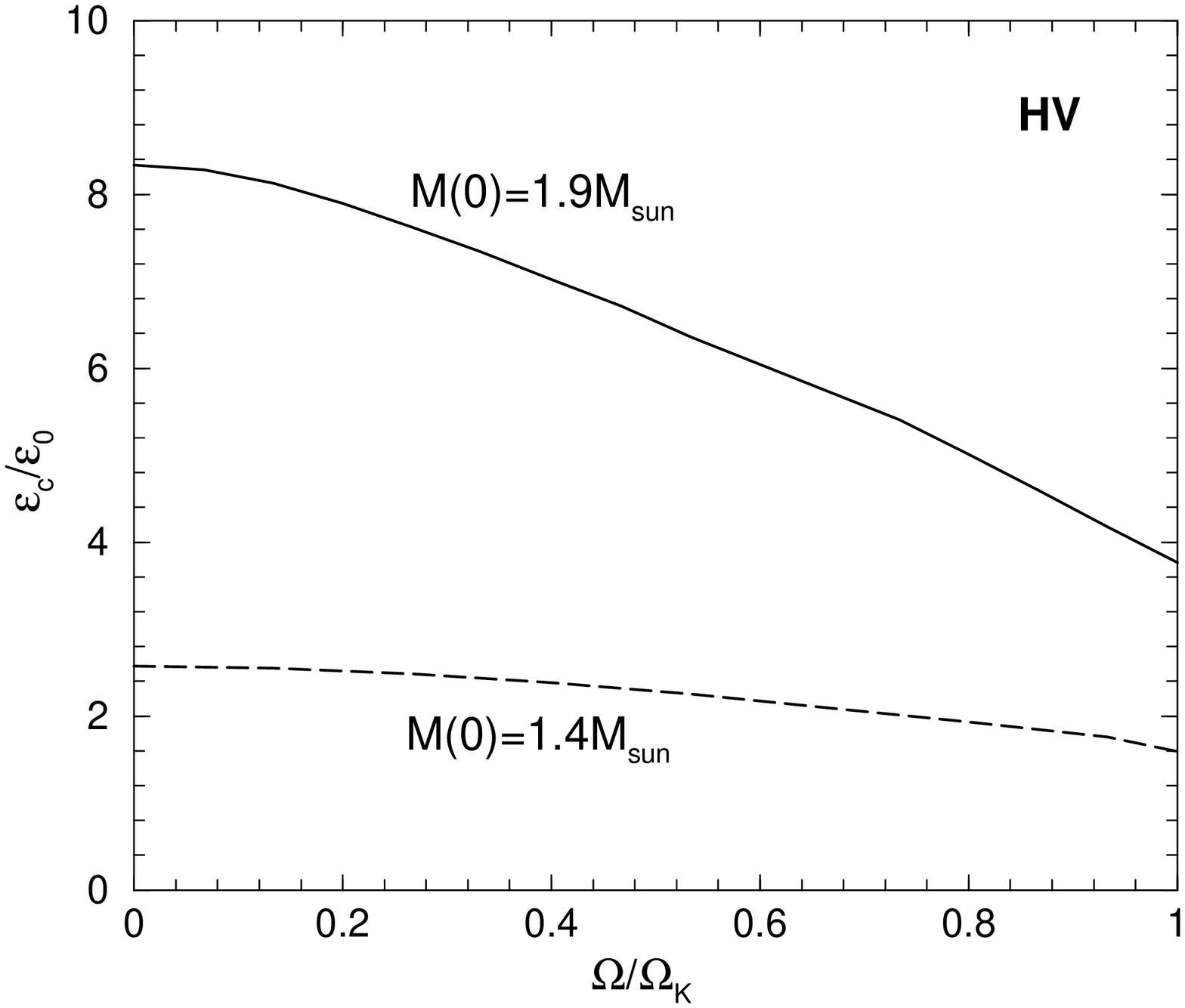}} \hskip 1.0cm
\rotatebox{-90}
{\includegraphics[scale=0.40]{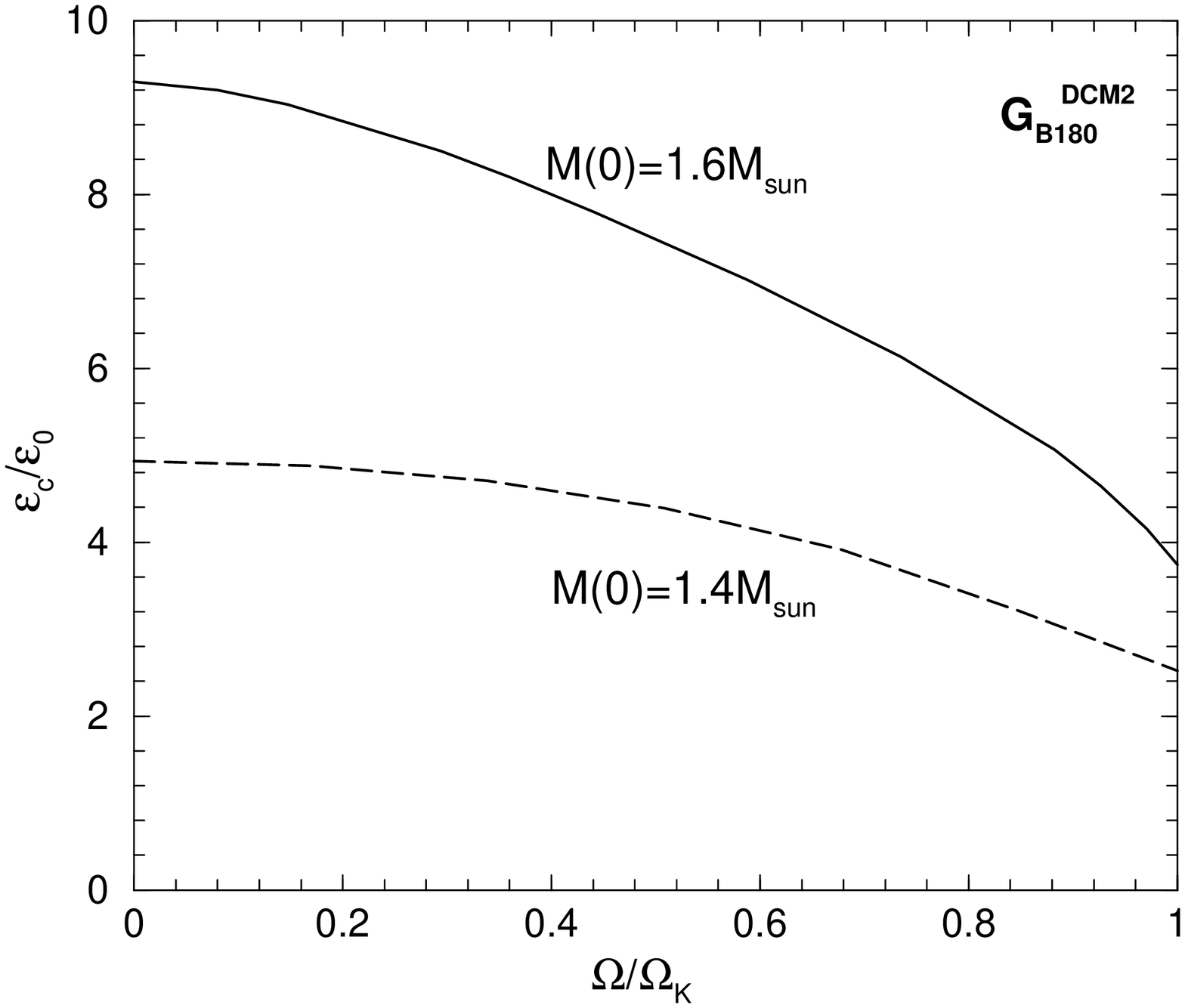}}}
\begin{minipage}[t]{16.5 cm}
\caption{Variation of central star density with rotational star frequency.}
\label{fig:ecvsomega}
\end{minipage}
\end{center}
\end{figure}
transform to strange quarks at about the same density at which up and
down quark deconfinement sets in \cite{glen91:pt,kettner94:b}.  Three
flavor quark matter could thus exist as a permanent component of
matter in the centers of neutron
stars \cite{weber99:book,glen97:book,weber99:topr}.

In passing we mention that the existence of a mixed phase depends
decisively on the unknow surface tension. Alford \etal \ studied the
CFL-nuclear mixed phase \cite{alford01:b} and found that if the surface
tension is above about $40~ {\rm MeV/fm}^2$ then the mixed phase does
not occur at all. This surface tension is completely unknown, but one
might expect the `natural scale' for it to be $(200~ {\rm MeV})^3 =
200~ {\rm MeV/fm}^2$. For what follows we will assume that the surface
tension is such that a mixed phase of quark matter and nuclear matter
exists above a certain density.

As pointed out by Glendenning \cite{glen91:pt,glen01:b}, in all earlier
work on the quark-hadron phase transition in neutron star matter,
assumed to be a first order transition, the possibility of reaching
the lowest energy state by rearranging electric charge between the
regions of confined hadronic matter and deconfined quark matter in
phase equilibrium was ignored.  This incorrectly yielded a description
of the possible quark-hadron transition in neutron stars as a constant
pressure one, which had the consequence of excluding the coexistence
phase of hadrons and quarks from neutron stars.  The microphysical
agent behind this preference for charge rearrangement is the
charge-symmetric nuclear force which acts to relieve the high isospin
asymmetry of neutron star matter as soon as it is in equilibrium with
quark matter. This introduces a net positive charge on the hadronic
regions and a compensatory net negative charge on quark matter. (The
nuclear matter and quark matter phases are thus not separately charge
neutral as assumed before Glendenning's work.) Because of this
preference for charge rearrangement exploited by a neutron star, the
pressure in the mixed quark-hadron phase varies as the proportions of
the phases.  Varying pressure in the mixed phase is of key importance
for the existence of the mixed phase inside neutron stars, because
hydrostatic equilibrium dictates that pressure drops monotonically
from the center toward the surface. For that reason the mixed phase is
not strictly--and incorrectly--excluded from neutron
stars \cite{glen91:pt,glen01:b,glen97:book}.

If the dense interior of a neutron star is converted to quark
matter \cite{weber99:book,glen91:pt,glen97:book,ellis91:a}, it must be
three-flavor quark matter (see Figs.\ \ref{fig:3.13.1} through
\ref{fig:3.16.1}) since it has lower energy than two-flavor quark
matter (see section \ref{sec:primerqm}). And just as for the hyperon
content of neutron stars, strangeness is not conserved on macroscopic
time scales which allows neutron stars to convert confined hadronic
matter to three-flavor quark matter until equilibrium brings this
process to a halt. Many of the earlier investigations have treated
neutron stars as containing only neutrons, and the quark phase as
consisting of the equivalent number of $u$ and $d$ quarks.  Pure
neutron matter, however, is not the ground state of a neutron star,
nor is a mixture of $u$ and $d$ quarks the ground state of quark
matter in compact stars. In fact the latter constitutes a highly
excited state of quark matter, which will quickly weak decay to an
approximate equal mixture of $u$, $d$ and $s$ quarks.  Several other
investigations have approximated the mixed phase as two components
which are separately charge neutral, which hides the possible
quark-hadron phase transition in neutron star
matter \cite{glen91:pt,glen01:b} because the deconfinement transition
is shifted to densities hardly reached in the cores of neutron stars
of average mass, $M\sim 1.4\, \msun$.

The Gibbs condition for phase equilibrium between quarks and hadrons
is that the two associated, independent chemical potentials, $\mu^n$
and $\mu^e$ respectively, and the pressure in the two phases be equal,
\begin{eqnarray}
  P_{\rm H}(\mu^n,\mu^e, \{ \phi \}, T) = P_{\rm Q}(\mu^n,\mu^e,T) \,
  ,
\label{eq:gibbs1}
\end{eqnarray} 
where $\mu^n$ and $\mu^e$ denote the chemical potentials of neutrons
and electrons, respectively, and the subscripts H and Q refer to
confined hadronic matter and deconfined quark matter.  The quantity
$\{\phi\}$ stands collectively for the particle fields and Fermi
momenta which characterize a solution to the equations of confined
hadronic matter discussed in section \ref{sec:confinedhm}.  As known
form Eq.\ (\ref{eq:mq}), the quark chemical potentials are related to
the baryon and charge chemical potentials in Eq.\ (\ref{eq:gibbs1}) as
\begin{eqnarray}
  \mu^u = \mu^c = {{1}\over{3}} \, \mu^n - {{2}\over{3}} \, \mu^e\,
  ,\qquad \mu^d = \mu^s = {{1}\over{3}} \, \mu^n + {{1}\over{3}} \,
  \mu^e \, .
\label{eq:cp.ChT}
\end{eqnarray} In accordance to what has bee said just above, Eq.\ 
(\ref{eq:gibbs1}) is to be supplemented with the conditions of baryon
charge conservation and electric charge conservation
 \cite{glen91:pt,glen01:b}.
\begin{figure}[tb]
\begin{center}
\epsfig{figure=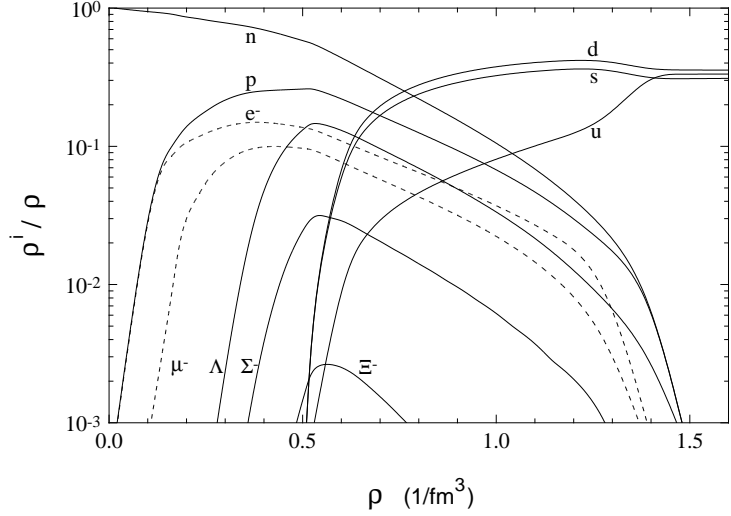,width=7.0cm,angle=-90}
\begin{minipage}[t]{16.5 cm}
\caption{Composition of chemically equilibrated, stellar quark-hadron
(hybrid star) matter as a function of baryon density. Hadronic matter
is described by the relativistic Hartree model
HV \cite{weber99:book,weber89:e}, the bag constant is $B =
250~\mevt$. (From Ref.\  \cite{weber99:book}.)}
\label{fig:3.13.1}
\end{minipage}
\end{center}
\end{figure}
Mathematically, the global conservation of baryon charge within an
unknown volume, $V$, containing $A$ baryons is expressed as
\begin{equation}
  \rho \equiv {A\over V} = (1-\chi) \, \rho_{\rm H}(\mu^n,\mu^e,T) +
  \chi \, \rho_{\rm Q}(\mu^n,\mu^e,T) \, ,
\label{eq:bcharge}
\end{equation} where $\chi\equiv V_{\rm Q}/V$ denotes the volume
proportion of quark matter, $V_{\rm Q}$, in the unknown volume $V$.
By definition, the parameter $\chi$ varies between 0 and 1,
determining how much confined hadronic matter exists as quark matter.
The global neutrality of electric charge within the volume $V$ is
mathematically expressed as \cite{glen91:pt,glen01:b}
\begin{equation}
  0 = {Q\over V} = (1-\chi) \, q_{\rm H}(\mu^n,\mu^e,T) + \chi \,
  q_{\rm Q}(\mu^n,\mu^e,T)  +  q_{\rm L} \, ,
\label{eq:echarge}
\end{equation}
where $q_{\rm H}=\sum_B q^\el_B \rho^B$ and $q_{\rm Q}=\sum_f q^\el_f
\rho^f$ denote the net electric charge carried by hadronic and quark
matter, respectively, and $q_{\rm L}=\sum_L q^\el_L \rho^L$ stands for
the electric charge density of the leptons (see sections
\ref{sec:confinedhm} and \ref{sec:primerqm}).  One sees that for a
given temperature $T$, Eqs.\ (\ref{eq:gibbs1}) through
(\ref{eq:echarge}) serve to determine the two independent chemical
potentials $\mu^n$ and $\mu^e$, and the volume $V$ for a specified
volume fraction $\chi$ of the quark phase in equilibrium with the
hadronic phase.  After completion, $V_{\rm Q}$ is obtained as $V_{\rm
Q}=\chi V$.  Through Eqs.\ (\ref{eq:gibbs1}) to (\ref{eq:echarge}) the
chemical potentials $\mu^n$ and $\mu^e$ obviously depend on $\chi$ and
thus on density $\rho$, which renders all properties that depend on
$\mu^n$ and $\mu^e$--from the energy density to the baryon and charge
densities of each phase to the common pressure--density dependent,
too.

Figures \ref{fig:3.13.1} through \ref{fig:3.16.1} show sample
quark-lepton populations computed for representative bag constants as
well as different many-body approximations employed to model confined
hadronic matter \cite{weber99:book,weber99:topr}.  Three features
emerge immediately from these populations. Firstly, one
\begin{figure}[tb]
\begin{center}
\leavevmode
\psfig{figure=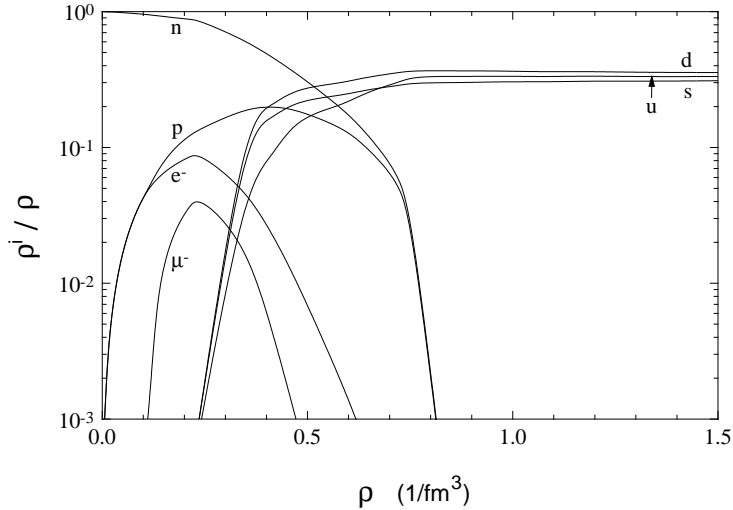,width=7.0cm,angle=-90}
\begin{minipage}[t]{16.5 cm}
\caption{Composition of chemically equilibrated, stellar quark-hadron
matter as a function of baryon density. Hyperons are artificially
suppressed. Hadronic matter is described by
HFV \cite{weber99:book,weber89:e}, the bag constant is $B =
150~\mevt$. (Fig.\ from Ref.\  \cite{weber99:book}.)}
\label{fig:3.15.1}
\end{minipage}
\end{center}
\end{figure} sees that the transition from pure hadronic matter to the
mixed phase occurs at rather low density of about $3\, \rho_0$ or even
somewhat less \cite{glen91:pt,glen01:b,hermann96:a}. Depending on the
bag constant and the underlying nuclear many-body approximation,
threshold values even as small as about $2\, \rho_0$ are possible.
\begin{figure}[tb]
\begin{center}
\epsfig{figure=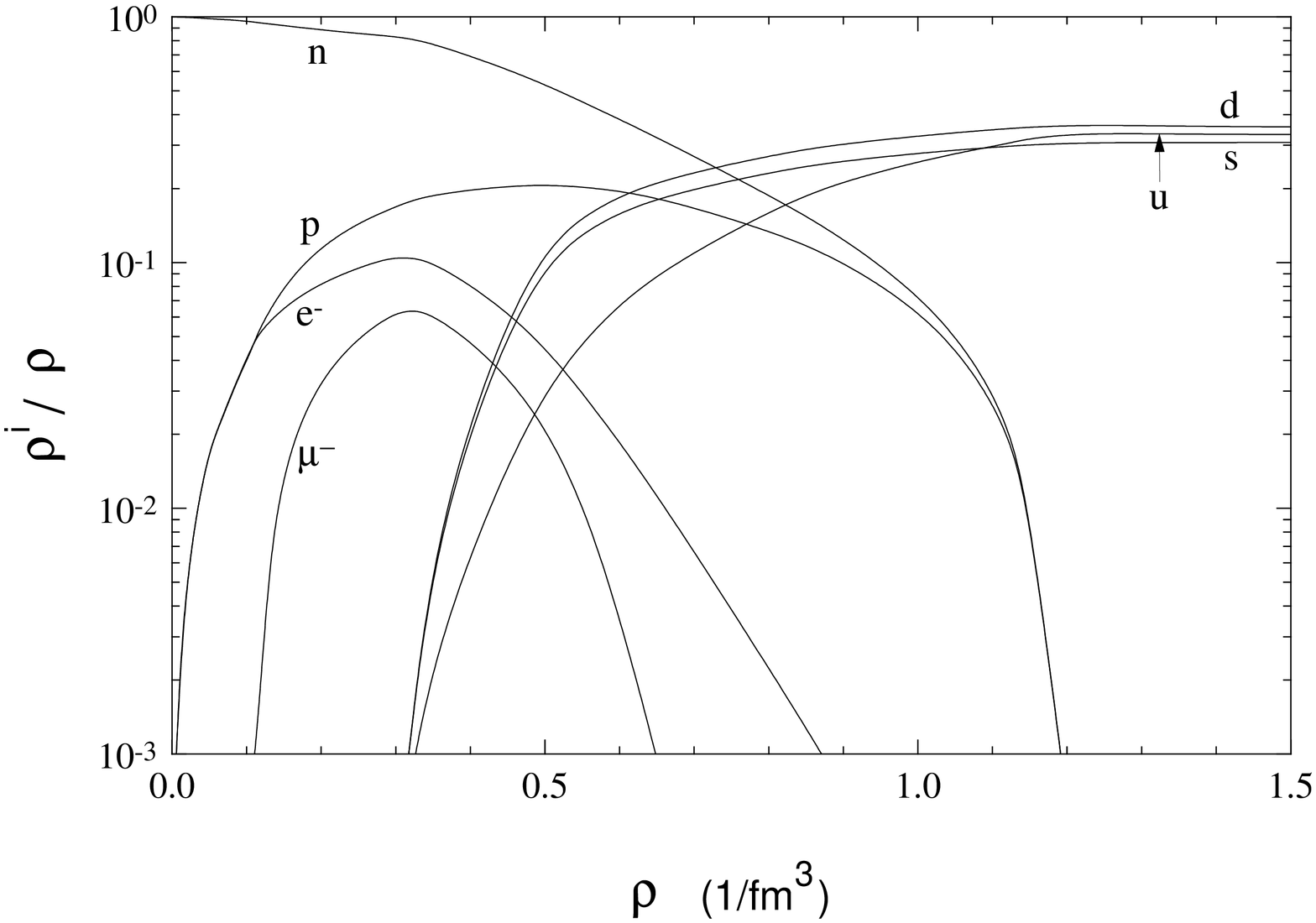,width=7.0cm,angle=-90}
\begin{minipage}[t]{16.5 cm}
\caption{Same as Fig.\ \ref{fig:3.15.1}, but for $B=
250~\mevt$. (Fig.\ from Ref.\  \cite{weber99:book}.)}
\label{fig:3.16.1}
\end{minipage}
\end{center}
\end{figure} Secondly, we emphasize the saturation of the number of 
electrons as soon as quark matter appears, for then electric charge
neutrality can be achieved more economically among the baryon-charge
carrying particles themselves. This saturation is of very great
importance for the possible formation of a $K^-$ condensate in neutron
stars \cite{li97:a,li97:b,brown97:a,brown96:a}, whose threshold
condition is given by $\mu^e = m^*_K$.  Figure~\ref{fig:Kmass} shows
that this condition may be fulfilled in neutron star matter, depending
on the underlying many-body approximation and the structure of
many-body background.  Thirdly, the presence of quark matter enables
the hadronic regions of the mixed phase to arrange to be more isospin
symmetric, i.e.\ closer equality in proton and neutron number is
achieved, than in the pure phase by transferring charge to the quark
phase in equilibrium with it. Symmetry energy will be lowered thereby
at only a small cost in rearranging the quark Fermi
surfaces. Electrons play only a minor role when neutrality can be
realized among the baryon-charge carrying particles. The stellar
implication of this charge rearrangement is that the mixed phase
region of the star will have positively charged regions of nuclear
matter and negatively charged regions of quark matter.

Because of the competition between the Coulomb and the surface
energies associated with the positively charged regions of nuclear
matter and negatively charged regions of quark matter, the mixed phase
will develop geometrical structures, similarly as it is expected of
the subnuclear liquid-gas phase
transition \cite{ravenhall83:a,ravenhall83:b,williams85:a}. This
competition establishes the shapes, sizes and spacings of the rarer
phase in the background of the other in order to minimize the lattice
energy \cite{glen91:pt,glen01:b,glen95:a}. As known from the
quark-hadron compositions shown above, the formation of quark (q)
drops may set in around $3\, \rho_0$. At a somewhat greater density
the drops are more closely spaced and slightly larger in size.  Still
deeper in the star, the drops are no longer the energetically favored
configuration but merge together to form quark rods of varying
diameter and spacing. At still greater depth, the rods grow together
to quark slabs. Beyond this density the forms are repeated in reverse
order until at the inner edge of the mixed phase hadronic (h) drops of
finite size but separated from each other are immersed in quark
matter.  At densities between six to ten times $\rho_0$, the hadronic
drops have completely dissolved into pure quark
matter \cite{glen91:pt,glen01:b,glen97:book}.  In all cases the
geometric forms lie between about 10 and
25~fm \cite{glen97:book,glen95:a}.  The change in energy accompanied by
developing such geometrical structures is likely to be very small in
comparison with the volume
energy \cite{glen91:pt,glen01:b,heiselberg92:a,heiselberg95:crete} and,
thus, cannot much affect the global properties of a neutron star.
Nevertheless, the geometrical structure of the mixed phase may be very
important for transport phenomena as well as irregularities (glitches)
in the timing structure of pulsar spin-down
 \cite{glen91:pt,glen01:b,glen97:book}.

We conclude this section with presenting a representative model for
the \eos of a quark-hybrid star, which is shown in
Fig.~\ref{fig:3.6.1} \cite{hermann96:a}.
\begin{figure}[tb]
\begin{center}
\epsfig{figure=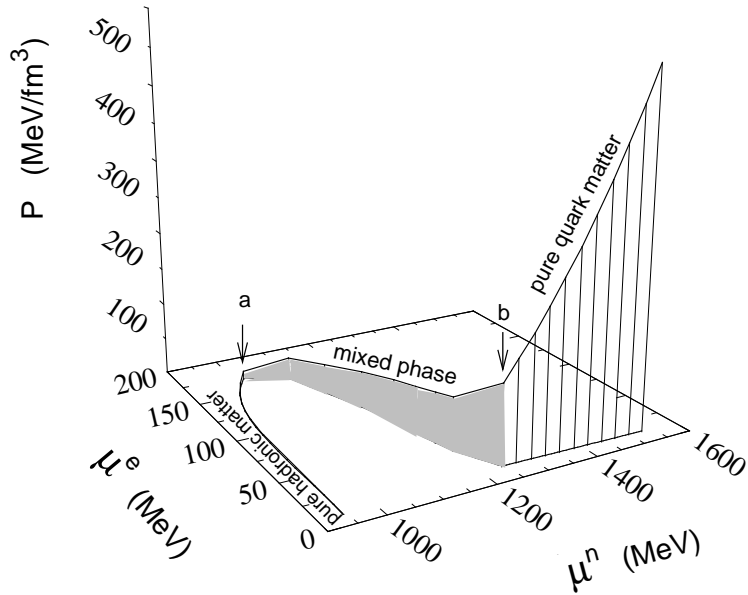,width=10.cm,angle=-90}
\begin{minipage}[t]{16.5 cm}
\caption{\Eos of neutron star matter accounting for quark
deconfinement \protect{ \cite{weber99:book}}.}
\label{fig:3.6.1}
\end{minipage}
\end{center}
\end{figure} The hadronic phase is modeled, as in
Fig.~\ref{fig:3.13.1}, in the framework of the relativistic Hartree
approximation (model HV of Refs.\
 \cite{weber99:book,weber89:e,weber99:topr}), the quark phase by the
bag model with a bag constant of $B=150~\mevt$ and a strange quark
mass of $150$~MeV.  The only difference with respect to Fig.\
\ref{fig:3.13.1} is the smaller bag constant which lowers the onset of
quark deconfinement from $3\, \rho_0$ to $2\,
\rho_0$ \cite{hermann96:a}.  Up to neutron chemical potentials of
$\mu^n \sim 10^3$~MeV, the matter stays in the pure hadronic
phase. The onset of quark deconfinement, which saturates the number of
electrons (cf.\ Fig.\ \ref{fig:3.13.1}), occurs at point $a$ in the
diagram where $\mu^e$ attains its maximum, $\mu^e \simeq 180$~MeV.  As
remarked just above, this values corresponds to a baryon number
density of about $2\, \rho_0$.  Beyond this density, $\mu^e$ decreases
toward rather small values because less and less electron are present
in dense quark-hybrid star matter.  The mixed phase region ($a$--$b$)
exists, in the direction of increasing density, for electron chemical
potential in the range $180~\mev \gs \mu^e \gs 25$~MeV, which
corresponds to neutron chemical potential of $10^3~\mev \ls \mu^n \ls
1.2\times 10^3~\mev$. For this range the volume proportion of quark
matter varies from $0 \leq \chi \leq 1$. The energy density in the
mixed phase is the same linear combination of the two phases as the
charge and baryon number \cite{glen91:pt,glen01:b,glen97:book}, namely
\begin{equation}
  \epsilon = (1-\chi) \, \epsilon_{\rm H}(\mu^n,\mu^e, \{\phi\}, T) +
  \chi \, \epsilon_{\rm Q}(\mu^n,\mu^e,T) \, .
\label{eq:eps.chi}
\end{equation} 
Most importantly, the pressure in the mixed phase region varies with
density rather than being constant, which would be the case if the
conservation of electron charge were ignored. The pure quark matter
phase ($\chi = 1$) sets in at $b$ where the density has grown to about
$6\, \rho_0$. It is characterized by a relatively steep increase of
pressure with density.  Whether or not this phase exists in neutron
stars constructed for this \eos depends of the star's central density
and thus on its mass.

\goodbreak
\subsubsection{\it H-dibaryons}\label{sec:hdibaryons}

A novel particle that could make its appearance in the center of a
neutron star is the so-called H-dibaryon, a doubly strange six-quark
composite with spin and isospin zero, and baryon number two
 \cite{jaffe77:a}. Since its first prediction in 1977, the H-dibaryon
has been the subject of many theoretical and experimental studies as a
possible candidate for a strongly bound exotic state.  In neutron
stars, which may contain a significant fraction of $\Lambda$ hyperons,
the $\Lambda$'s could combine to form H-dibaryons, which could give
way to the formation of H-dibaryon matter at densities somewhere above
$\sim 3\, \epsilon_0$ \cite{tamagaki91:a,sakai97:a,glen98:a} depending
on the in-medium properties of the H-dibaryon.  For an attractive
optical potential, $U_{\rm H}$, of the H-dibaryon at normal nuclear
density the equation of state is softened considerably, as shown in
Fig.\ \ref{fig:eosH}.  H-dibaryon matter could thus exist in the cores
of moderately dense neutron stars.  H-dibaryons with a vacuum mass of
about 2.2~GeV and a moderately attractive potential in the medium of
\begin{figure}[tb]
\begin{center}
\includegraphics[height=0.3\textheight]{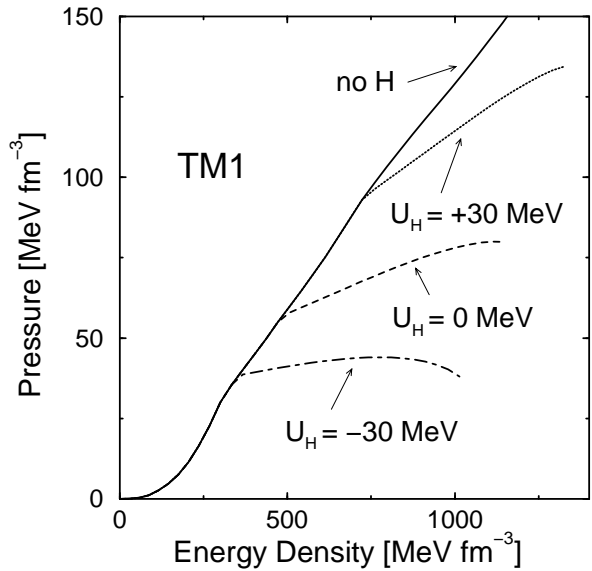}
\begin{minipage}[t]{16.5 cm}
\caption{Equations of state of neutron star matter accounting for a
  H-dibaryon condensate \cite{glen98:a}. $U_{\rm H}$ is the optical
  potential of the H-dibaryon at normal nuclear density.  Reprinted
  figure with permission from N. K. Glendenning and
  J. Schaffner-Bielich, Phys.~Rev.~C 58 (1998) 1298.  Copyright 1998
  by the American Physical Society.}
\label{fig:eosH}
\end{minipage} 
\end{center}
\end{figure}
about $U_{\rm H} = - 30$~MeV, for instance, could go into a boson
condensate in the cores of neutron stars if the limiting star mass is
about that of the Hulse-Taylor pulsar PSR~1913+16, $M=1.444\, \msun$
 \cite{glen98:a}. Conversely, if the medium potential were moderately
repulsive, around $U_{\rm H} = + 30$~MeV, the formation of H-dibaryons
may only take place in heavier neutron stars of mass $M\gs 1.6\,
\msun$. If formed, however, H-matter may not remain dormant in neutron
stars but, because of its instability against compression could
trigger the conversion of neutron stars into hypothetical strange
stars \cite{sakai97:a,faessler97:a,faessler97:b}.

\goodbreak
\subsection{\it Strange Stars}\label{sec:sstars}

\subsubsection{\it General Properties}\label{sec:ss_general}

If the strange matter hypothesis is true, a new class of compact stars
called strange stars should exist. Possible strange star candidates as
compiled in table \ref{tab:ss_candidates}.
\begin{table}[b]
\caption{Possible strange star candidates.}\label{tab:ss_candidates}
\begin{center}
\begin{tabular}{l|l|l} \hline
Compact object                 &Peculiar feature                &References \\ \hline
\rxj1856                       &Small radius                    & \cite{turolla04:a,drake02:a,gondek02:a,haensel01:a,xu02:b}   \\
\eu1728                        &Small radius                    & \cite{bombaci03:a} \\
\sax                           &Small radius                    & \cite{li99:a}            \\
Her~X-1                        &Small radius                    & \cite{dey98:a}          \\
\1e1207                        &Peculiar timing                 & \cite{xu04:a}           \\
\p0943                         &Microstorms                     & \cite{xu99:a}             \\
\3c58 (J0205+6449)             &Low temperature                 & \cite{slane02:a} \\
\groj1744                      &X-ray burst features            & \cite{cheng98:a} \\
\sgr0526, \sgrnineteenoo, \sgreighteenosix  &X-ray burst features            & \cite{usov01:a,zhang00:a,cheng96:a,cheng02:a} \\
\hline
\end{tabular}
\end{center}
\end{table}
They would form a distinct and disconnected branch of compact stars
and are not a part of the continuum of equilibrium configurations that
include white dwarfs and neutron stars (see Fig.\ \ref{fig:mr145}). In
principle both strange and neutron stars could coexist. However if
strange matter is the true ground state, the galaxy is likely to be
contaminated by strange quark nuggets which, depending
\begin{figure}[tb]
\begin{center}
\epsfig{figure=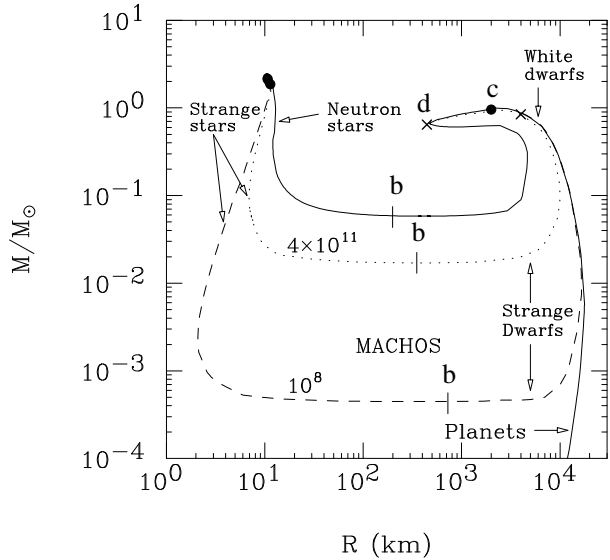,width=8.0cm,angle=0}
\begin{minipage}[t]{16.5 cm}
\caption{Mass versus radius of strange-star configurations with
nuclear crusts (dashed and dotted curves) and gravitationally bound
neutron stars and white dwarfs (solid curve). The strange stars carry
nuclear crusts with chosen inner densities of $\ecrusti = 4\times
10^{11}~\gcmt$ and $\ecrusti= 10^{8}~\gcmt$, respectively.  Crosses
denote the termination points of strange-dwarfs sequences, whose quark
matter cores have shrunk to zero. Dots refer to maximum-mass stars,
minimum-mass stars are located at the vertical bars labeled `b'.
(Fig.\ from Ref.\  \cite{weber99:book}.)}
\label{fig:mr145}
\end{minipage}
\end{center}
\end{figure}
on their velocities \cite{madsen03:paper+reply}, could convert neutron
stars to strange stars \cite{madsen98:b,glen91:a,caldwell91:a}. This
would mean that the objects known to astronomers as pulsars would be
rotating strange stars (see Fig.\ \ref{fig:Sstar}) rather than
rotating neutron stars.  Another peculiar
\begin{figure}[tb]
\begin{center}
\epsfig{figure=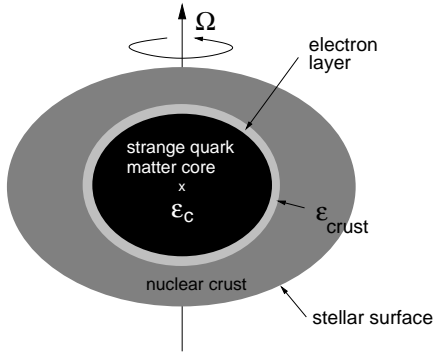,width=6.cm,angle=0}
\begin{minipage}[t]{16.5 cm}
\caption{Schematic illustration of the cross section of a rotating
strange star carrying a nuclear crust. The quantities $\epsilon_c$ and
$\ecrusti$ denote the star's central density and the density at the
base of the crust, respectively. (Fig.\ from Ref.\
 \cite{weber99:book}.)}
\label{fig:Sstar}
\end{minipage}
\end{center}
\end{figure}
consequence of the hypothesis could be the existence of an entirely
new class of dense white-dwarf-like strange stars, called strange
dwarfs, plus an expansive range of planetary-like strange matter stars
referred to as strange MACHOS. These objects could carry nuclear
crusts that are between several hundred and several thousand
kilometers thick \cite{glen94:a}.  The situation is graphically
illustrated in Fig.\ \ref{fig:mr145}.  The important astrophysical
implication of the existence of strange MACHOS would be that they
occur as natural stellar candidates which effectively hide baryonic
matter, linking strange quark matter to the fundamental dark matter
problem, which is currently one of the problems of greatest importance
to astrophysics and cosmology.  Observationally, the strange MACHOS
could be seen by the gravitational microlensing
experiments \cite{alcock00:a}, provided such objects exist abundantly
enough.

As described in section \ref{sec:primerqm}, strange quark matter is
expected to be a color superconductor which, at extremely high
densities, should be in the CFL phase. This phase is rigorously
electrically neutral with no electrons
required \cite{rajagopal01:b}. For sufficiently large strange quark
masses, however, the low density regime of strange quark matter is
rather expected to form other condensation patterns (e.g. 2SC,
CFL-$K^0$, CFL-$K^+$, CFL-$\pi^{0,-}$) in which electrons are
present \cite{rajagopal01:a,alford01:a}. The presence of electrons
causes the formation of an electric dipole layer on the surface of
strange matter, as illustrated schematically in Fig.\
\ref{fig:Core_crust},
\begin{figure}[tb]
\begin{center}
\epsfig{figure=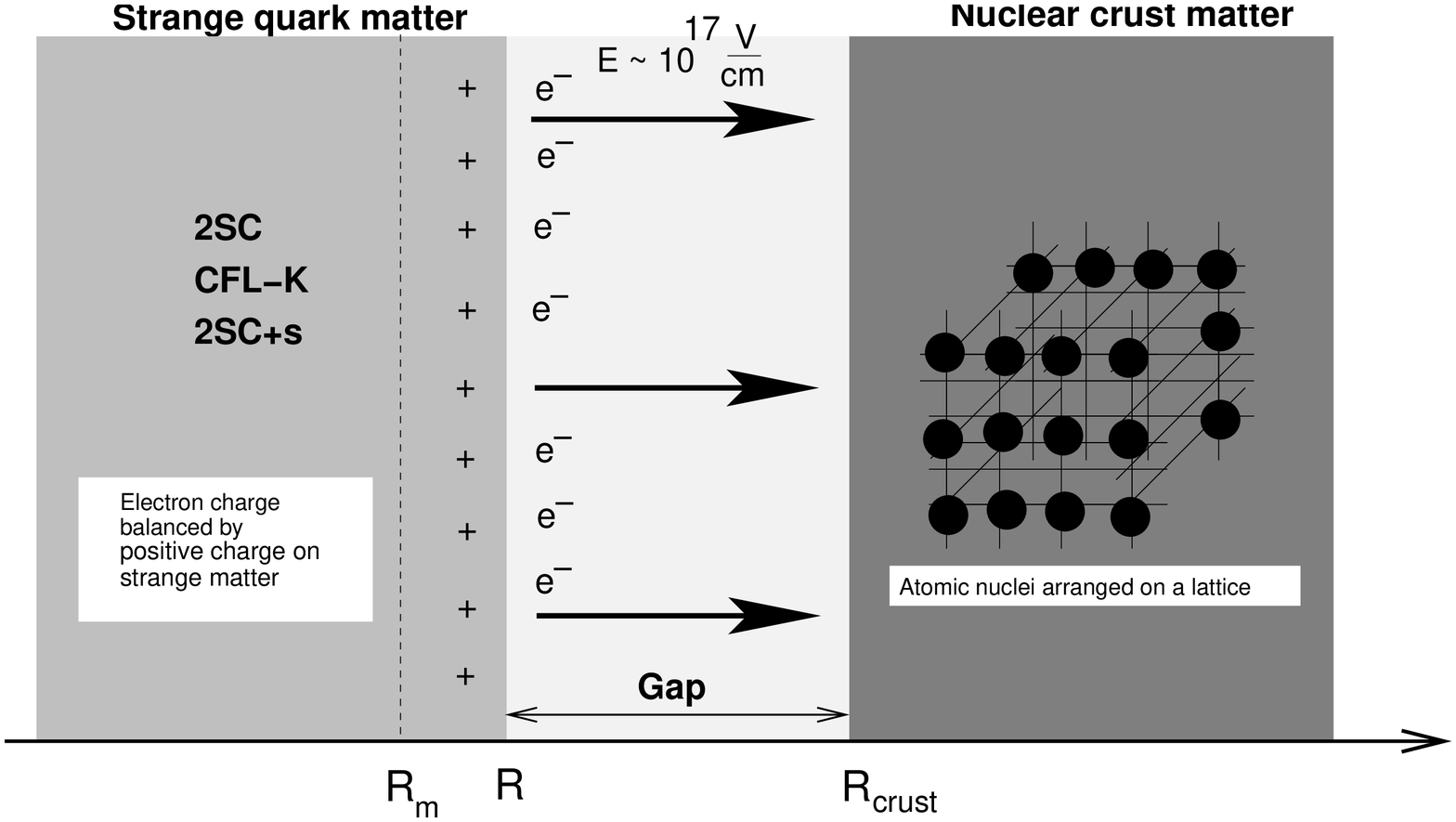,width=10.0cm,angle=0} 
\begin{minipage}[t]{16.5 cm}
\caption{Surface region of strange quark matter.  The region between
$R\leq r \leq \rcrust$ is filled with electrons that are bound to
strange matter but extend beyond its surface, $R$, leading to a
deficit of electrons in the range $R_{\rm m} \leq r \leq R$ and
therefore a net positive charge in this region. The associated
electric field, $E \sim 10^{17}$~V/cm, is sufficiently strong
to avoid contact between atomic matter and strange matter, enabling
strange matter to be enveloped by ordinary atomic matter. (Fig.\ from Ref.\
 \cite{weber99:book}.)}
\label{fig:Core_crust}
\end{minipage}
\end{center}
\end{figure} 
which enables strange quark matter stars to carry nuclear
crusts \cite{alcock86:a,alcock88:a,kettner94:b}.  An analytical
expression for the electron number density and the electric field on
the surface of strange quark matter can be derived using the
Thomas-Fermi model.  One finds \cite{xu99:b} that
\begin{eqnarray}
n_e \sim {{9.5 \times 10^{35}} \over {(1.2 \, z_{11} + 4)^3}} ~{\rm
cm}^{-3} \, , \quad E \sim {{7.2 \times 10^{18}} \over {(1.2 \, z_{11}
+ 4)^2}} ~{\rm V} ~ {\rm cm}^{-1} \, ,
\label{eq:ne+E}
\end{eqnarray}
where $z$ is a measured height above the quark surface, $z_{11} = z /
(10^{-11} ~ {\rm cm})$. From Eq.\ (\ref{eq:ne+E}) one sees that very
strong electric fields, on the order of $\sim 10^{17}$~V/m, may be
expected near the quark surface. This makes it possible for a
non-rotating star to support a nuclear crust with a mass up to $\sim
10^5\, \msun$ \cite{alcock86:a}. The maximal possible density at the
base of the crust, called the inner crust density, is determined by
neutron drip, which occurs at about $4\times 10^{11}~\gcmt$.  This
somewhat complicated situation of the structure of strange matter
enveloped in a nuclear crust can be represented by a proper choice for
the \eos shown in Fig.\ \ref{fig:eos.ss}. The \eos is characterized by
a discontinuity in density between strange quark matter and nuclear
\begin{figure}[tb]
\begin{center}
\epsfig{file=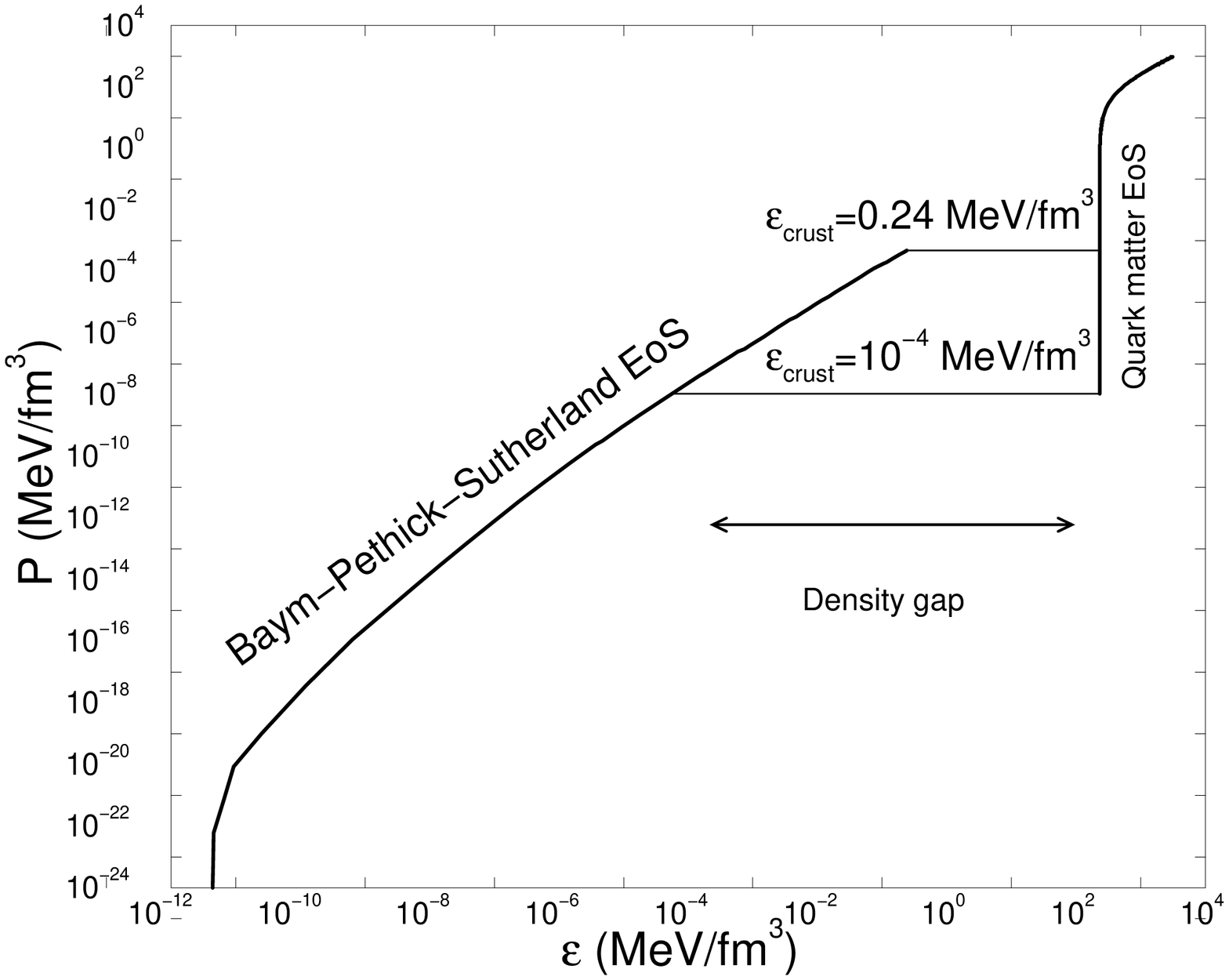,width=6cm,angle=-90}
\begin{minipage}[t]{16.5 cm}
\caption{Equation of state of strange quark matter surrounded by a
nuclear crust \cite{glen92:crust}. The maximal possible nuclear matter
density is determined by neutron drip which occurs at $\ecrusti =
0.24~\mevt$ ($4.3\times 10^{11}~\gcmt$). Any nuclear density that is
smaller than neutron drip is possible. As an example, we show the \eos
for a sample density of $\ecrusti = 10^{-4}~\mevt$ ($10^8~\gcmt$).}
\label{fig:eos.ss}
\end{minipage}
\end{center}
\end{figure}
crust matter across the electric dipole gap where the pressure of the
nuclear crust at its base equals the pressure of strange matter at its
surface \cite{weber99:book,glen92:crust}.

Since the nuclear crust surrounding the quark matter core of a strange
star is bound to the core by the gravitational force rather than
confinement, the mass-radius relationship of strange matter stars with
nuclear crust is qualitatively similar to the one of neutron stars and
white dwarfs, which are exclusively bound by gravity.  The
strange-star sequences in Fig.\ \ref{fig:mr145} are computed for the
maximal possible inner crust density set by neutron drip ($\ecrusti =
4\times 10^{11}~\gcmt$) as well as for a considerably smaller sample
density of $10^8~\gcmt$.  Of course there are other possible sequences
of strange stars with any smaller value of inner crust
density \cite{weber99:book,glen92:crust,weber93:b}. From the
maximum-mass star (solid dots), the central star density decreases
monotonically through the sequence in each case.  The fact that
strange stars with nuclear crusts possess smaller radii than neutron
stars leads to smaller Kepler (mass shedding) periods, $\pkgr$, for
strange stars. This is indicated by the classical expression $\pkgr
\propto \sqrt{R^3/M}$ and has its correspondence in the general
relativistic expression for $\pkgr$ derived in Eq.\
(\ref{eq:okgr}). Since the qualitative dependence of $\pkgr$ on mass
and radius remains valid \cite{glen93:drag}, one finds that the
complete sequence of strange stars, and not just those close to the
mass peak as is the case for neutron stars, can sustain extremely
rapid rotation.  In particular, a strange star with a typical pulsar
mass of $\sim 1.4\,\msun$ can rotate at Kepler periods as small as $\pkgr
\sim 0.5$~ms, depending on crust thickness and the modeling of
strange quark matter \cite{weber99:book}.  This is to be compared with
the larger limiting value of $\pkgr\sim 1$~ms obtained for neutron
stars of the same mass \cite{weber99:book}. Exceptions to this,
however, are possible if the nuclear matter exhibits a very strong
softening at intermediate densities \cite{burgio03:a}.

Of considerable relevance for the viability of the strange matter
hypothesis is the question of whether strange stars can exhibit
glitches in rotation frequency. Pulsar glitches are sudden changes in
the rotational frequency of a rotating neutron star which otherwise
decreases very slowly with time due to the loss of rotational energy
through the emission of electromagnetic dipole radiation and an
electron-positron wind. They occur in various pulsars at intervals of
days to months or years, and in some pulsars are small (Crab), and in
others large (Vela) and infrequent ($\Delta \Omega/\Omega \sim
10^{-8}-10^{-6}$, respectively).  Glitches have been attributed to
several causes related to the assumed structure of neutron stars.  One
such is the crust quake in which an oblate solid crust in its present
shape slowly comes out of equilibrium with the forces acting on it as
the period of rotation changes, and fractures when the built up stress
exceeds the sheer strength of the crust
material \cite{ruderman69:a,baym71:c}.  The stellar frequency and rate
of change of frequency, $\Omega$ and $\dot \Omega$ respectively,
slowly heal to the trend preceding the glitch as the coupling between
crust and core reestablishes their corotation.  The compatibility of
pulsar glitches with the strange matter hypothesis will have a
decisive impact on the question of whether or not strange matter is
the true ground state of strongly interacting
matter \cite{glen92:crust,alpar87:a}.  From the study performed in
Refs.\  \cite{glen92:crust,zdunik01:a} it is known that the ratio of
the crustal moment of inertia to the total moment of inertia,
$\icrust/\itotal$, varies between $10^{-3}$ and $\sim 10^{-5}$.  If
the angular momentum of the pulsar is conserved in a stellar quake
then the relative frequency change and moment of inertia change are
equal, and one arrives for the change of the star's frequency at
 \cite{glen92:crust}
\begin{equation}
       {{\Delta \Omega}\over{\Omega}}  =  {{|\Delta I|}\over
       {I_0}}  >  {{|\Delta I|}\over {I}}  \equiv  f \,
       {\icrust\over I} \sim  (10^{-5} - 10^{-3})\, f \, , ~{\rm
       with} \quad 0 < f < 1\, .
\label{eq:delomeg}
\end{equation}
Here $I_0$ denotes the moment of inertia of that part of the star
whose frequency is changed in the quake. It might be that of the crust
only, or some fraction, or all of the star. The factor $f$ in Eq.\
(\ref{eq:delomeg}) represents the fraction of the crustal moment of
inertia that is altered in the quake, i.e., $f \equiv |\Delta I|/
\icrust$.  Since the observed glitches have relative frequency changes
$\Delta \Omega/\Omega = (10^{-9} - 10^{-6})$, a change in the crustal
moment of inertia of $f\ls 0.1$ would cause a giant glitch even in the
least favorable case \cite{glen92:crust}. Moreover it turns out that
the observed range of the fractional change in the spin-down rate,
$\dot \Omega$, is consistent with the crust having the small moment of
inertia calculated and the quake involving only a small fraction $f$
of that, just as in Eq.\ (\ref{eq:delomeg}).  For this purpose we
write \cite{glen92:crust}
\begin{equation} 
        { {\Delta \dot\Omega}\over{\dot\Omega } }  = 
        { {\Delta \dot\Omega /  \dot\Omega} \over 
          {\Delta    \Omega  /     \Omega }  } \,
        { {|\Delta I |}\over{I_0} }  = 
        { {\Delta \dot\Omega /  \dot\Omega} \over  
          {\Delta    \Omega  /     \Omega }  } \, f \,
         {\icrust\over {I_0} }  >  
       (10^{-1}~ {\rm to} ~ 10) \, f \, , 
\label{eq:omdot}
\end{equation} 
where use of Eq.\ (\ref{eq:delomeg}) has been made. Equation
(\ref{eq:omdot}) yields a small $f$ value, i.e., $f < (10^{-4} ~ {\rm
to} ~ 10^{-1})$, in agreement with $f \ls 10^{-1}$ established just
above. Here measured values of the ratio $(\Delta \Omega/\Omega) /
(\Delta \dot \Omega / \dot\Omega) \sim 10^{-6}$ to $10^{-4}$ for the
Crab and Vela pulsars, respectively, have been used.

\goodbreak
\subsubsection{\it \sax}

As discussed just above, strange stars are self-bound objects at zero
external pressure, which would exist stably even if gravity were
switched off. The latter makes such objects even more compact. The
radii of compact quark stars are thus expected to be smaller than
those of neutron stars of comparable mass, which are bound solely by
gravity.  Figures \ref{fig:mrad2}, \ref{fig:mr145} and
\ref{fig:bombaci_MR} show that this difference in radius may be three
to four kilometers for stars of canonical mass, $\sim 1.4 \, \msun$,
and even bigger than that for less massive objects. One neutron star
that may have such an unusually small radius is the transient X-ray
burst source \sax\  \cite{li99:a}, which was discovered in September
1996 by the BeppoSAX satellite.  Two bright type-I X-ray bursts were
detected, each one lasting less than 30 seconds. An analysis of the
bursts in \sax\ indicates that it is 4~kpc distant and has a peak
X-ray luminosity of $6\times 10^{36}$~erg/s in its bright state, and a
X-ray luminosity lower than $10^{35}$~erg/s in
quiescence \cite{zand98:a}.  Coherent pulsations at a period of 2.49
milliseconds were also discovered \cite{wijnands98:a}.  The binary
nature of \sax\ was firmly established with the detection of a 2 hour
orbital period \cite{chakrabarty98:a} as well as with the optical
identification of the companion star.  \sax\ is the first pulsar that
shows both coherent pulsations in its persistent emission and X-ray
bursts.

Li \etal\  \cite{li99:a} extracted a mass-radius relationship for the
compact star in \sax\ from the following two requirements on the
geometry of the stellar binary system (see also section \ref{sec:spin}
where mass accretion onto a compact star is discussed in more
detail). Firstly, detection of X-ray pulsations requires that the
inner radius, $r_{\rm m}$, of the accretion disk should be larger than
the stellar radius $R$. Secondly, the inner radius must be smaller
than the disk's co-rotation radius, $r_{\rm c}$. Otherwise accretion
will be inhibited by a centrifugal barrier.  From these two conditions
one finds that $r_{\rm m} \ls r_{\rm c} = (M P^2/(4\pi^2))^{1/3}$,
with $M$ and $P$ the star's mass and rotational period,
respectively. Expressing the location of the inner disk in terms of
the Alfv\'en radius, $r_{\rm A}$, which is that distance from the
neutron star at which the accreting matter is pulled by the magnetic
fields to the poles of the star which happens when the kinetic energy
density is comparable to the magnetic energy density \cite{heuvel91:a},
one arrives at $r_{\rm m} = \xi r_{\rm A} = \xi (B^2 R^6 / \dot{M}
(2M)^{1/2})^{2/7}$. The symbols $B$ and $\dot{M}$ are the surface
magnetic field and the mass accretion rate of the rotating neutron
star, respectively.  The quantity $\xi\sim 1$ is a parameter which
depends very weakly on the accretion rate \cite{li97:c}.  Denoting the
minimum and maximum accretion rates of \sax\ as $\dmmin$ and $\dmmax$,
the conditions discussed above can be expressed as
\begin{equation}
r \ls r_{\rm m} (\dmmax) < r_{\rm m} (\dmmin)\ls r_{\rm c} \, .
\label{eq:r_inequality}
\end{equation}
To connect this relation to the observed data, let us assume that the
mass accretion rate is proportional to the X-ray flux $F$ observed
with the Rossi X-ray Timing Explorer (RXTE). This is supported by the
fact that the X-ray spectrum of \sax\ was remarkably stable and that
\begin{figure}[tb]
\begin{center}
\includegraphics[scale=0.65,angle=0]{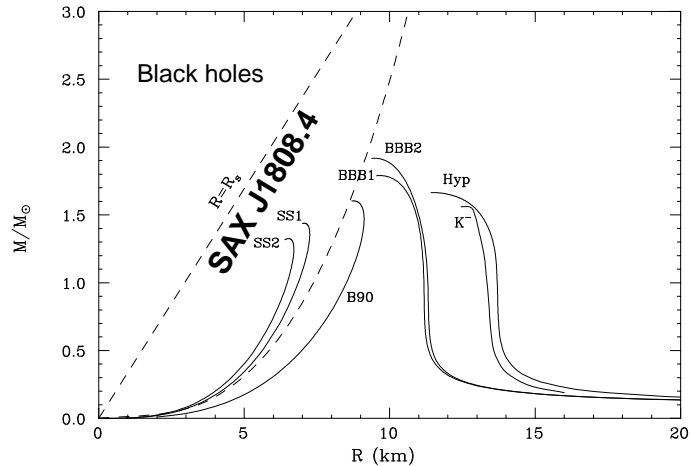}
\begin{minipage}[t]{16.5 cm}
\caption{Comparison of the mass-radius relationship of \sax \
determined from RXTE observations (i.e.\ Eq.\ (\ref{eq:MR-sax})) with
theoretical models of neutron stars and strange
stars \cite{li99:a}. The solid curves represents theoretical
mass-radius relationships for neutron stars and strange stars.
Reprinted figure with permission from X. D. Li \etal,
Phys.~Rev.~Lett. 83 (1999) 3776.  Copyright 1999 by the American
Physical Society.}
\label{fig:bombaci_MR}
\end{minipage}
\end{center}
\end{figure}
there was only a slight increase in the pulse amplitude when the X-ray
luminosity varied by a factor of $\sim 100$ during the 1998 April/May
outburst \cite{gilfanov98:a,cui98:a,psaltis99:a}. From this, an upper
limit for the stellar radius of $R < (F_{\rm min}/F_{\rm max})^{2/7}
r_{\rm c}$ was derived in \cite{li99:a}. This relation can be
conveniently written as
\begin{eqnarray} 
R < 27.5 \bigg({{F_{\rm min}} \over {F_{\rm max}}}\bigg)^{2/7}
\bigg({{P} \over {2.49~{\rm ms}}}\bigg)^{2/3} \bigg({{M} \over
{M_\odot}}\bigg)^{1/3} ~{\rm km} \, ,
\label{eq:MR-sax}
\end{eqnarray} where $\fmax$ and $\fmin$ denote the X-ray fluxes
measured during high and low X-ray emission states, respectively.
Adopting a flux ratio of $\fmax / \fmin ~\sim 100$, Eq.\
(\ref{eq:MR-sax}) constrains the mass-radius values of \sax \ to
values that lie between the dashed curves in Fig.\
\ref{fig:bombaci_MR}, which suggests that \sax \ could be a strange
star.  The dashed line labeled $R = R_{\rm s}$ $(= 2 M R)$ denotes the
Schwarzschild limit on the radius of a compact object set by
gravitational collapse to a black hole. The curves labeled BBB1, BBB2,
Hyp, and $K^-$ denoted the mass-radius relationships of conventional
neutron stars computed for different models for the \eosp.
Constraints on the mass-radius relationship which, if robust, are
better described in terms of strange stars than neutron stars have
also been obtained for the compact star in the X-ray source 4U~1728-34
($M < 1.0~M_\odot$, $R < 9 ~{\rm km}$)  \cite{li99:b}, for the isolated
neutron star \rxj1856  \cite{turolla04:a}, which will be discussed in
the next section, as well as for the X-ray pulsar Her X-1 ($M = 1.1-
1.8\, \msun$, $R = 6.0 - 7.7~{\rm km}$) \cite{dey98:a}. (See also table
\ref{tab:ss_candidates}.)

In passing, we mention the recent discovery of significant absorption
lines in the spectra of 28 bursts of the low-mass X-ray binary
\exo \cite{cottam02:a}.  These lines have been identified with iron and
oxygen transitions, all with a gravitational redshift of $z=0.35$. As
shown in Ref.\  \cite{cottam02:a}, for a stellar mass range of $M\sim
1.3 - 2.0\, \msun$ such a $z$ value is completely consistent with
conventional neutron star models, made of normal nuclear matter, and
excludes even some models in which neutron stars are made of more
exotic matter.

\goodbreak
\subsubsection{\it \rxj1856}\label{sec:rxj1856}

Aside from X-ray emission from neutron stars in binaries, discussed in
the previous section, the thermal emission from the surface of an
isolated neutron star (INS) is of key importance for the determination
of the mass and radius of a neutron star, too. ROSAT was the first
satellite with a sufficient sensitivity in the X-ray band to perform a
systematic search and study of such objects. The nearby neutron star
\rxj1856, discovered in 1992  \cite{walter96:a}, is the brightest INS
in X-rays. It does not show any signs of activity such as variability
or pulsation.  Since its discovery \rxj1856 was studied in great
detail in X-rays, UV and the optical band using and variety of
different astrophysical observatories (ROSAT, EUVE, ASCA, HST,
Chandra, XMM Newton, VLA). Detailed Chandra observations of \rxj1856
have shown that this neutron star has a featureless thermal spectrum
for which a simple blackbody distribution seems to provide a better
fit to X-ray data than more sophisticated atmospheric
models \cite{burwitz01:a,burwitz03:a}.  Several sets of observations
taken with ROSAT, EUVE, NTT, Keck, HST and Chandra have yielded a
proper motion of 330~mas/yr and parallax 8.5~mas, which corresponds to
a transverse stellar velocity of about 200~km/s and a distance from
Earth of $D=120$~pc  \cite{kaplan01:a,walter02:a}. These measurements
could indicate that the star originated $\sim 10^6$~years ago in the
Upper Scorpio association at about the same time that a supernova
ejected the runaway O star $\zeta$ Ophiuchus, suggesting that \rxj1856
and $\zeta$ Ophiuchus could have been members of a binary system and
that \rxj1856 may be the remnant of the star that exploded.
\begin{figure}[tb]
\begin{center}
\epsfig{file=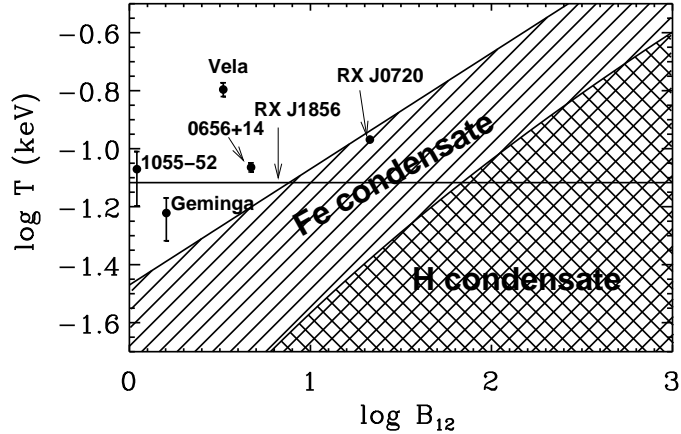,width=10.0cm,angle=0}
\begin{minipage}[t]{16.5 cm}
\caption{The critical temperature for hydrogen (H) and iron (Fe) as a
function of magnetic field, $B$ \cite{turolla04:a}. A phase transition
to a solid condensate is possible in the shaded region for Fe and in
the cross-hatched region for H. The positions of five cool, isolated
neutron stars (see table \ref{tab:observations1}) is shown. The
horizontal line is drawn in correspondence to the color temperature of
\rxj1856. Reprinted figure with permission from R. Turolla, S. Zane,
and J. J. Drake, Astrophys.~J.~603 (2004) 265. Copyright 2004 by the
Astrophysical Journal.}
\label{fig:f1_Turolla}
\end{minipage}
\end{center}
\end{figure}
The apparent (redshifted) radius of \rxj1856 can be calculated
from \cite{turolla04:a}
\begin{eqnarray}
R^\infty = 4.25 f^{-1/2}_A f^{-1/2}_E \gamma^2 D_{100} \left( T_{\rm
bb}^\infty / 60 ~{\rm eV} \right)^{-2} ~ {\rm km} \, .
\label{eq:rad_radius}
\end{eqnarray}
The true (coordinate) radius of the star, $R$, is connected to the
apparent radius through
\begin{equation}
R = R^\infty (1 - 2M/R)^{1/2} \, .
\label{eq:r_true}
\end{equation}
The quantity $f_A$ in Eq.\ (\ref{eq:rad_radius}) denotes the fraction
of the stellar surface responsible for the observed emission, $f_E$ is
the ratio of emitted power to blackbody power, $\gamma$ stands for the
spectral hardening, $D_{100} \equiv D / (100~{\rm pc})$ is the star's
distance, and $T_{\rm bb} = T_{\rm bb}^\infty (1 - 2M/R)^{-1/2}$ its
blackbody temperature. If the emission comes from the entire stellar
surface ($f_A=1$) and the temperature distribution is uniform ($\gamma
= 1$), Eq.\ (\ref{eq:rad_radius}) simplifies to
\begin{eqnarray}
R^\infty = 4.25 f^{-1/2}_E D_{100} \left( T_{\rm bb} / 60 ~{\rm eV}
\right)^{-2} ~ {\rm km} \, .
\label{eq:rad_radius2}
\end{eqnarray}
Based on this relation Drake {\it et al.}  \cite{drake02:a} reported an
apparent radius for \rxj1856 of $R^\infty / D_{100} = 4.12 \pm
0.68$~km, where the quoted uncertainty represents the combined
temperature determination uncertainty ($\pm 1$~eV) and the dominant
absolute effective area uncertainty of Chandra's Low Energy
Transmission Grating (LETG) and High Resolution Camera spectroscopic
plate detector array (HRC-S) combination ($\pm 15\%$).  Depending on
the star's distance, which still appears to be an open
issue \cite{kaplan03:KIAS}, this implies a rather small apparent
stellar radius in the range between about $R^\infty=4$~km to 8~km.
The latest distance measurement of 175~pc  \cite{kaplan03:KIAS}
combined with specific assumptions about the star's surface
composition (see below) could favor $R^\infty \simeq
8$~km \cite{drake03:KIAS}, which corresponds to a true radius of about
6~km. As can be seem from Figs.\ \ref{fig:mrad2} and
\ref{fig:bombaci_MR}, such a value would be too small for
conventional neutron star equations of state \cite{drake02:a} which
predict $12 ~{\rm km} \ls R^\infty \ls 17 ~{\rm
km}$ \cite{lattimer01:a}.  Proposed explanations for such a small
radius include a reduced X-ray emitting region (such as a heated polar
cap), or the presence of a more compact object such as a strange quark
star \cite{gondek02:a,haensel01:a,xu02:b}.  An alternative possibility,
suggested in \cite{turolla04:a}, could be cold neutron stars ($T \ls
10^6$~K) endowed with rather strong magnetic fields ($B \gs
10^{13}$~G) and with metal dominated outer layers. Depending on $T$
and $B$ such stars may undergo a phase transition to a solid
condensate in the outermost layers (Fig.\ \ref{fig:f1_Turolla}),
resulting in an X-ray spectrum that is featureless as observed for
\rxj1856. The observed UV-optical enhanced emission could be explained
by the presence of a gaseous, thin hydrogen shell on top of the iron
condensate, where the optical flux is reprocessed.  This model
predicts a value for the apparent radius of \rxj1856 in the range of
$R^\infty \sim 8 - 12$~km, depending on whether one assumes uniform or
meridional variation temperature distributions on the star's surface.

For a canonical neutron star mass of $1.4\, \msun$, apparent radii in
the range $R^\infty \sim 10 - 12$~km (true radii of $R \sim 7 - 9$~km
for a $1.4\, \msun$ neutron star) can be accommodated by conventional
models for the equations of state of superdense matter which exhibit a
soft behavior at high densities \cite{turolla04:a}.  This is different
for $R^\infty \ls 8$~km which may indicate that such stars are made of
self-bound hadronic matter, of which strange quark matter may be the
most plausible state of matter. Sequences of such stars are shown in
Figs.\ \ref{fig:mrad2} and \ref{fig:bombaci_MR}. While a strange star
may be an intriguing conceivable option, present observations of
\rxj1856 do not necessarily demand this interpretation and more
conventional interpretations involving conventional neutron stars are
certainly possible \cite{turolla04:a,thoma03:a}.  A conventional
interpretation is also supported by models which fit the X-ray and
optical data of \rxj1856 with a two-component blackbody
model \cite{burwitz03:a}. The latter are best fitted with
$T^\infty_{\rm bb,X} \simeq 63.5$~eV and $R^\infty_{\rm bb,X} \simeq
4.4 (D/120~{\rm pc})$~km for the hot X-ray emitting region, and
$T^\infty_{\rm bb,opt} < 33$~eV and $R^\infty_{\rm bb,opt} > 17
(D/120~{\rm pc})$~km for the rest of the neutron star surface
responsible for the optical flux \cite{burwitz03:a}.

\goodbreak
\subsubsection{\it The Neutron Star in \3c58}\label{sec:3c58}

Murray {\it et al.} discovered the 65~ms pulsars \psrotwoofive \ in
the supernova remnant \3c58 located in the constellation
Cassiopeia \cite{murray02:a}. Historical evidence strongly suggests an
association of the remnant with supernova SN~1181, which went off in
1181~AD. This renders \psrotwoofive \ younger than B0531+21, the
pulsar in the Crab nebula, born in a supernova explosion recorded by
Chinese astronomers in 1045~AD. Using data provided by the Chandra
X-ray observatory, Slane {\it et al.} inferred an upper limit on the
effective surface temperature of \psrotwoofive of only $T^\infty <
1.08 \times 10^6$~K which falls below predictions from standard
cooling models \cite{slane02:a}. Based on its low temperature, it was
suggested that \psrotwoofive\ may be a strange quark star rather than
a neutron star \cite{slane02:a}. We will discuss this issue in greater
detail in section \ref{sec:cooling}.

\goodbreak
\subsubsection{\it X-ray, Gamma-Ray Burst and SGR  Associations}

X-ray bursts are sudden increases in the X-ray flux of X-ray sources,
with rise times of $\le$ 1s, and subsequent decay with characteristic
times ranging from 10s to a few minutes.  They are classified into two
classes: type-I and type-II X-ray bursts. Type-I X-ray bursts are
characterized by the relatively long burst intervals (hours to days)
and the significant spectral softening during the burst decay as
compared to their type-II counterparts. The differences between
type-I(II) bursts and the HXRBs is discussed by \cite{lewin96:a}.

Gamma ray bursts (GRBs) are most intense transient gamma ray events in
the sky when they are on but the nature of gamma ray bursts is still a
mystery since their discovery \cite{klebesadel73:a}. The isotropic and
inhomogeneous distribution of GRBs detected with BATSE and the
identification of counterparts in radio, optical, and X-ray by
BeppoSAX and other ground based telescopes supports that they are
located at cosmological distances, which make them the most energetic
events ever known \cite{fishman95:a,piran99:a}.  Gamma ray bursts at
truly cosmological distances could be due to collisions of two neutron
or two strange stars in binary systems \cite{haensel91:a}, depending
on the true ground state of strongly interacting matter, and/or may
also involve black holes \cite{brown00:a}. Binary neutron star or
strange star collisions could release $\sim 10^{52}$~ergs in the form
of gamma rays over a time period of about 0.2~s. The central engine
that powers gamma ray bursts should be capable of releasing a total
energy of $\sim 10^{53}$~erg, which may or may not be possible in
stellar collisions.  This may be different for the conversion of a
neutron star to a strange star \cite{bombaci00:a}.  Depending on the
model for the \eos the total energy given off in such a conversion is
$(1-4) \times 10^{53}$~erg \cite{bombaci00:a}.

Bare strange stars have also been associated with soft gamma repeaters
(SGRs).  Unlike gamma ray bursts, which emit large amounts of high
energy gamma radiation, SGRs have a larger proportion of lower-energy
X-ray radiation. Also, in contrast to gamma ray bursts, which can
rumble on for many minutes, SGRs pop off in a fraction of a second.
Conventional models associate SGRs with young neutron stars energizing
a large cloud of gas cast off in a supernova explosion, or X-ray
binary stars that accrete matter at irregular intervals and emit gamma
rays when the accreted matter hits the surface. Most strikingly, the
intensity of the outbursts is between $10^3$ and $10^4$ times the
Eddington limit, $L_{\rm Edd}$. The latter is defined as the critical
luminosity at which photon radiation pressure from the surface of a
star of mass $M$ equals gravity,
\begin{equation}
 L_{\rm Edd} = 1.3 \times 10^{38} \bigl( M / \msun \bigr) ~{\rm erg}
   ~{\rm s}^{-1} \, .
\label{eq:Eddington}
\end{equation}
The Eddington limit does not apply to bare strange stars since quark
matter is held in place by the strong interaction and not gravity (see
section \ref{sec:spss}).  For that reason the bursting activity of
SGRs can be comfortably explained by fast heating of the surfaces of
bare quark stars up to temperatures of $\sim (1 - 2) \times 10^9$~K
and its subsequent thermal emission \cite{usov01:a,usov01:c}.  The
fast heating mechanism of SGRs may be either impacts of comets onto
bare strange stars \cite{usov01:a,zhang00:a,usov01:b} or fast decay of
superstrong magnetic fields \cite{sgr_magn_field,ouyed04:a}.

A very high-luminosity flare took place in the Large Magellanic Cloud
(LMC), some 55~kpc away, on 5~March 1979.  Another giant flare was
observed on 27 August 1998 from \sgrnineteenoo.  The inferred peak
luminosities for both events is $\sim 10^7$ (i.e. $\sim
10^{45}~\ergs$) times the Eddington limit for a solar mass object, and
the rise time is very much smaller than the time needed to drop $\sim
10^{25}$~g (about $10^{-8}\,\msun$) of normal material onto a neutron
star.  Alcock \etal\  \cite{alcock86:a} suggested a detailed model for
the 5~March 1979 event burst which involves the particular properties
of strange matter (see also Horvath \etal\
 \cite{usov01:a,horvath93:a}). The model assumes that a lump of strange
matter of $\sim 10^{-8}\,\msun$ fell onto a rotating strange
star. Since the lump is entirely made up of self-bound high-density
matter, there would be only little tidal distortion of the lump, and
so the duration of the impact can be very short, around $\sim
10^{-6}\rms$, which would explain the observed rapid onset of the
gamma ray flash. The light curves expected for such giant bursts
 \cite{usov01:c,usov01:b,usov98:a,cheng03:a} should posses
characteristic features that are well within the capabilities of ESA's
INTErnational Gamma-Ray Astrophysics Laboratory (INTEGRAL
 \cite{integral02:a}) launched by the European Space Agency in October
of 2002.

\goodbreak
\subsubsection{\it Rotational Instabilities}\label{sec:r-mode}

An absolute limit on rapid rotation is set by the onset of mass
shedding from the equator of a rotating star.  However, rotational
instabilities in rotating stars, known as gravitational radiation
driven instabilities, set a more stringent limit on rapid stellar
rotation than mass shedding.  These instabilities originates from
counter-rotating surface vibrational modes which at sufficiently high
rotational star frequencies are dragged forward. In this case
gravitational radiation, which inevitably accompanies the aspherical
transport of matter, does not damp the instability modes but rather
drives them. Viscosity plays the important role of damping these
instabilities at a sufficiently reduced rotational frequency such that
the viscous damping rate and power in gravity waves are comparable.
\begin{figure}[tb]
\begin{center} 
\epsfig{figure=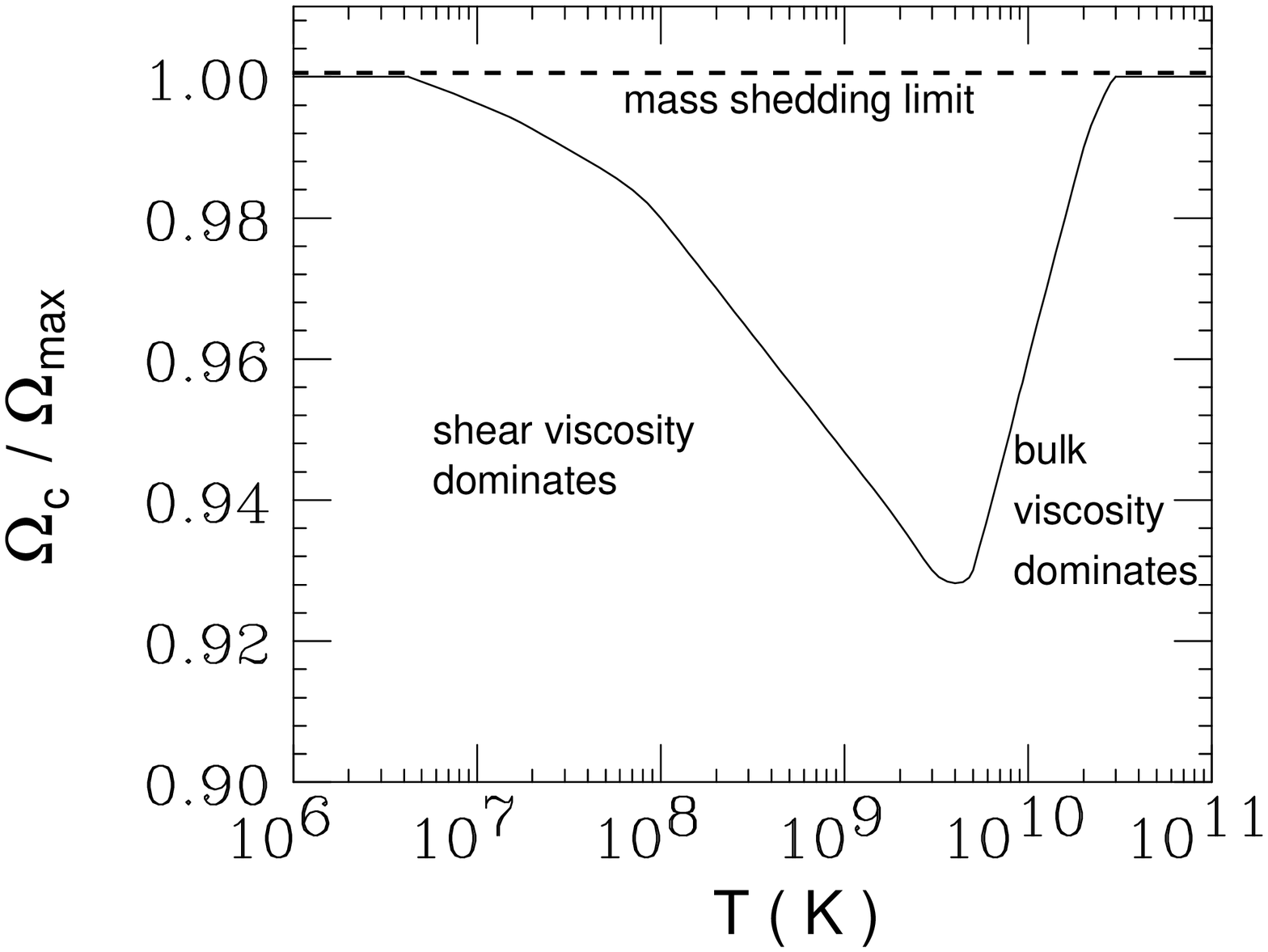,width=8.0cm,angle=0} \hskip 1.0cm
\epsfig{figure=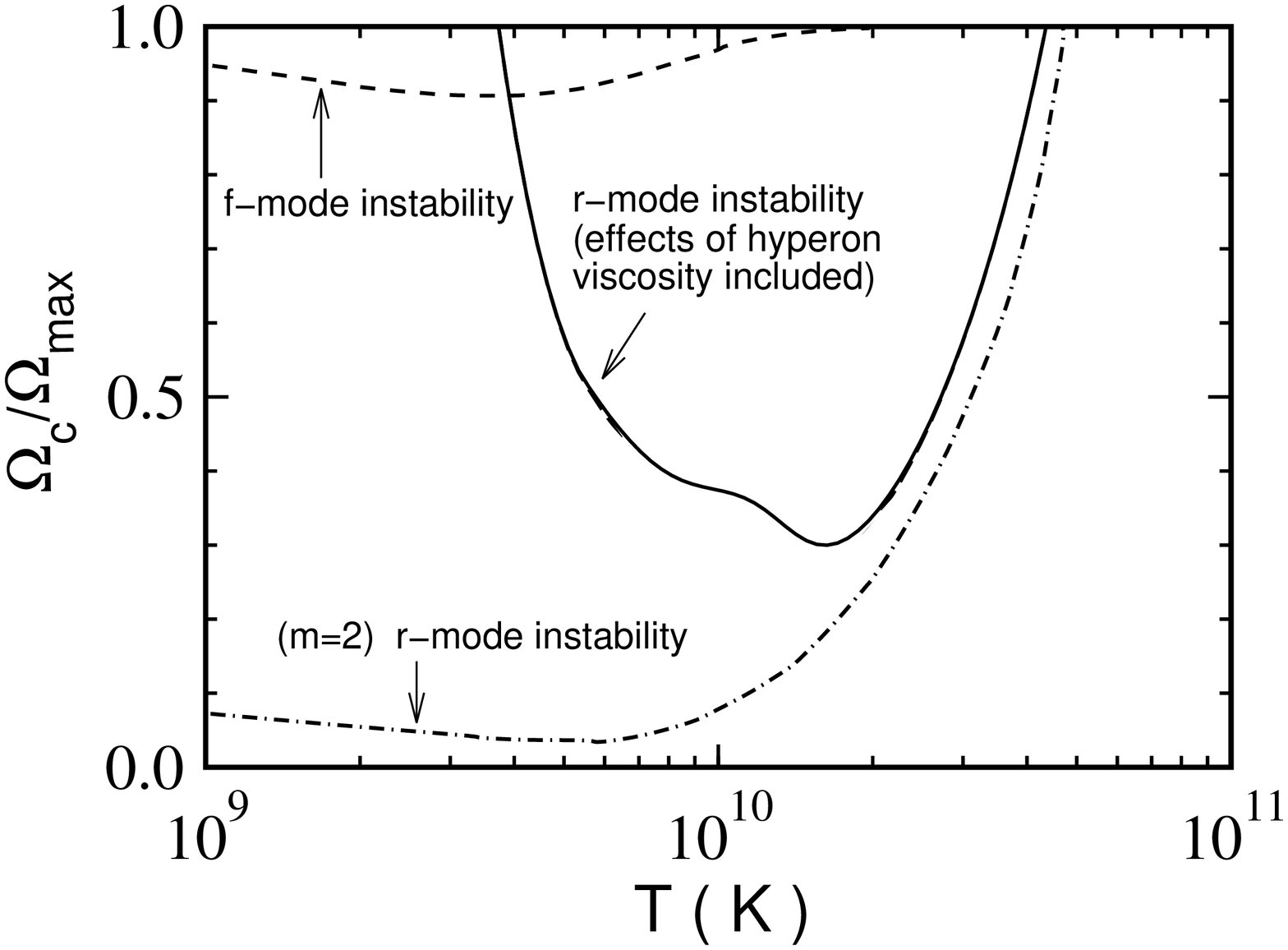,width=8.0cm,angle=0} 
\begin{minipage}[t]{16.5 cm}
\caption{Temperature dependence of the critical angular velocity
$\Omega{\rm c}$ of rotating neutron stars. The left figure shows the
gravitational radiation driven $f$-mode instability suppressed by
shear and bulk viscosity (Fig.\ from Ref.\  \cite{weber99:book}). Right
figure: comparison of $f$-mode instability with $r$-mode instability.
(Data from Refs.\  \cite{lindblom01:proc,lindblom02:a}.)}
\label{fig:OgrrvsT}
\end{minipage}
\end{center}
\end{figure}
The most critical instability modes that are driven unstable by
gravitational radiation are $f$-modes and $r$-modes.  Figure
\ref{fig:OgrrvsT} shows the stable neutron star frequencies if only
$f$-modes were operative in neutron star. One sees that hot as well as
cold neutron stars can rotate at frequencies close to mass shedding,
because of the large contributions of shear and bulk viscosity,
respectively, for this temperature regime.  The more recently
discovered $r$-mode instability may change the picture completely, as
can be seen from Fig.\ \ref{fig:OgrrvsT}.  These modes are driven
unstable by gravitational radiation over a considerably wider range of
angular velocities than the $f$-modes (cf.\ dashed curve labeled
($m=2$) $r$-mode instability). In stars with cores cooler than $\sim
10^9$~K, on the other hand, the $r$-mode instability may be completely
suppressed by viscous phenomena so that stable rotation would be
limited by the $f$-mode instability again \cite{lindblom02:a}.

Figures \ref{fig:cfl} and \ref{fig:2sc} are the counterparts to Fig.\
\ref{fig:OgrrvsT} but calculated for strange stars made of CFL and 2SC
quark matter, respectively \cite{madsen98:a,madsen00:b}. The $r$-mode
instability seems to rule out that pulsars are CFL strange stars, if
the characteristic time scale for viscous damping of $r$-modes are
exponentially increased by factors of $\sim \Delta/T$ as calculated in
 \cite{madsen98:a}. An energy gap as small as $\Delta = 1$~MeV was
assumed. For much larger gaps of $\Delta \sim 100$~MeV, as expected
for color superconducting quark matter (see section \ref{sec:color}),
the entire diagram would be $r$-mode unstable.  The full curve in
Fig.\ \ref{fig:cfl} is calculated for a strange quark mass of $m_s =
200$~MeV, the dotted curve for $m_s=100$~MeV.  The box marks the
positions of most low mass X-ray binaries (LMXBs) \cite{klis00:a}, and
the crosses denote the most rapidly rotating millisecond pulsars
known. All strange stars above the curves would
\begin{figure}[tb]
\begin{center} 
\parbox[t]{7.5cm} {\epsfig{file=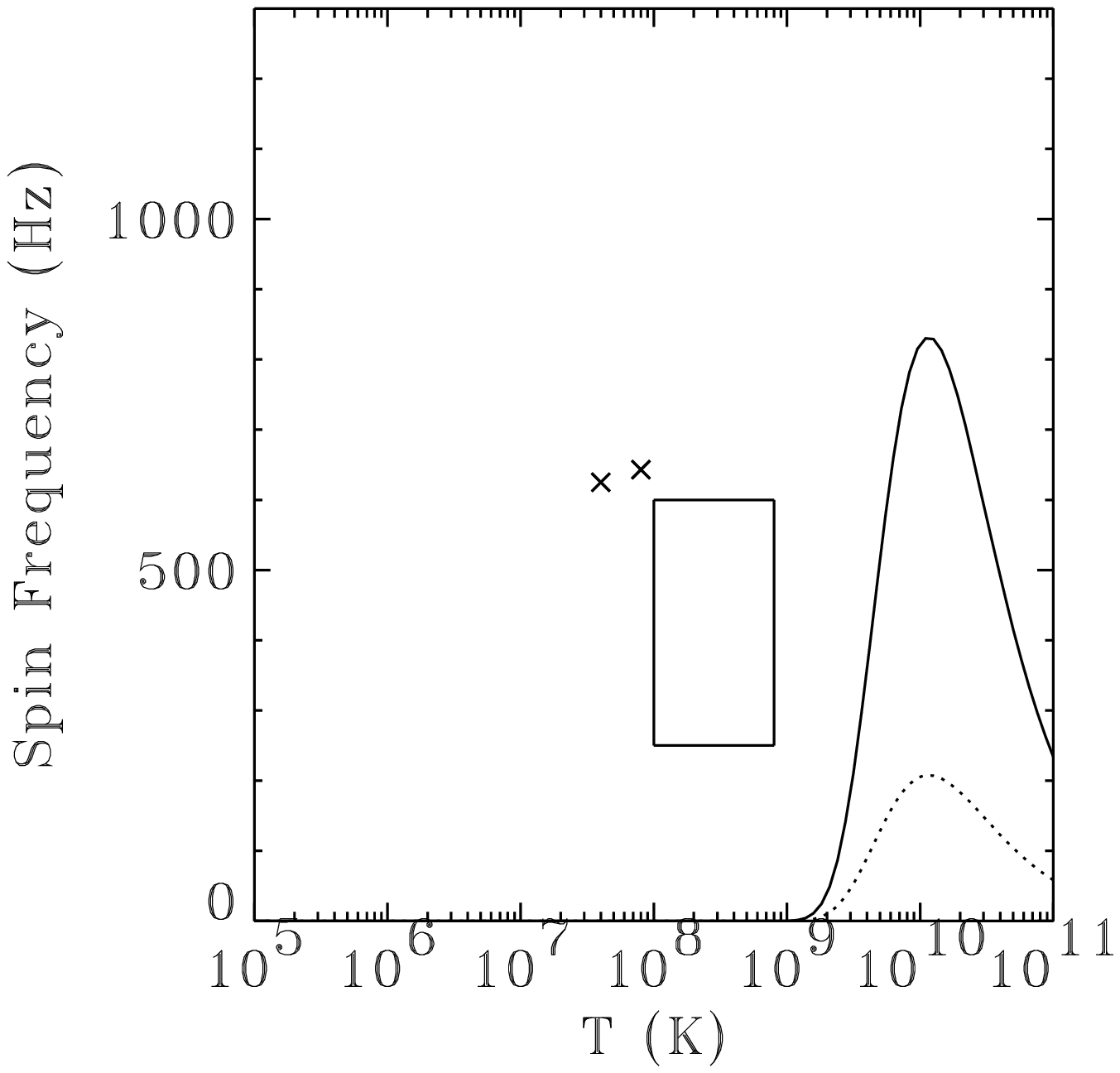,width=7.0cm}
{\caption{Critical rotation frequencies versus stellar temperature for
CFL strange stars \cite{madsen00:b}. Reprinted figure with permission
from J. Madsen, Phys. Rev. Lett. 85 (2000) 10. Copyright 2000 by the
American Physical Society.}
\label{fig:cfl}}}
\ \hskip 1.5 cm \ 
\parbox[t]{7.5cm} {\epsfig{file=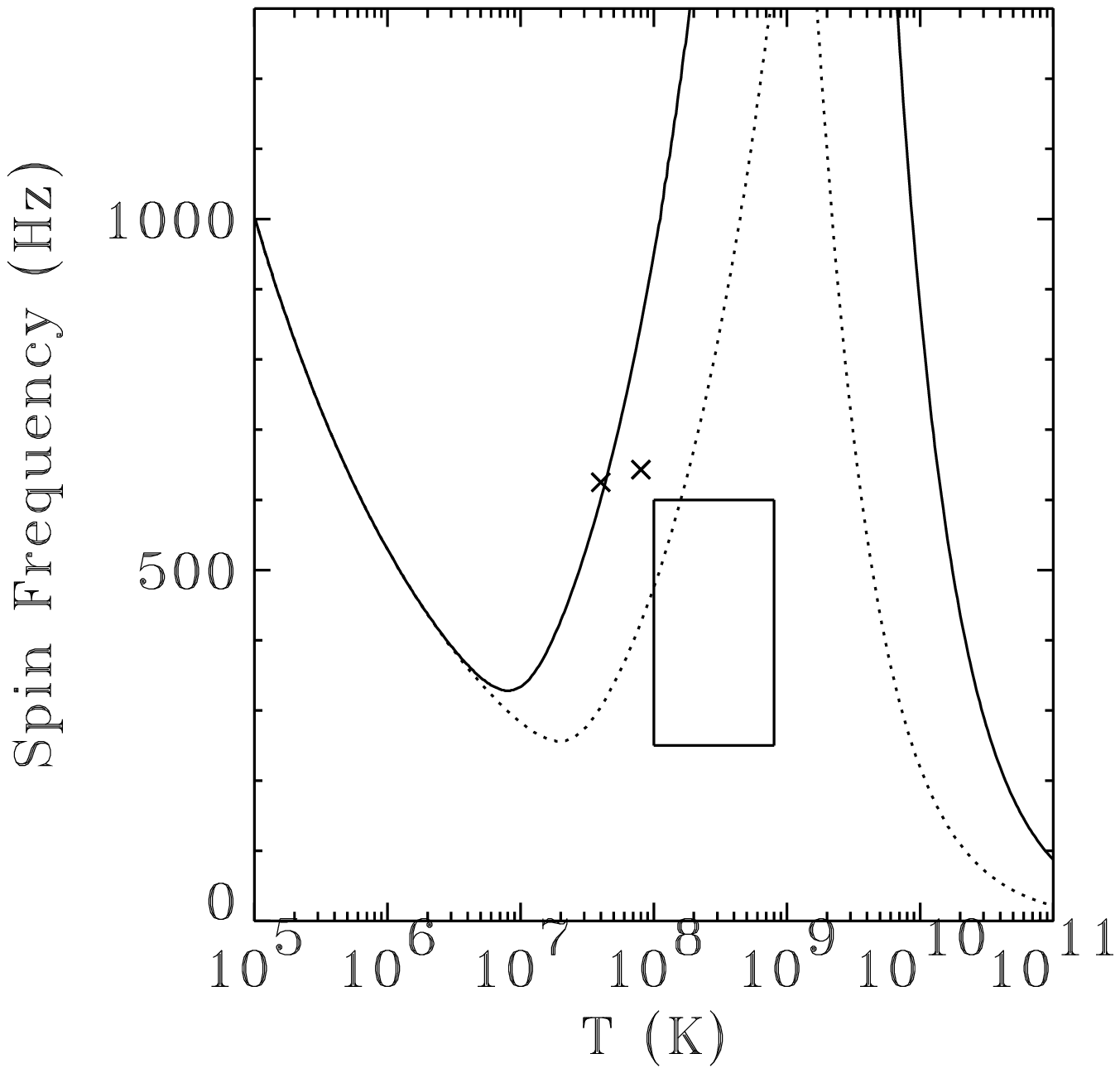,width=7.0cm}
{\caption{Same as Fig.\ \ref{fig:cfl}, but for 2SC quark
stars \cite{madsen00:b}.  Reprinted figure with permission from
J. Madsen, Phys. Rev. Lett. 85 (2000) 10. Copyright 2000 by the
American Physical Society.}
\label{fig:2sc}}}
\end{center}
\end{figure}
spin down on a time scale of hours due to the $r$-mode instability, in
complete contradiction to the observation of millisecond pulsars and
LMXBs, which would rule out CFL quark matter in strange stars (see,
however, Ref.\ \cite{manuel04:a}). Figure \ref{fig:2sc} shows the
critical rotation frequencies of quark stars as a function of internal
stellar temperature for 2SC quark stars.  For such quark stars the
situation is less conclusive.  Rapid spin-down, driven by the $r$-mode
gravitational radiation instability, would happen for stars above the
curves.

\goodbreak
\subsubsection{\it Surface Properties of Strange Stars}\label{sec:spss}

Strange quark matter with a density of about two times the density of
nuclear matter may exist up to the surface of a strange
star \cite{alcock86:a}. Such a bare strange star differs qualitatively
from a neutron star which has a density at the surface of about 0.1 to
$1~\gcmt$. As know from section \ref{sec:ss_general} the thickness of
the quark surface would be just $\sim 1~\fm$, the length scale of the
strong interaction.  Electrons are held to quark matter
electrostatically, and the thickness of the electron surface is
several hundred fermis.  Since neither component, electrons and quark
matter, is held in place gravitationally, the Eddington limit (Eq.\
(\ref{eq:Eddington})) to the luminosity that a static surface may emit
does not apply, and thus the object may have photon luminosities much
greater than $10^{38}~\ergs$.  It was shown by Usov \cite{usov98:a}
that this value may be exceeded by many orders of magnitude by the
luminosity of $e^+ e^-$ pairs produced by the Coulomb barrier at the
surface of a hot strange star. For a surface temperature of $\sim
10^{11}$~K, the luminosity in the outflowing pair plasma was
calculated to be as high as $\sim 3 \times 10^{51}~\ergs$.  Such an
effect may be a good observational signature of bare strange
stars \cite{usov01:c,usov01:b,usov98:a,cheng03:a}. If the strange star
is enveloped by a nuclear crust however, which is gravitationally
bound to the strange star, the surface made up of ordinary atomic
matter would be subject to the Eddington limit. Hence the photon
emissivity of such a strange star would be the same as for an ordinary
neutron star.  If quark matter at the stellar surface is in the CFL
phase the process of $e^+ e^-$ pair creation at the stellar quark
matter surface may be turned off, since cold CFL quark matter is
electrically neutral so that no electrons are required and none are
admitted inside CFL quark matter \cite{rajagopal01:b}. This may be
different for the early stages of a hot CFL quark star \cite{vogt03:a}.

\goodbreak
\subsection{\it Strange Dwarfs}\label{sec:sdwarfs}

The strange white dwarfs constitute the strange counterparts of
ordinary white dwarfs. They consist of a strange quark matter core in
the star's center which is enveloped by a nuclear curst made up of
ordinary atomic matter \cite{alcock86:a}.  The crust is suspended out
of contact with the quark core due to the existence of an electric
dipole layer on the core's surface
 \cite{alcock86:a,kettner94:b}, which insulates the crust from
conversion to quark matter. Even so, the maximum density of the crust
is strictly limited by the neutron drip density ($\edrip=4\times
10^{11}~\gcmt$), at which neutrons begin to drip out of the nuclei and
would gravitate into the core where they would be dissolved into
strange matter.

Strange white dwarfs comprise a largely unexplored consequence of the
strange matter hypothesis. Depending on the amount of crust mass,
their properties may differ considerably from those of ordinary white
dwarfs. For instance, it is well known that the maximum density
attained in the limiting-mass white dwarf is about $\epswd=10^9~\gcmt$
 \cite{harrison65:a,baym71:a}.  Above this density, the electron
pressure is insufficient to support the star, and there are no stable
equilibrium configurations until densities of the order of nuclear
density $(\epsilon \gs 10^{14}~\gcmt)$ are reached, which are neutron
or strange stars. One class of strange dwarfs can be envisioned as
consisting of a core of strange matter enveloped within what would
otherwise be an ordinary white dwarf. They would be practically
indistinguishable from ordinary white dwarfs.  Of greater interest is
the possible existence of a new class of white dwarfs that contain
nuclear material up to the neutron drip density, which would not exist
without the stabilizing influence of the strange quark core
 \cite{glen94:a}.  The density at the inner edge of the nuclear crust
carried by these strange dwarfs could fall in the range of $\epswd <
\ecrusti < \edrip$. The maximum inner crust density therefore could be
about 400 times the central density of the limiting-mass white dwarf
and $4\times 10^4$ times that of the typical $0.6\,\msun$ white dwarf.
An investigation of the stability of such very dense dwarf
configurations to acoustical (radial) vibrations, which will
be discussed below, reveals stability over an expansive mass range
from $\sim 10^{-3}\,\msun$ to slightly more than $1\,\msun$.  This is
the same range as ordinary white dwarfs except that the lower mass
limit is smaller by a factor of $\sim 10^{-2}$.  This is because of
the influence of the strange core, to which the entire class owes its
stability.  Whether or not a star is stable against radial
oscillations is determined by a stability analysis against radial
oscillations \cite{chandrasekhar64:a,bardeen66:a}.  The adiabatic
motion of a star in its \nth normal eigenmode ($n=0$ is the
fundamental mode) is expressed in terms of an amplitude $\xi_n(r)$
given by
\begin{eqnarray}
  \delta r(r,t) = e^{2\, \Phi(r)} \, \xi_n(r) \, e^{i \, \omega_n
    \, t} \, r^{-2} \, ,
\label{eq:delr}
\end{eqnarray} where $\delta r(r,t)$ denotes small Lagrangian
perturbations in $r$. The quantity $\omega_n$ is the star's
oscillation frequency, which we want to compute. The eigenequation for
$\xi_n(r)$, which governs the normal modes, is of Sturm-Liouville
type,
\begin{eqnarray}
  {d\over{d r}} \left( \Pi(r) \, {{d\xi_n(r)}\over{d r}} \right) +
  \left( Q(r) + \omega^2_n \, W(r) \right) \, \xi_n(r) = 0 \, ,
\label{eq:eigen}
\end{eqnarray} which implies that the eigenvalues $\omega_n^2$ are all
real and form an infinite, discrete sequence $\omega_0^2 <
\omega_{1}^2 < \omega_{2}^2 < \ldots$. Another consequence is that the
eigenfunction $\xi_n$ corresponding to $\omega_n^2$ has $n$ nodes in
the radial interval $0 < r < R$. Hence, the eigenfunction $\xi_0$ is
free of nodes in this interval.  The functions $\Pi(r)$, $Q(r)$, and
$W(r)$ are expressed in terms of the equilibrium configurations of the
star by
\begin{eqnarray}
  \Pi &=& e^{2\, \Lambda + 6 \, \Phi} \, r^{-2} \, \Gamma
  \, P \, ,
\nonumber \\ %
Q &=& - 4 \, e^{2\, \Lambda  + 6 \, \Phi} \, r^{-3} \,
{{d P}\over{d r}} - 8 \, \pi \, e^{6 \, (\Lambda +  2 \, \Phi)}
\, r^{-2} \, P \, \left(
\epsilon  +  P \right)
+  e^{2\, \Lambda + 6 \, \Phi} \, r^{-2} \, \left(
\epsilon  +  P \right)^{-1}  \left({{d P}\over{d r}}\right)^2  ,
\label{eq:q} \\ %
W &=& e^{6\, \Lambda + 2\, \Phi} \, r^{-2} \, \left(\epsilon + P
 \right) \, .  \nonumber
\end{eqnarray} The quantities $\epsilon$ and $P$ denote the energy density 
and the pressure of the stellar matter. The pressure gradient, $d P/
 d r$, is obtained from the Tolman-Oppenheimer-Volkoff
 Eq.~(\ref{eq:f28}). The symbol $\Gamma$ denotes the adiabatic
 index at constant entropy, $s$, given by
\begin{eqnarray}
  \Gamma = { {\partial\ln P(\rho,s)}\over{\partial\ln \rho} } = {{(
      \epsilon + P )}\over P} \, {{\partial P(\epsilon,s)}\over
    {\partial \epsilon}} \, ,
\label{eq:adiabatic}
\end{eqnarray} which varies throughout the star's interior. 
 Solving (\ref{eq:eigen}) subject to the boundary conditions $\xi_n
\propto r^3$ at the star's origin and $d\xi_n/d r = 0$ at the star's
surface leads to the ordered frequency spectrum $\omega^2_n <
\omega_{n+1}^2$ ($n=0,1,2,\ldots$) of the normal radial stellar
modes. If any of these is negative for a particular star, the
frequency is imaginary to which there corresponds an exponentially
\begin{figure}[tb]
\begin{center}
\parbox[t]{7.5cm} {\epsfig{figure=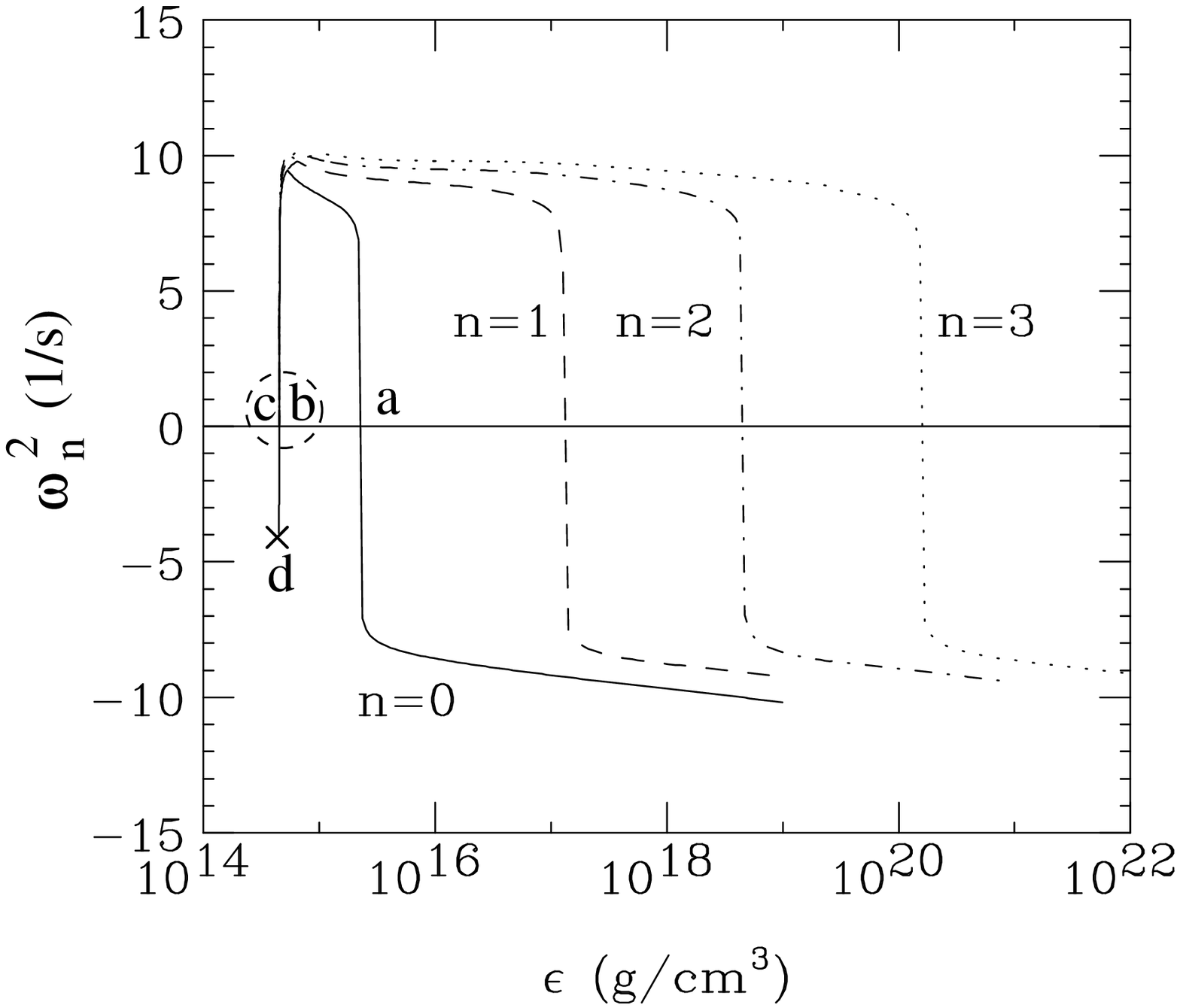,width=7.0cm,angle=0}
{\caption{Oscillation frequencies of the lowest four ($n=0,1,2,3$)
normal radial modes of strange stars with $\ecrusti=\edrip$ as a
function of central star density . The cross at `d' refers to the
termination point of the strange dwarf sequence shown in
Fig.~\ref{fig:mr145}. (Fig.\ from Ref.\  \cite{weber93:b}.)}
\label{fig:puls1}}}
\ \hskip 1.0cm   \
\parbox[t]{7.5cm} {\epsfig{figure=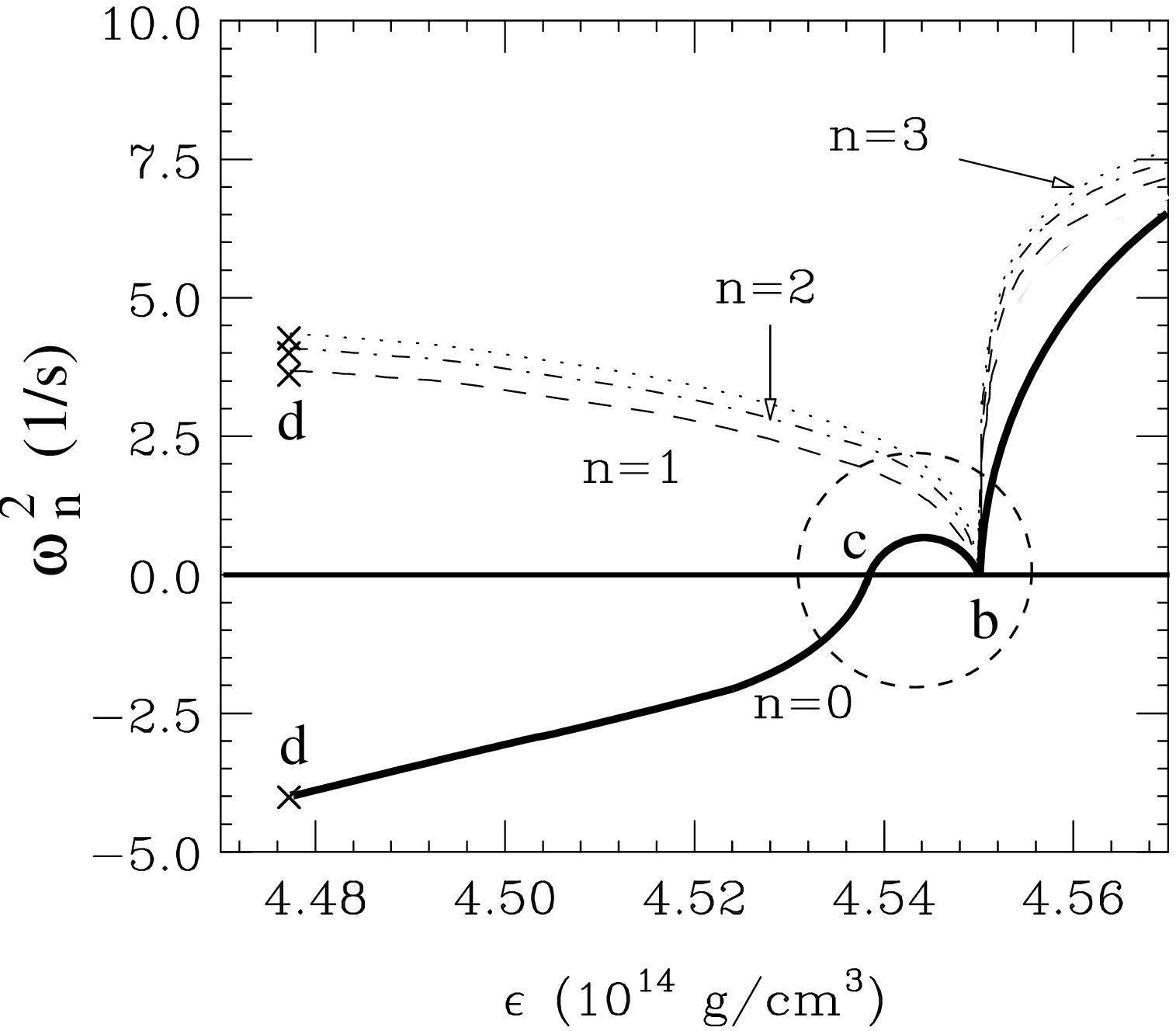,width=7.0cm,angle=0}
{\caption{Same as Fig.~\ref{fig:puls1}, but for strange dwarfs in the
vicinity of the termination point, `d', of the sequence. The labels
`b' and `c' refer to the lightest and heaviest star, respectively,
marked in Fig.\ \ref{fig:mr145}. (Fig.\ from Ref.\  \cite{weber93:b}.)}
\label{fig:puls2}}}
\end{center}
\end{figure}
growing amplitude of oscillation. Such stars are unstable.  Figures
\ref{fig:puls1} and \ref{fig:puls2} show the solutions to Eq.\
(\ref{eq:eigen}) for the strange star sequence in Fig.\
\ref{fig:mr145} whose inner crust density is equal to neutron drip.
It follows that all stars between `c' and `d' are unstable against
radial pulsations since for them $\omega_0^2 < 0$. This is different
for the stars to the left of `c' whose eigenfrequencies are all
positive \cite{weber93:b}.

At present there is neither a well studied model for the formation of
hypothetical strange dwarfs, nor exists a study that determines their
abundance in the universe. One possible scenario would be the
formation of strange dwarfs from main sequence progenitors that have
been contaminated with strange nuggets over their lifetimes. The
capture of strange matter nuggets by main sequence stars is probably
an inevitable consequence if strange matter were more stable than
hadronic matter \cite{madsen88:a} because then the Galaxy would be
filled with a flux of strange nuggets which would be acquired by every
object they come into contact with, like planets, white dwarfs,
neutron stars and main sequence stars. Naturally, due to the large
radii of the latter, they arise as ideal large-surface long
integration time detectors for the strange matter
flux \cite{madsen98:a}.  Nuggets that are accreted onto neutron stars
and white dwarfs, however, never reach their centers, where the
gravitational potential is largest, because they are stopped in the
lattice close to the surface due to the large structural energy
density there. This prevents such stars from building up a core of
strange matter. The situation is different for main sequence stars
which are diffuse in comparison with neutron stars and white
dwarfs. In this case the accreted nuggets may gravitate to the star's
core, accumulate there and form a strange matter core that grows with
time until the star's demise as a main sequence
star \cite{weber93:b,glen94:a}. An upper limit on the baryon number of
strange matter that may accumulate in a main sequence star is given
by \cite{madsen98:b,madsen88:a}
\begin{equation}
 A = 1.6\times 10^{47} \bigl( M / \msun \bigr)^{-0.15} {}
v_{250}^{-1}\, \rho_{24} \Bigl( 1 + 0.164\, v_{250}^{2} \bigl( {M / \msun}
\bigr)^{-0.25} \Bigr) \, ,
\label{eq:acmm} 
\end{equation} 
with $v_{250} \equiv v_{\infty } / (250~{\rm km/s})$ and $\rho_{24}
\equiv \rho _{\infty } / (10^{-24}\gcmt)$. The quantities $v_{\infty
}$ and $\rho_{\infty }$ are the nugget speed and contribution to the
density of the galactic halo far from the star, respectively.  If one
assumes that all the dark matter in the halo of our galaxy would
consist of strange nuggets (certainly a crude overestimate) then
$v_{250}^{-1}\, \rho_{24} \sim 1$, and Eq.\ (\ref{eq:acmm}) leads for
typical progenitor star masses of $\sim 1\, \msun$ to the upper limit
of $A \sim 10^{48}$. The mass and radius of such a strange core are
$\sim 2\times 10^{21}~{\textrm{kg}}$ ($\sim 10^{-9}\, \msun$ or about
$10^{-3}$ times the mass of the Earth) and $\sim 11~{\textrm{m}}$.
These values follows from the approximate relations $R = 4.35\times
10^{26}(M/{\textrm{kg}})$ and $R/{\textrm{km}} = 1.12\times 10^{-18} A
^{1/3}$ for strange matter with $\bag =145$ MeV.

Another plausible mechanism has to do with primordial strange matter
bodies.  Such bodies of masses between $10^{-2}$ and about $1\,\msun$
may have been formed in the early universe and survived to the present
epoch \cite{cottingham94:a}. Such objects will occasionally be captured
by a main sequence star and form a significant core in a single and
singular event.  The core's baryon number, however, cannot be
significantly larger than $\sim 5 \times 10^{31}~(M/\msun)^{-1.8}$
where $M$ is the star's mass.  Otherwise a main sequence star is not
capable of capturing the strange matter core \cite{madsen98:b}.
Finally we mention the possibility that in the very early evolution of
the universe primordial lumps \cite{alcock88:a,madsen86:a} of hot
strange matter will have evaporated nucleons which are plausibly
gravitationally bound to the lump. The evaporation will continue until
the quark matter has cooled sufficiently. Depending on the original
baryon number of the quark lump, a strange star or dwarf, both with
nuclear crusts could have been formed.

For many years only rather vague tests of the theoretical mass-radius
relationship of white dwarfs were possible. Recently, however, the
quality and quantity of observational data on the mass-radius relation
of white dwarfs has been reanalyzed and profoundly improved by the
availability of Hipparcos parallax measurements of several white
dwarfs  \cite{provencal98:a}.  In that work Hipparcos parallaxes were
used to deduce luminosity radii for 10 white dwarfs in visual binaries
of common proper-motion systems as well a 11 field white dwarfs.
Since that time complementary follow-up HST observations have been
made \cite{provencal02:a,kepler00:a} to better determine the
spectroscopy for Procyon~B and pulsation of G226-29.  Procyon~B at
first appeared as a rather compact star which, however, was later
confirmed to lie on the normal mass-radius relation of white dwarfs.
Stars like Sirius~B and 40~Erin~B, fall nicely on the expected
mass-radius relation too.  Several other stars of this sample
(e.g. GD~140, EG~21, EG~50, G181-B5B, GD~279, WD2007-303, G238-44)
however appear to be unusually compact and thus could be strange dwarf
candidates  \cite{mathews04:a}.

\goodbreak
\section{Neutrino Emission and Stellar Cooling}\label{sec:cooling}

The detection of thermal photons from the surfaces of neutron stars
via X-ray observatories serves as the principal window on the
properties of such objects. The surface temperatures of neutron stars
are derivable from the measured photon flux and spectrum.  The
predominant cooling mechanism of hot (temperatures of several $\sim
10^{10}$~K) newly formed neutron stars immediately after formation is
neutrino emission. Immediately after the birth of a (proto) neutron
star in a supernova explosion, neutrinos are trapped inside the star
because their mean free paths are shorter than the stellar
radius. About 10 seconds after birth most of the neutrinos have left
the star by diffusion.  As shown in
 \cite{carter00:a,steiner01:a,reddy02:a} the possible presence of quark
matter in the core of a neutron star could modify the diffusion rate
slightly. Depending on distance and stellar mass the neutrino bursts
will be detectable by terrestrial neutrino detectors such as SuperK,
IMB, Kamioka, SNO, and UNO \cite{prakash03:a}. The $\bar\nu_e$ count
\begin{figure}[tb]
\begin{center}
\includegraphics[scale=0.7,angle=0]{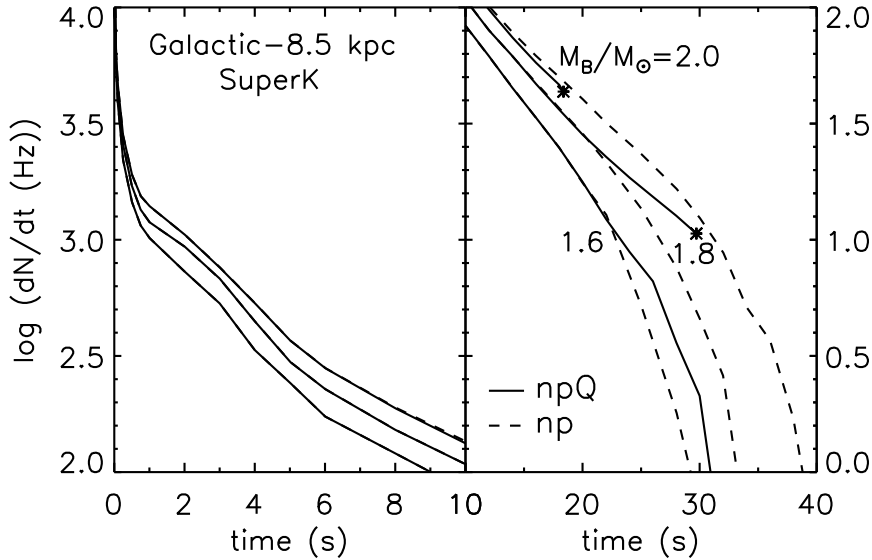}
\begin{minipage}[t]{16.5 cm}
\caption{A comparison of $\bar\nu_e$ count rates expected in SuperK
from a proto-neutron star containing either $np$ or $np+Q$
matter \cite{pons01:a}.  Reprinted figure with permission from
J. A. Pons \etal, Phys.~Rev.~Lett. 86 (2001) 5223.  Copyright 2001 by
the American Physical Society.}
\label{fig:dndt_vs_t}
\end{minipage}
\end{center}
\end{figure}
rates expected for SuperK from proto-neutrons star containing either
only nucleons ($n p$) or nucleons plus quark matter ($n p + Q$) are
shown in Fig.\ \ref{fig:dndt_vs_t}, where the left panel shows times
less than 10~s, while the right panel shows times greater than 10~s.
Fig.\ \ref{fig:lum1} shows the total neutrino luminosity for
proto-neutron stars containing quark matter in their centers. The
\begin{figure}[tb]
\begin{center}
\includegraphics[scale=0.4]{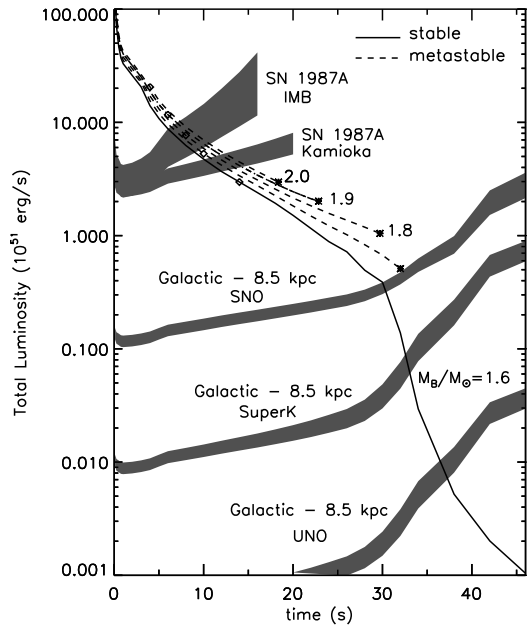}
\begin{minipage}[t]{16.5 cm}
\caption{Total neutrino luminosity for proto neutron stars made of
nucleons and quarks \cite{pons01:a}.  Reprinted figure with permission
from J. A. Pons \etal, Phys.~Rev.~Lett. 86 (2001) 5223.  Copyright
2001 by the American Physical Society.}
\label{fig:lum1}
\end{minipage}
\end{center}
\end{figure}
shaded bands illustrate the limiting luminosities corresponding to a
count rate of 0.2~Hz, assuming a supernova distance of 50~kpc for IMB
and Kamioka, and 8.5~kpc for SNO and SuperK. The widths of the shaded
regions represent uncertainties in the average neutrino energy from
the use of a diffusion scheme for neutrino transport (for details, see
Refs.\  \cite{pons01:a,pons98:a}). Observable effects of quarks only
become apparent for stars older than 10 to 20 s.  Sufficiently massive
stars containing negatively charged, strongly interacting, particles
(such as quarks, but also including hyperons and kaon condensates) may
collapse to black holes during the first minute of evolution.  Since
the neutrino flux vanishes when a black hole forms, this would
constitute an obvious signal that quarks (or other types of strange
matter) have appeared.  The collapse timescales for stars containing
quarks are predicted to be intermediate between those containing
hyperons and kaon condensates.

Already a few hours after birth, the internal neutron star temperature
drops to $\sim 10^9$~K. The cooling of the star is primarily dependent
for the next several thousand years on the neutrino emissivity of the
core's composition.  Photon emission overtakes neutrinos only when the
internal temperature has fallen to $\sim 10^8$~K, with a corresponding
surface temperature roughly two orders of magnitude smaller. Being
sensitive to the adopted nuclear equation of state, the neutron star
mass, the assumed magnetic field strength, the possible existence of
superfluidity, meson condensates and quark matter, theoretical cooling
calculations provide most valuable information about the interior
matter and neutron star structure. The thermal evolution of a neutron
star also yields information about such temperature sensitive
properties as transport coefficients, transition to superfluid states,
crust solidification, and internal pulsar heating mechanisms such as
frictional dissipation at the crust-superfluid
interfaces \cite{schaab99:b}.  In general, the possible existence of
meson condensates and (certain phases of) quark matter enhance the
neutrino emissivity from the core, leading to a more rapid early
cooling. An overview of neutrino emitting processes from the core and
crust of a neutron star is given in table \ref{tab:emis.over}.
\begin{table}[tb]
\begin{center}
\begin{minipage}[t]{16.5 cm}
\caption{Overview of neutrino emitting processes relevant for neutron
  star cooling \cite{weber99:book}.}
\label{tab:emis.over}
\end{minipage}
\begin{tabular}{l|l|c|c} \hline
Name     &Processes   &Emissivity &Efficiency                  \\ \hline
Modified Urca       &$n+n \rightarrow n+p+e^-+\bar\nu_e$ 
&$\sim 10^{20} T_9^8$  &slow  \\
                    &$n+p+e^- \rightarrow n+n+\nu_e$      &    & \\
Direct Urca         &$n \rightarrow p+e^-+\bar\nu_e$  
&$\sim 10^{27} T_9^6$  &fast \\
                    &$p+e^- \rightarrow n+\nu_e$          &    & \\
Quark modified Urca &$d+u+e^- \rightarrow d+d+\nu_e$      
&$\sim 10^{20} T_9^8$  &slow \\
                    &$u+u+e^- \rightarrow u+d+\nu_e$      &    & \\
                    &$d+u+e^- \rightarrow d+s+\nu_e$      &    &  \\
                    &$u+u+e^- \rightarrow u+s+\nu_e$      &    & \\
Quark direct Urca  &$d \rightarrow u+e^-+\bar\nu_e$  
&$\sim 10^{26} T_9^6$ &fast \\
                   &$u+e^- \rightarrow d+\nu_e$           &    &\\
                   &$s \rightarrow u+e^-+\bar\nu_e$       &    &\\
                   &$u+e^- \rightarrow s+\nu_e$           &    & \\
$\pi^-$ condensate &$n+<\pi^-> \rightarrow n+e^-+\bar\nu_e$  
&$\sim 10^{26} T_9^6$  &fast \\
$K^-$ condensate   &$n+<K^-> \rightarrow n+e^-+\bar\nu_e$  
&$\sim 10^{26} T_9^6$ &fast \\
Quark bremsstrahlung &$Q_1+Q_2 \rightarrow Q_1+Q_2+\nu+\bar\nu$ 
&$\sim 10^{20} T_9^8$ &slow  \\
Core  bremsstrahlung &$n+n \rightarrow n+n+\nu_e+\bar\nu_e$  
&$\sim 10^{19} T_9^8$ &slow \\
                     &$n+p \rightarrow n+p+\nu_e+\bar\nu_e$     &   & \\
                     &$e^-+p \rightarrow e^-+p+\nu_e+\bar\nu_e$ &   & \\
Crust bremsstrahlung &$e^-+(A,Z) \rightarrow e^-+(A,Z)$  
& &slow \\ 
                     &$+\nu_e+\bar\nu_e$ & & \\
\hline
\end{tabular}
\end{center}
\end{table}
Superfluidity of nucleons, on the other hand, has the opposite effect
on cooling. Quantitative constraints on cooling have been hampered by
the relatively small number of young pulsars known, the complication
that several of them also display non-thermal, beamed X-ray emission
from their magnetospheres, and uncertainties in distance and
interstellar absorption. Table~\ref{tab:observations1} summarizes
\begin{table}[tb]
\begin{center}
\begin{minipage}[t]{16.5 cm}
\caption{Surface temperatures of neutron stars at infinite distance
 from the stars.  $\tau$ denotes the age of the star}
\label{tab:observations1}
\end{minipage}
\begin{tabular}{l|c|c|l|l} \hline
 Source                 &P (ms)  &${\rm log}_{10}~\tau$ (y)   &${\rm log}_{10}~T^\infty_{\rm s}$ (K)      &Refs.  \\ \hline
  B1706--44             &102.45  &4.23                  &$5.91^{+0.01}_{-0.23}$        & \cite{mcgowan04:a} \\
  B1823--13             &101.45  &4.33                  &$6.01\pm 0.02$                & \cite{finley93:b} \\
  2334+61               &495.24  &4.61                  &$5.92^{+0.15}_{-0.09}$        & \cite{becker93:b} \\
  B0531+21 (Crab Pulsar) &33.40  &2.97                  &$< 6.3$                       & \cite{weisskopf04:a}\\
  B1509--58 (MSH 15-52)  &150.23 &3.19                  &$6.11\pm 0.1$                 & \cite{seward83:a,trussoni90:a} \\
  0540--69              &50.37   &3.22                  &$6.77 ^{+0.03}_{-0.04}$       & \cite{finley93:a} \\
  1951+32 (CTB 80)      &39.53   &5.02                  &$6.14 ^{+0.03}_{-0.05}$       & \cite{safiharb95:a} \\
  1929+10               &226.51  &6.49                  &5.52                          & \cite{oegelman95:a,yancopoulos93:a} \\
  0950+08               &253.06  &7.24                  &$4.93 ^{+0.07}_{-0.05}$       & \cite{seward88:a} \\
  J0437--47             &5.75    &8.88                  &$5.36\pm 0.1$                 & \cite{becker93:c} \\
  0833--45 (Vela Pulsar)  &89.29 &4.05                  &$6.24 \pm 0.03$               & \cite{pavlov01:a} \\
  0656+14               &384.87  &5.05                  &$5.98\pm 0.05$                & \cite{possenti96:a} \\
  0630+18 (Geminga)     &237.09  &5.53                  &$5.75^{+0.05}_{-0.08}$        & \cite{halpern97:a} \\
  B1055--52             &197.10  &5.73                  &$5.90^{+0.06}_{-0.12}$        & \cite{greiveldinger96:a} \\
  J0205+6449 (3C58)     &65.86   &2.91                  &$< 6.04$                      & \cite{slane02:a} \\
  J0822--4300           &not known  &3.3--3.7           &6.20--6.28                    & \cite{zavlin99:a} \\
  1E~1207.4--5209       &424.13  &$\gs 3.85$            &6.04--6.18                    & \cite{zavlin98:a} \\
  J1856.4--3754         &$\sim 220$  &5.7               &$< 5.7$                       & \cite{pavlov03:a}  \\
  J0720.4--3125         &8391.11 &$\sim 6.11$           &$\sim 5.7$                    & \cite{motch03:a} \\
\hline
\end{tabular}
\end{center}
\end{table}
the temperatures of a collection of neutron stars.  To compare the
observations with theory, one needs mainly the neutron star effective
temperatures $T_{\rm s}$ and ages $\tau$, which are compiled in
table~\ref{tab:observations1} for a representative collection of
neutron stars. The thermal photon luminosity in the local reference
frame of the star is given by $L_\gamma = 4 \pi R^2 \sigma T_{\rm
s}^4$ with $\sigma$ the Stefan-Boltzmann constant. The apparent
(redshifted) effective surface temperature $T_{\rm s}^\infty$ and
luminosity $L_\gamma^\infty$, as detected by a distant observer, are
given by
\begin{equation}
    T_{\rm s}^\infty = T_{\rm s}\,\sqrt{1 - {R_{\rm s} / R}} \quad
{\rm and} \quad L_\gamma^\infty = 4 \pi (R^\infty)^2 \sigma
(T^\infty_{\rm s})^4 = L_\gamma \, (1 - R_{\rm s} / R),
\label{apparent}
\end{equation}
where $R_{\rm s}=2 M = 2.95\, M/M_\odot$~km is the Schwarzschild
radius.  The coordinate radius $R$ is connected to the apparent
radius according to Eq.\ (\ref{eq:r_true}).

As already mentioned in section \ref{sec:3c58}, recent Chandra X-ray
observations have identified pulsar \psrotwoofive \ at the center of
the young Crab-like supernova remnant 3C58. Historical evidence
suggests an association of the remnant with supernova SN~1181, which
makes 3C58 younger than Crab (see table \ref{tab:observations1}).  The
Chandra observation indicate that the thermal component must be very
small, since the radiation is nearly completely fit by a power-law
spectrum \cite{slane02:a}. The temperature of a possible residual
thermal component is thereby limited to an effective black-body value
of $T^\infty < 95$~eV (surface temperature of $ < 6.0334$)
 \cite{slane02:a} which, as can be seen from Fig.~\ref{fig:cool}, falls
well below predictions of standard cooling calculations
 \cite{slane02:a}. 
\begin{figure}[tb]
\begin{center}
\epsfig{figure=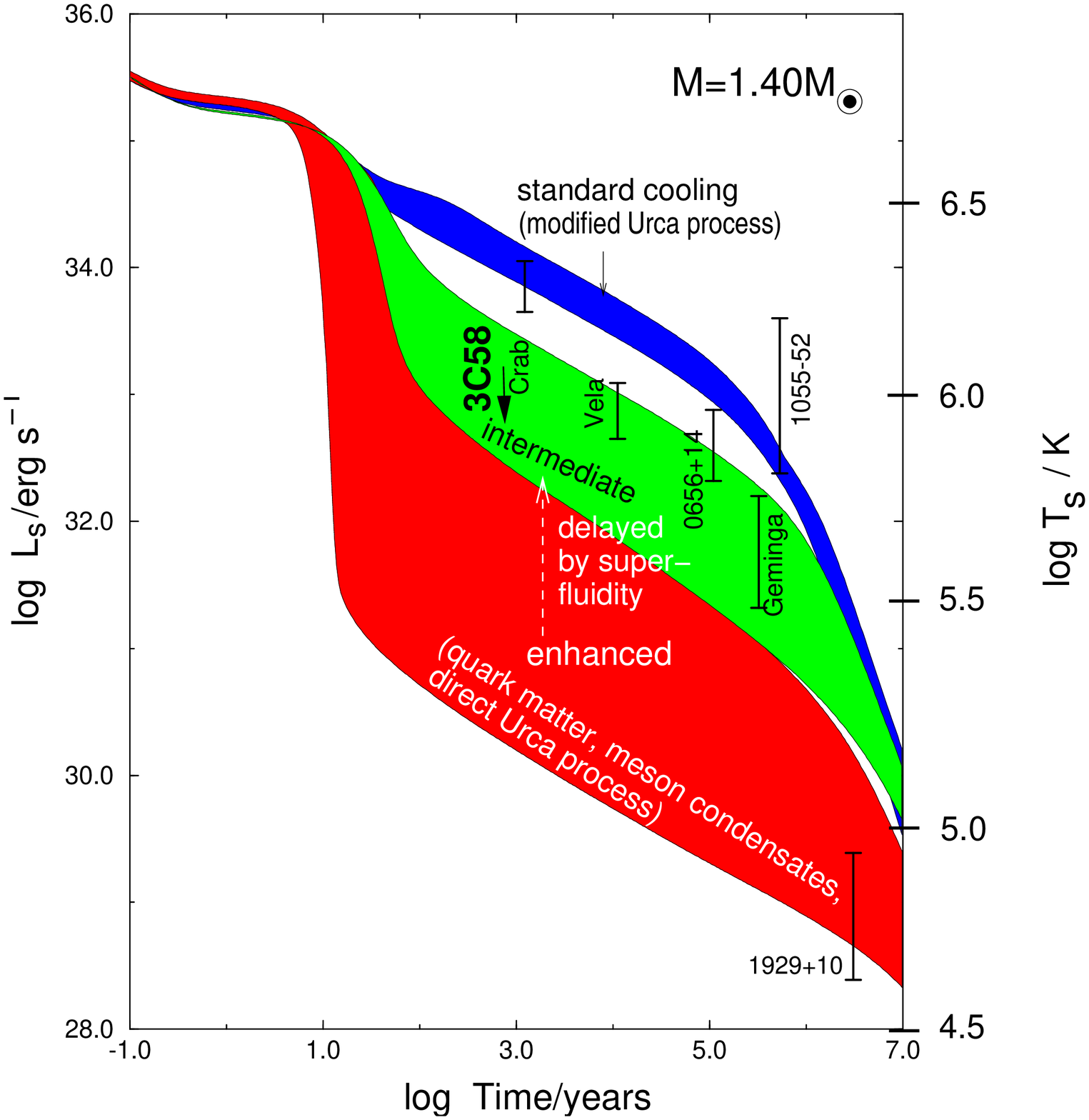,width=8.0cm}
\begin{minipage}[t]{16.5 cm}
\caption{Cooling behavior of a $1.4\,\msun$ neutron star for competing
assumptions about the properties of superdense matter. Three distinct
cooling scenarios, referred to as `standard', `intermediate', and
`enhanced' can be distinguished. The band-like structures reflects
the uncertainties inherent in the equation of state of superdense
matter \cite{weber99:book}.}
\label{fig:cool} 
\end{minipage}
\end{center}
\vskip -0.4cm
\end{figure}
As pointed out by Prakash \etal\  \cite{prakash03:a}, this upper limit
can be fit with standard neutrino cooling (like $n+n \rightarrow
n+p+e^{-}+\bar\nu_e$) plus pair-breaking and formation, but the
luminosity and ages of other neutron stars cannot be simultaneously
fit using the same \eosp \cite{yakovlev04:a,yakovlev02:a} which has the
interesting consequence that more exotic, rapid cooling processes may
exist in the core of \psrotwoofive.  Physical processes which would
enable such a rapid drop in temperature range from the presence of
meson condensates \cite{brown96:a,page98:a}, to quark
matter \cite{blaschke01:a,page02:a,blaschke03:KIAS}, to the direct Urca
process \cite{lattimer91:a}. (For a very recent overview on neutron
star cooling, see Ref.\  \cite{yakovlev04:a}.)

Before the discovery of color superconductivity of quark matter, it
was believed that depending on the density of electrons in quark
\begin{figure}[tb]
\begin{center}
\parbox[t]{7.5cm} {\epsfig{figure=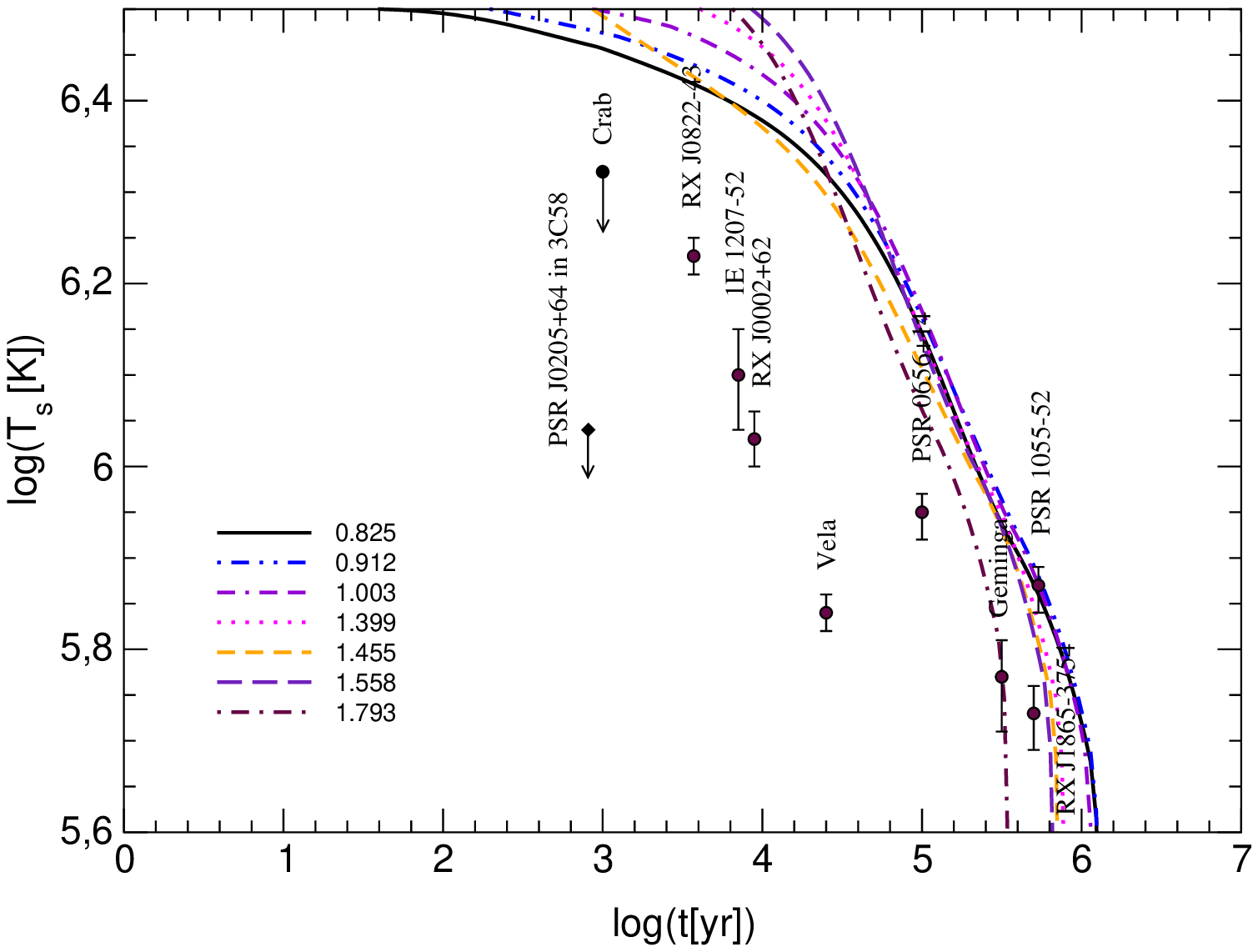,width=9.5cm}
{\caption{Cooling curves of 2SC quark hybrid stars for a quark pairing
gap of 1~MeV. (From Ref.\  \cite{blaschke03:KIAS}.)}
\label{fig:csl1}}}
\ \hskip 1.0cm   \
\parbox[t]{7.5cm} {\epsfig{figure=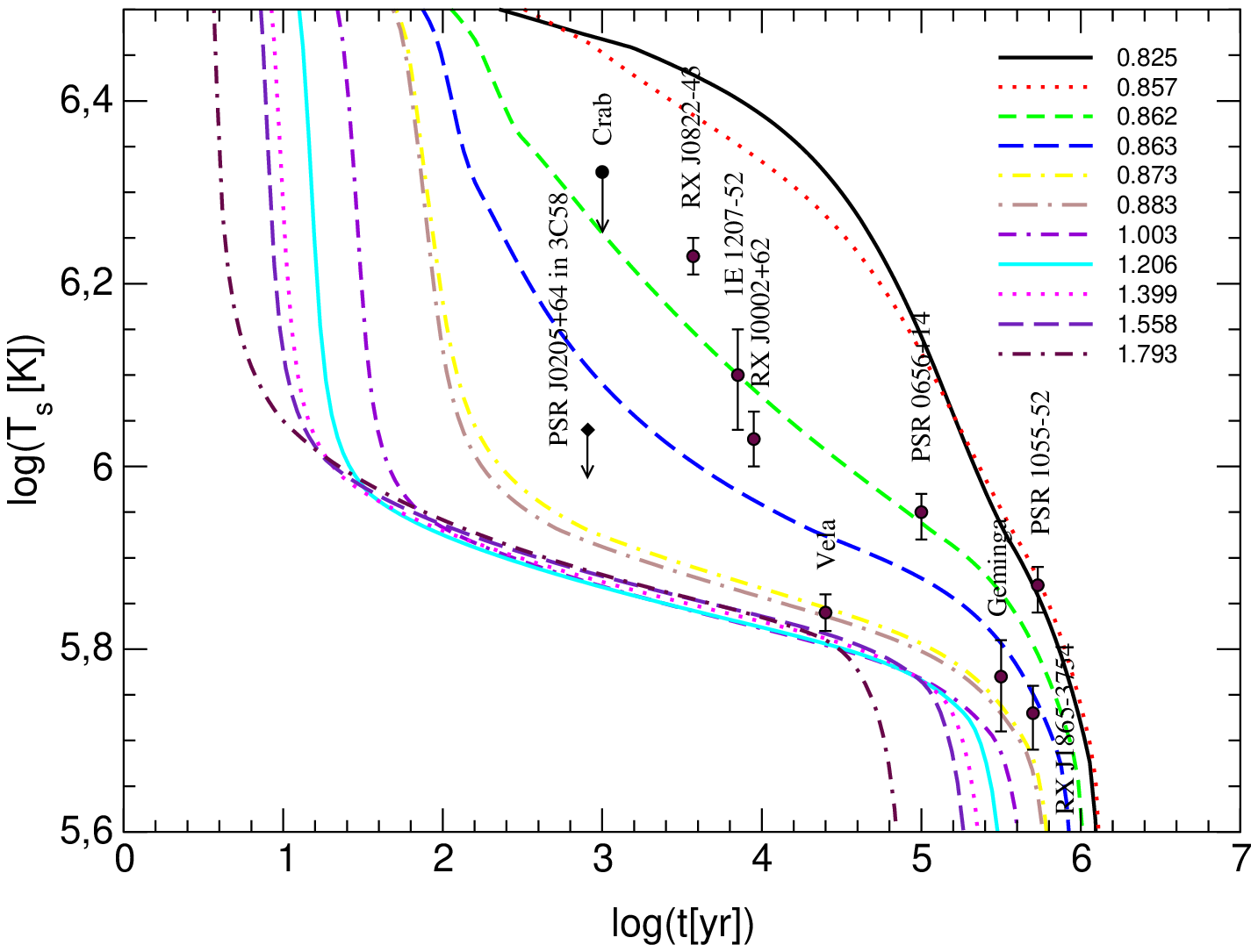,width=9.5cm}
{\caption{Same as Fig.\ \ref{fig:csl1} but for a quark pairing gap of
50~keV. (From Ref.\  \cite{blaschke03:KIAS}.)}
\label{fig:csl005}}}
\end{center}
\vskip -0.4cm
\end{figure}
matter, the temperature of quark stars could drop much more rapidly
than for neutron stars \cite{weber99:book,schaab95:a,schaab97:a}. The
density of electrons is crucial since the fast quark direct Urca
processes (see table \ref{tab:emis.over}),
\begin{eqnarray}
  d \rightarrow u + e^- + \bar{\nu}_e \, , \qquad u + e^- \rightarrow
d + \nu_e \, , \qquad s \rightarrow u + e^- + \bar{\nu}_e \, , \qquad
u + e^- \rightarrow s + \nu_e \, ,
\label{eq:quark2}
\end{eqnarray} 
are only possible if the electron Fermi momentum in quark matter is
sufficiently high so that energy and momentum conservation in the
above reactions is guaranteed. If the electron Fermi momentum is too
small for this to happen, a bystander quark is needed to ensure energy
and momentum conservation in the scattering process.  The emissivity
in the latter case is considerably smaller than the emissivities
associated with the direct Urca processes in (\ref{eq:quark2}),
because of the different phase spaces associated with two-quark
scattering and quark decay.  If the electron fraction in quark matter
vanishes entirely both the quark direct and the quark modified Urca
processes become unimportant.  The neutrino emission is then entirely
dominated by bremsstrahlung,
\begin{equation}
  Q_1 + Q_2 \longrightarrow Q_1 + Q_2 + \nu + \bar\nu \, ,
\label{eq:q1q2.brems}
\end{equation} where $Q_1$, $Q_2$ denote any pair of quark flavors.
In this case quark star cooling would proceed rather slowly, at about
the same rate as conventional neutron
stars \cite{weber99:book,schaab95:a,schaab97:a}. The same were the case
if quark matter would be superfluid with gaps on the order of a few
MeV, as described in the paper by Bailin and
Love \cite{bailin79:a,bailin84:a}. The neutrino emissivities would then
be suppress by a factor of $\exp(-\Delta/T)$, with $\Delta$ the gap
energy.

The situation is more complicated if quark matter forms a color
superconductor \cite{rajagopal01:a}.  If, as in the CFL phase, all
quarks have a gap $\Delta \gg T$ then both the heat capacity $C_V$
and neutrino luminosity $L_\nu$ are suppressed by $\sim
\exp(-\Delta/T)$ which would render quarks in the centers of compact
stars invisible. Vanishingly small quark gaps, on the other hand,
would lead to cooling behaviors indistinguishable from those of
ordinary neutron stars made of either nucleons or nucleons and
hyperons.  In Ref.\  \cite{jaikumar02:a}, the photon and neutrino
emission rates from the decay of photons and Nambu-Goldstone (NG)
bosons associated with the spontaneous breaking of baryon number,
$U(1)_B$, in the CFL phase were calculated.  The emission rates were
found to be very small so that they would be inefficient for the core
cooling of neutron stars containing quark matter in the CFL phase,
rendering quark pairing in the CFL phase invisible to
telescopes. This finding is in accordance with the quantitative
determination of the mean free paths and thermal conductivities of
photons ($\gamma$) and NG bosons ($\phi$) in the CFL phase performed
in \cite{shovkovy02:a}.  The total conductivity associated with these
particles was found to be $\kappa_{\rm CFL} = \kappa_{\phi} +
\kappa_\gamma \simeq (2 \pi^2/9) T^3 R_0$ which can be conveniently
written as
\begin{equation}
\kappa_{\rm CFL} \simeq 1.2 \times 10^{32} \left( T/{\rm MeV}
\right)^3 \left( R_0 /{\rm km} \right) ~ {\rm erg} ~ {\rm cm}^{-1} ~
{\rm sec}^{-1} ~ {\rm K}^{-1} \, ,
\label{eq:tcd}
\end{equation}
where $R_0$ is the radius of the quark matter core.  This expression
reveals that the thermal conductivity of the CFL phase from photons
and NG bosons is many orders of magnitude larger than the thermal
conductivity of regular nuclear matter in a neutron star.  The cooling
of the quark matter core in the center of a compact star thus arises
primarily from the heat flux across the surface of direct contact with
the nuclear matter enveloping the CFL quark matter core. Since the
thermal conductivity of the neighboring layer is also high, the entire
interior of the star should be nearly
isothermal \cite{shovkovy02:a}. The results in Ref.\
 \cite{shovkovy02:a} confirm that the cooling time of neutron stars
with CFL quark matter cores is similar to that of conventional neutron
stars.

Finally we mention briefly the cooling behavior of compact stars
hiding color superconducting 2SC quark matter in their cores. The
cooling of such stars is complicated by the fact that up and down
quarks may pair with a gap $\Delta \sim 100$~MeV that is orders of
magnitude larger than the stellar temperature, $\ls 1$~MeV, and are
therefore inert with respect to the star's temperature evolution.  In
contrast to the CFL phase, where diquark condensation produces gaps
for quarks of all three flavors and colors, there exits quark pairing
channels that lead to weak pairing with gaps on the order of several
keV to about 1~MeV, which is on the same order of magnitude as the
star's temperature. These quarks may thus not pair but, instead, may
radiate neutrinos rapidly via the quark direct Urca process shown in
table~\ref{tab:emis.over}. If this is the case, the 2SC quark matter
core will cool rapidly and determine the cooling history of the
star \cite{rajagopal01:a,blaschke03:KIAS}. Examples of cooling curves
of neutron stars containing quark matter in the 2SC phase are shown in
Figs.\ \ref{fig:csl1} and \ref{fig:csl005} for different star masses.
A quark gap of 1~MeV, as chosen in Fig.\ \ref{fig:csl1}, leads to too
slow a cooling, while a reduced gap of 50~keV reproduces the observed
data quite well.

\goodbreak
\section{Signals of Quark Matter in Rotating Neutron Stars}
\label{sec:mqdec}

In this section we explore possible signals of quark deconfinement in
neutron stars, assuming that the densities in the centers of such
objects are high enough so that quark deconfinement occurs. A
convincing discovery of the type of signals described in this section
\begin{figure}
\begin{center}
\parbox[t]{7.5cm} {\epsfig{figure=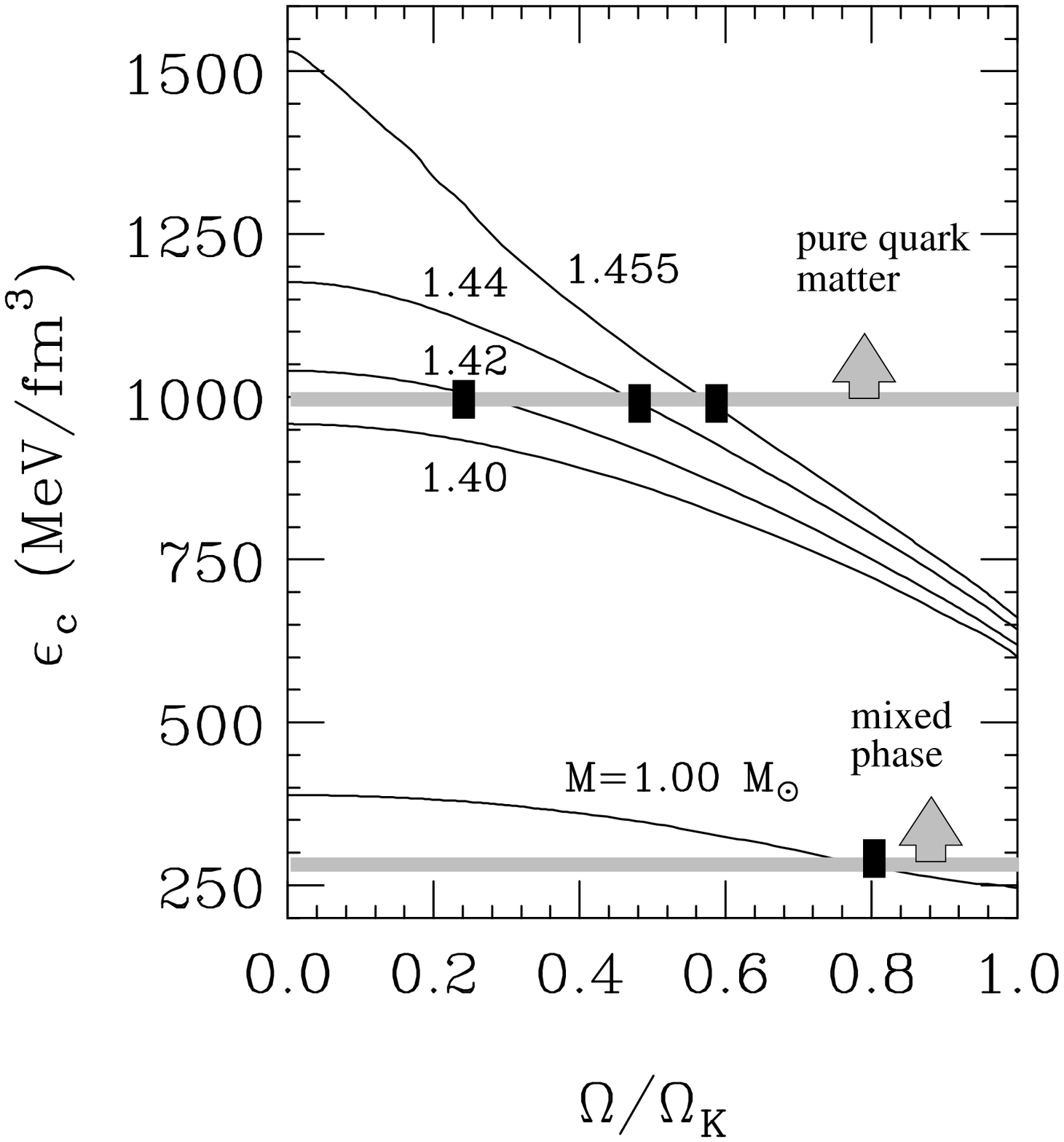,width=7.0cm,angle=-90}
{\caption{Central star density versus rotational frequency for
neutron stars of different masses. The vertical bars mark the density
where quark matter is produced. (From Ref.\  \cite{weber99:book}.)}
\label{fig:ec24B18}}}
\ \hskip 1.0cm   \
\parbox[t]{7.5cm}
{\epsfig{figure=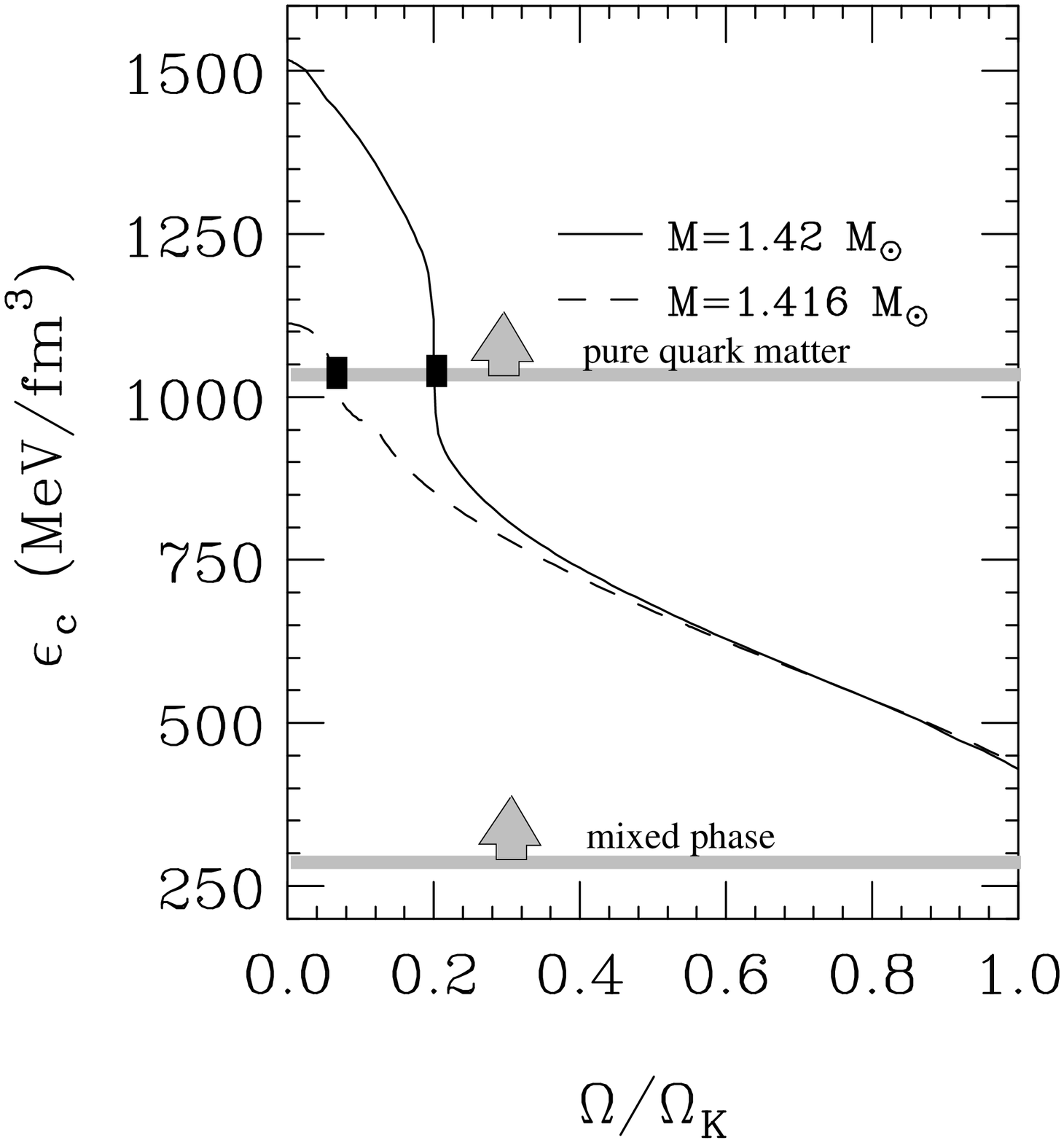,width=7.0cm,angle=-90}
{\caption{Same as Fig.\ \protect{\ref{fig:ec24B18}}, but for \eos
$\KBt$ (From Ref.\  \cite{weber99:book}.)}
\label{fig:ec3B18}}}
\end{center}
\vskip -0.4cm
\end{figure} 
could indicate that strange quark matter is not absolutely stable,
ruling out the absolute stability of strange quark matter and the
existence of strange quark stars, since it is impossible for
(quark-hybrid) neutron stars and strange quark stars to both coexist
stably. The signals described below require rather pronounced
modifications of the \eos caused by quark confinement and depend on
the rate at which the mixed phase of quarks and hadrons gives way to
pure quark matter. In addition great care is to be taken with respect
to the numerical modeling \cite{spyrou02:a} of rotating stars as well
as the properties of the nuclear curst \cite{cheng02:b}.  We shall
begin our discussion with isolated rotating neutron stars, which spin
down because of the loss of rotational energy caused by the emission
of an electron-positron wind from the star and by the emission of
electromagnetic dipole radiation. This is followed by a discussion of
accreting neutron stars in binary system. The spin period of such
neutron stars increases over time. The densities inside both neutron
stars that are spinning down as well as neutron stars that are being
spun up by mass accretion changes dramatically, which could lead to
signals of quark matter in observable data.

\goodbreak
\subsection{\it Isolated Pulsars}\label{sec:isolpuls}

It is known from Figs.\ \ref{fig:ec1445fig} and \ref{fig:ecvsomega}
that the weakening of centrifuge accompanied by the slowing down of a
\begin{figure}[tb]
\begin{center} 
\parbox[t]{7.5cm}
{\epsfig{figure=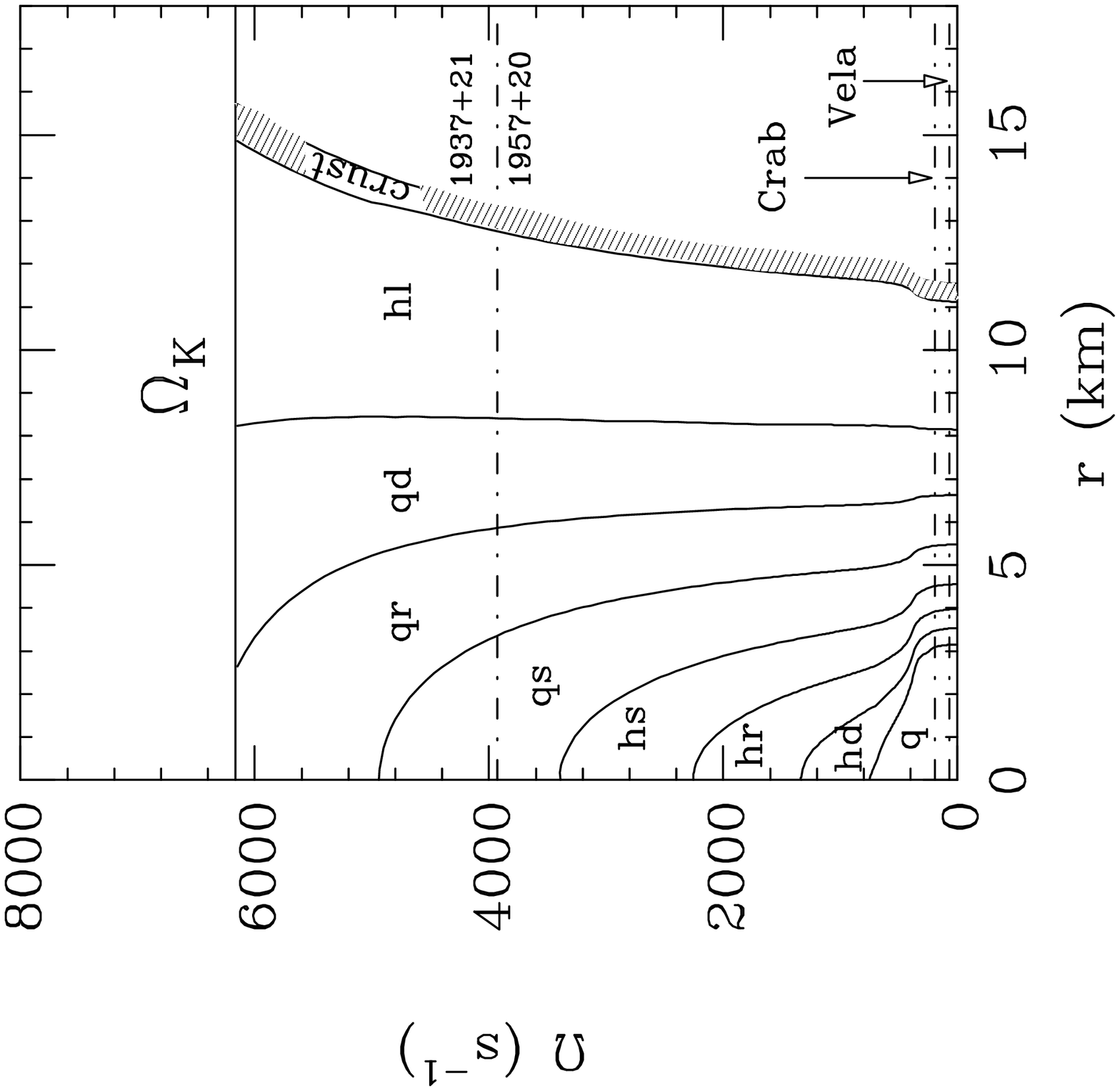,width=7.0cm,angle=-90}
{\caption{Frequency dependence of quark structures in equatorial star
direction for \eos $\KBt$ and a non-rotating star mass of
$1.416\,\msun$ \protect{ \cite{weber99:book}}.}
\label{fig:OkEq_1.416_G3B18}}}
\ \hskip 1.0cm   \
\parbox[t]{7.5cm}
{\epsfig{figure=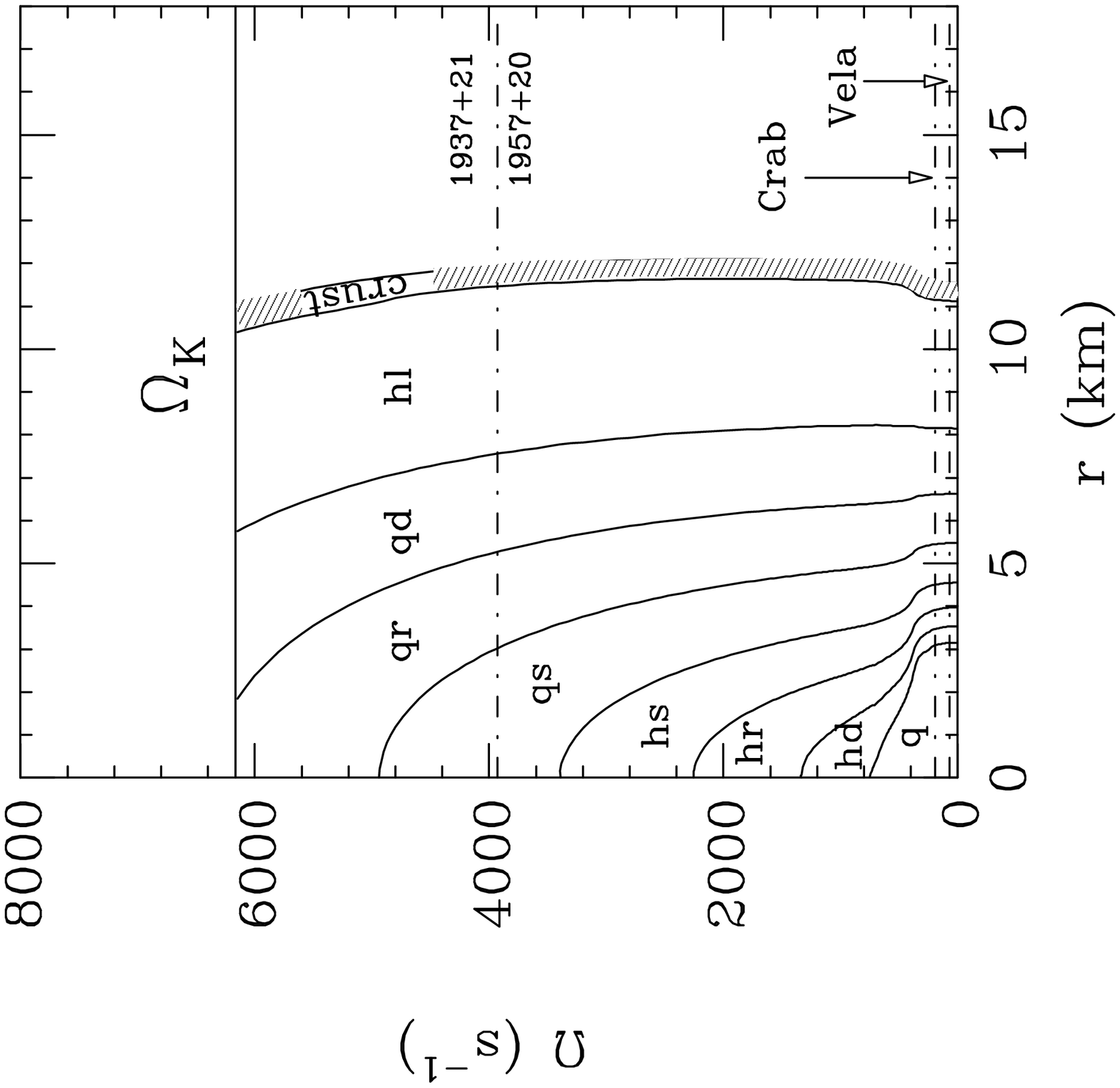,width=7.0cm,angle=-90} {\caption{Same
as Fig.\ \protect{\ref{fig:OkEq_1.416_G3B18}}, but in polar direction
\protect{ \cite{weber99:book}}.}
\label{fig:OkPo_1.416_G3B18}}}
\end{center}
\vskip -0.4cm
\end{figure}
rotating neutron star causes a significant increase of its central
density. From Fig.\ \ref{fig:ec1445fig}, for instance, one reads off
that the central density of a neutron star model of mass $M=1.42\,
\msun$, computed for a soft \eosp ($\KBt$ in the present
case \cite{glen97:a}), increases from about $450~\mevt$ for rotation at
the mass shedding frequency, $\okgr$, to more than $1500~\mevt$ for at
zero rotation, which is a $\sim 66$\% effect.  Such dramatic changes
in the interior density of a neutron star driven by changes in
frequency modify the stellar composition considerably.  If the mass
and initial rotational frequency of a pulsar is such that during its
slowing down phase the interior density rises from below to above the
critical density for the quark-hadron phase transition, first at the
center where the density is highest (Figs.\ \ref{fig:ec24B18},
\ref{fig:ec3B18}, \ref{fig:OkEq_1.416_G3B18} through
\ref{fig:OkPo_1.42_G3B18}) and then in a region expanding in the
radial outward direction away from the star's center, matter will be
gradually converted from the relatively incompressible nuclear matter
phase to the more compressible quark matter phase, as shown in
Fig.~\ref{fig:Mph142055}.  The tremendous weight of the overlaying
layers of nuclear matter tends to compress the quark matter core,
which causes the entire star to shrink on a length scale of several
hundred meters, as shown in Figs.\ \ref{fig:OkEq_1.42_G3B18} through
\ref{fig:OkPo_1.416_G3B18}. The mass concentration in the core will be
further enhanced by the increasing gravitational attraction of the
\begin{figure}[tb]
\begin{center} 
\parbox[t]{7.5cm}
{\epsfig{figure=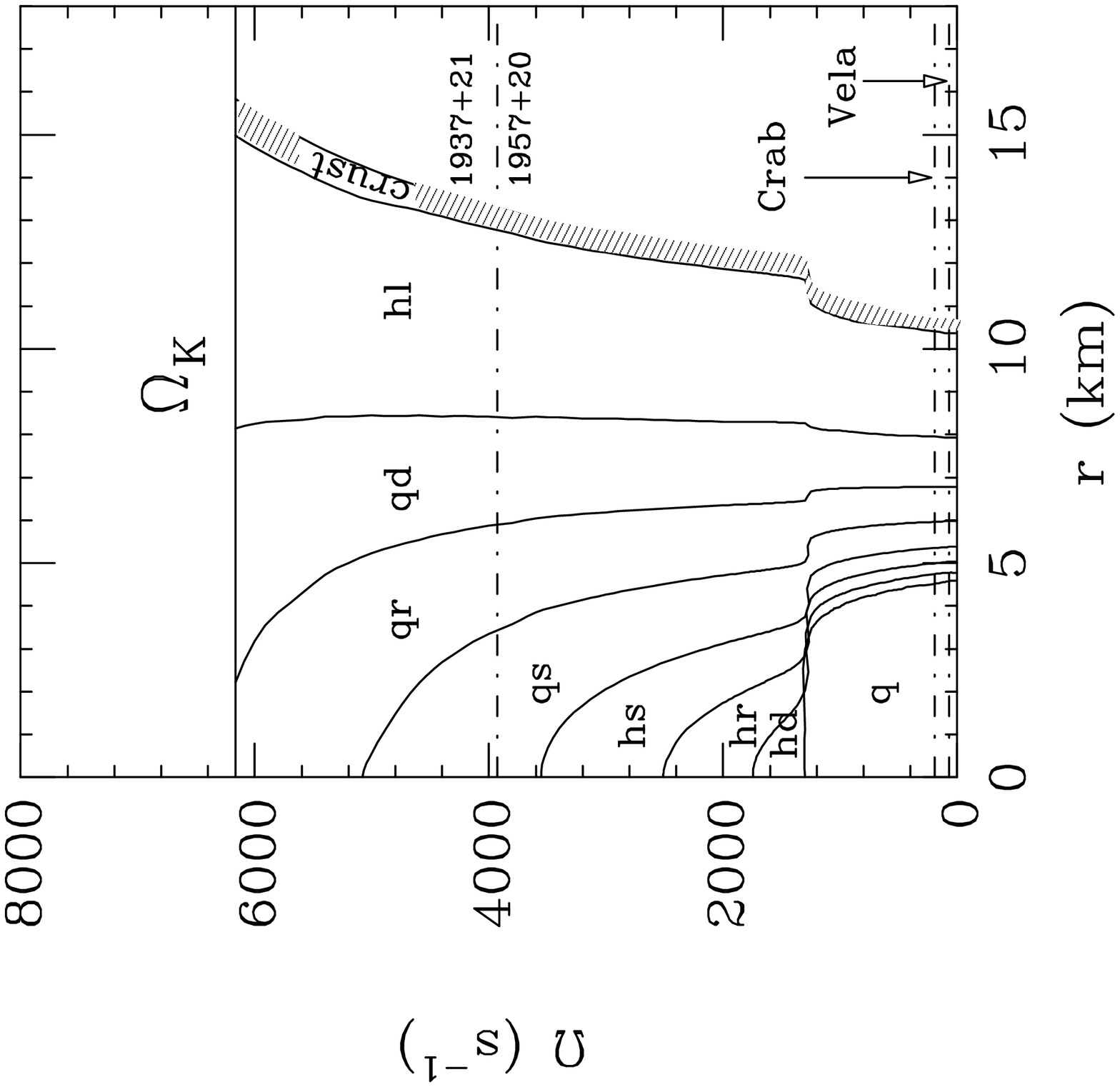,width=7.0cm,angle=-90}
{\caption[Frequency dependence of quark structure]{Frequency
dependence of quark structure in equatorial neutron star direction
computed for $\KBt$ and a non-rotating star mass of $1.42\,\msun$
\protect{ \cite{weber97:jaipur}}.}
\label{fig:OkEq_1.42_G3B18}}}
\ \hskip 1.0cm   \
\parbox[t]{7.5cm}
{\epsfig{figure=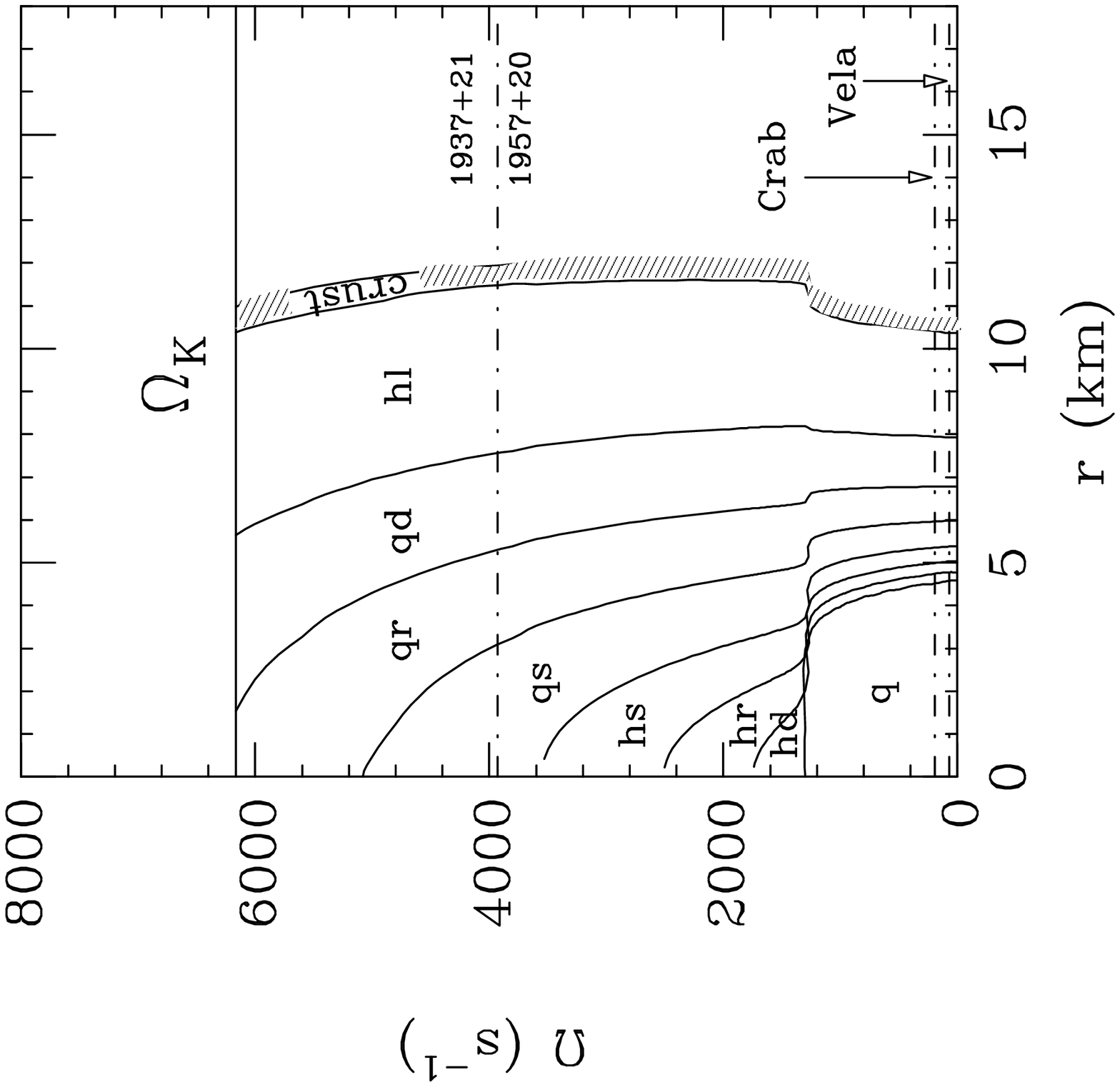,width=7.0cm,angle=-90}
{\caption[Frequency dependence of quark structure]{Same as Fig.\
\protect{\ref{fig:OkEq_1.42_G3B18}}, but in polar direction
\protect{ \cite{weber97:jaipur}}.}
\label{fig:OkPo_1.42_G3B18}}}
\end{center}
\vskip -0.4cm
\end{figure} 
quark core on the overlaying nuclear matter.  The moment of inertia
\begin{figure}[tb] 
\begin{center}
\epsfig{figure=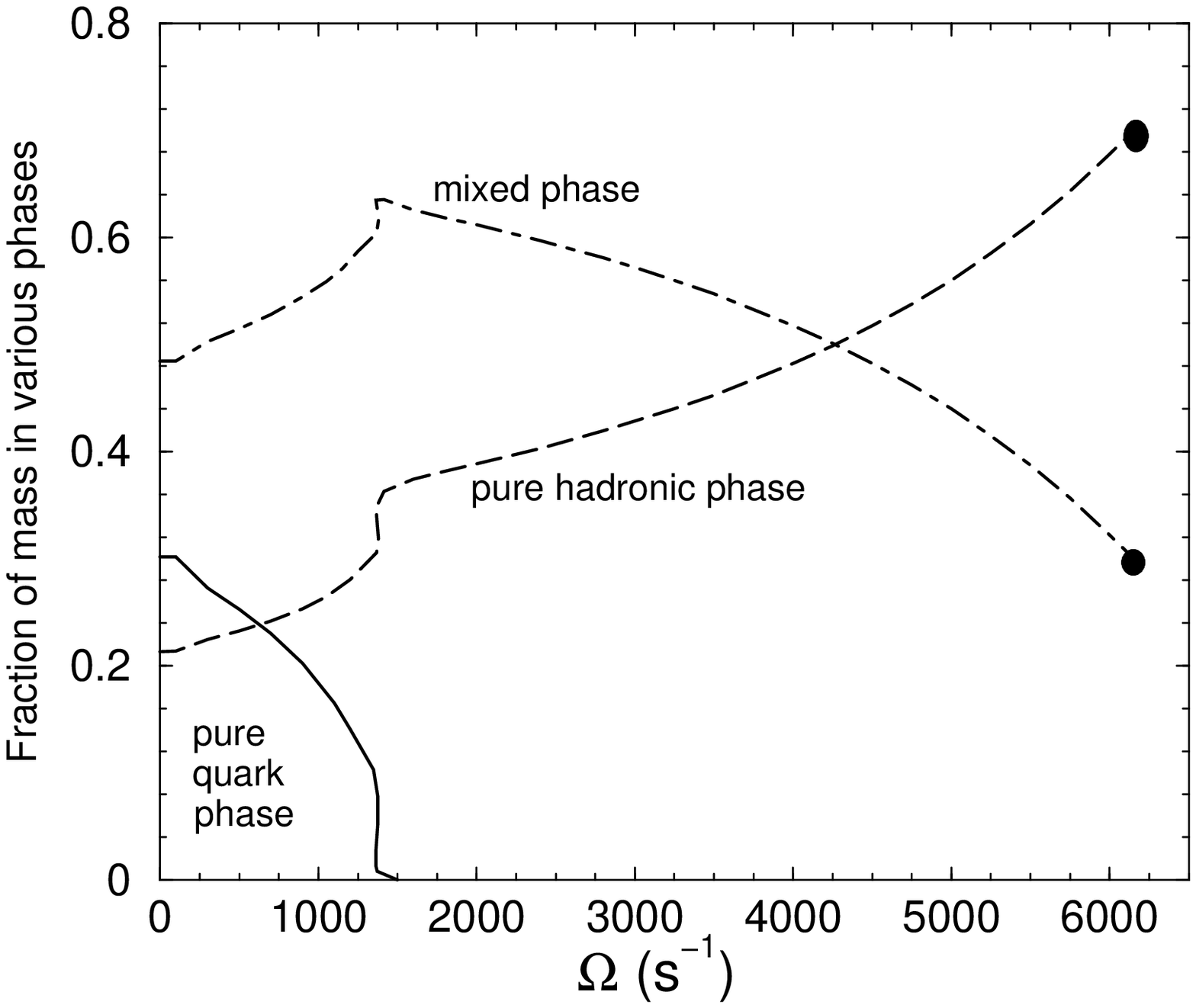,width=7.0cm,angle=0}
\begin{minipage}[t]{16.5 cm}
\caption{Fraction of mass existing in the form of pure quark matter,
pure hadronic matter, and in the mixed phase of quarks and hadrons for
the star of Fig.\ \ref{fig:OkEq_1.42_G3B18}. (From Ref.\
 \cite{weber99:book}.)}
\label{fig:Mph142055}
\end{minipage}
\end{center}
\vskip -0.4cm
\end{figure}
thus decreases anomalously with decreasing rotational frequency as the
new phase slowly engulfs a growing fraction of the
star \cite{glen97:a}, as can be seem from Figs.\ \ref{fig:moi1} and
\ref{fig:moi2}.
\begin{figure}[tb]
\begin{center} 
\parbox[t]{7.5cm} {\epsfig{figure=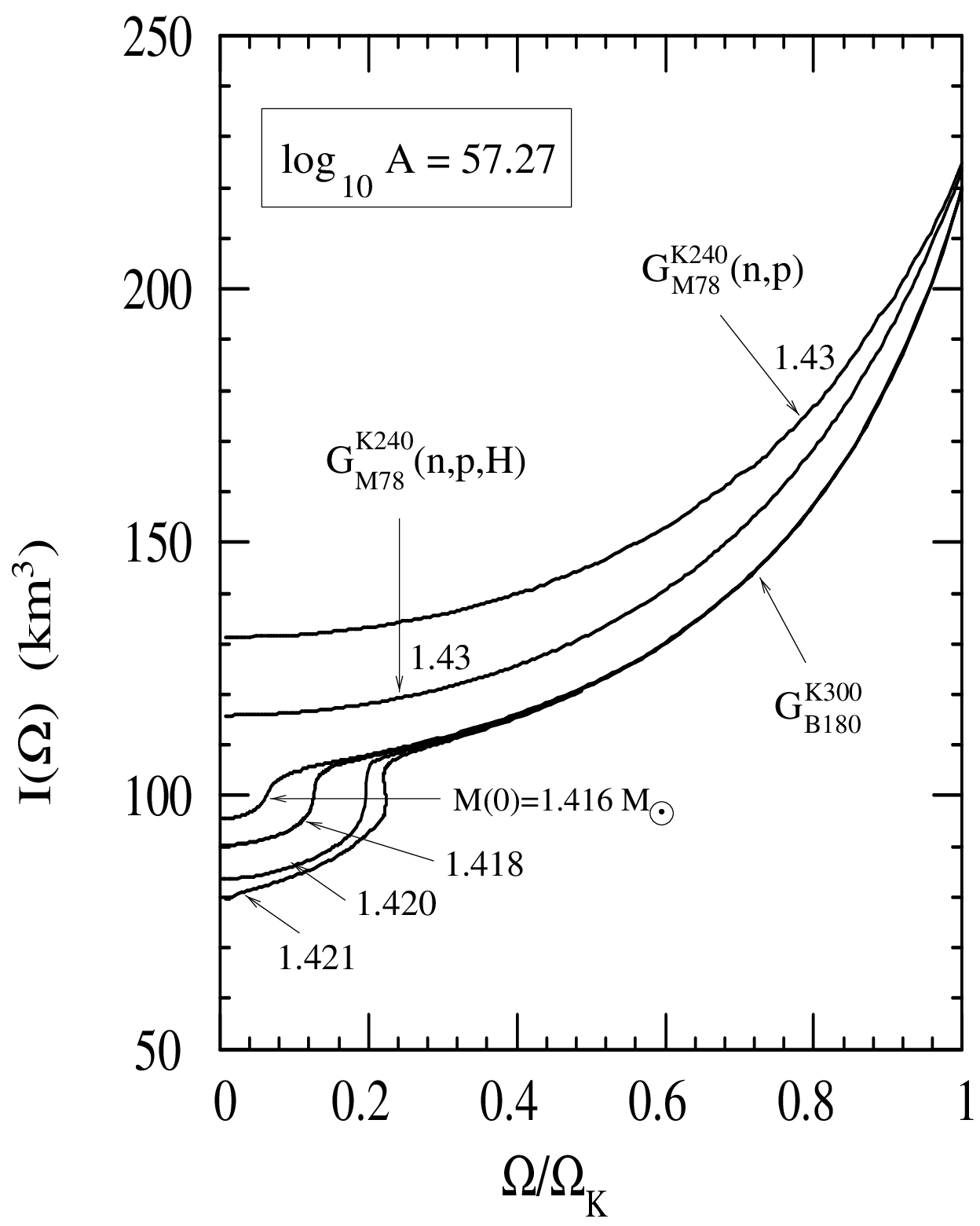,width=7.0cm,angle=-90}
{\caption{Moment of inertia versus rotational frequency of neutron
stars having the same baryon number, $A$, but different internal
constitutions. The dips at low $\Omega$'s are caused by quark
deconfinement  \cite{weber99:topr}.}
\label{fig:moi1}}}
\ \hskip 1.5 cm \ 
\parbox[t]{7.5cm} {\epsfig{figure=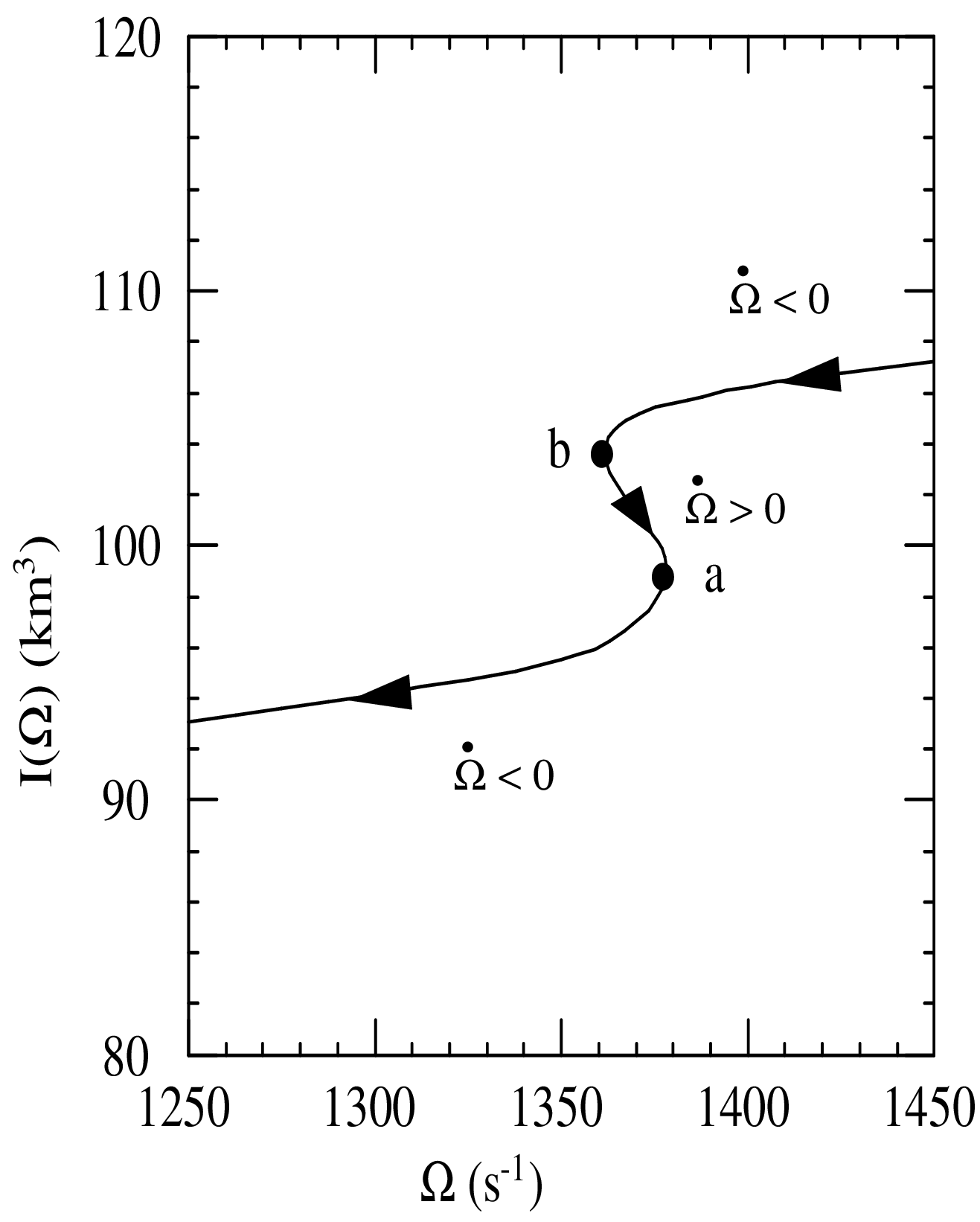,width=7.0cm,angle=-90}
{\caption{Enlargement of the lower-left portion of Fig.\
\protect{\ref{fig:moi1}} for the quark-hybrid star of mass $M=1.421\,
\msun$, which is characterized by a backbending of $I$ for frequencies
between `a' and `b'  \cite{weber99:book,weber99:topr,glen97:a}.}
\label{fig:moi2}}}
\end{center}
\end{figure} Figure~\ref{fig:moi1} shows the moment of inertia, $I$,
computed self-consistently from Eq.~(\ref{eq:11.71bk}) of several
sample stars having the same baryon number but different internal
constitutions \cite{weber97:jaipur}.  The curve labeled $M=1.420\,
\msun$, computed for $\KBt$, shows the moment of inertia of the
\begin{figure}[tb] 
\begin{center}
\parbox[t]{7.5cm} {\epsfig{figure=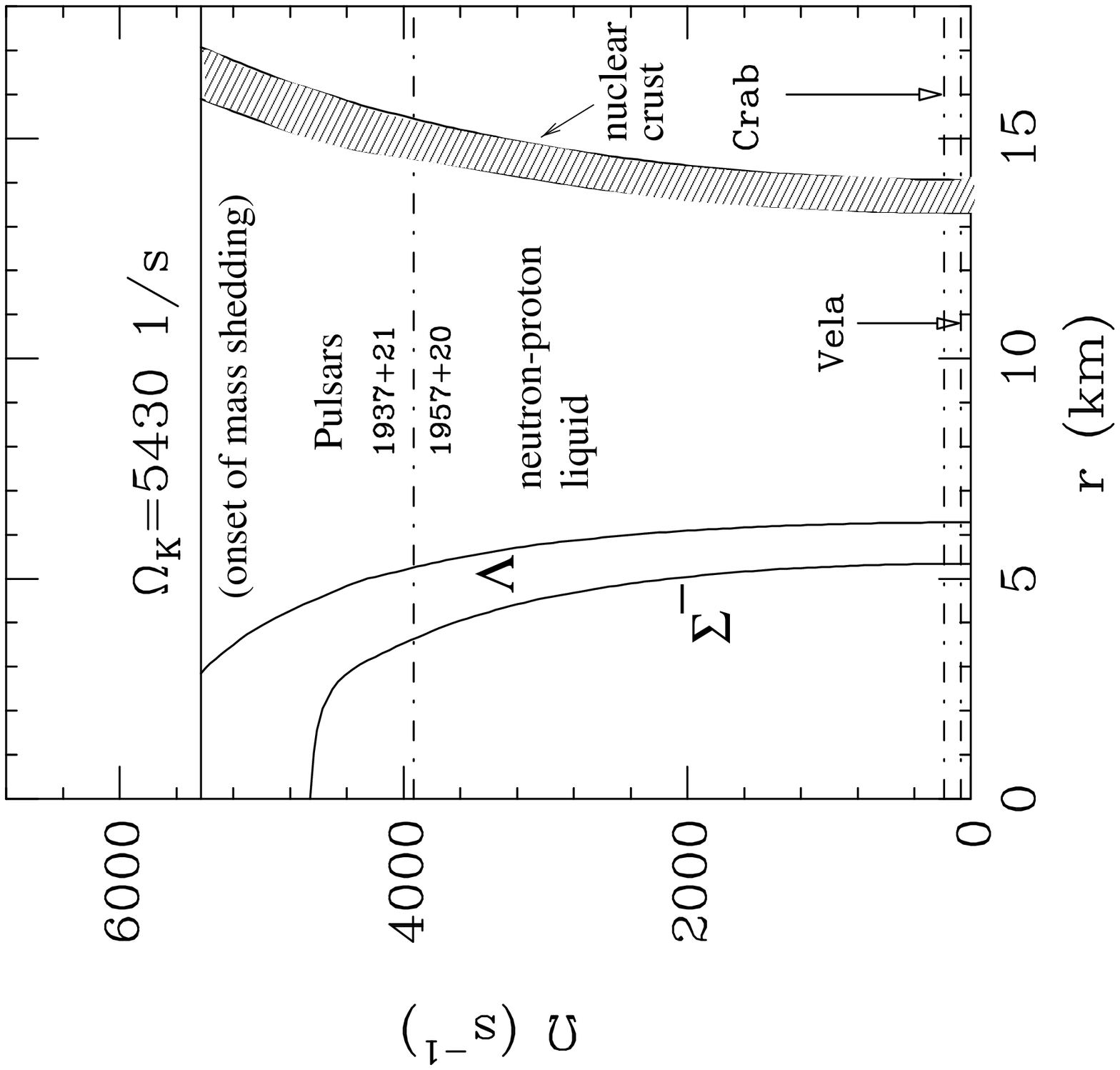,width=7.0cm,angle=-90}
{\caption{Frequency dependence of hyperon thresholds in equatorial
neutron star direction computed for HV. The star's non-rotating mass
is $1.40\,\msun$ \protect{ \cite{weber99:book}}.}
\label{fig:freqhp1}}}
\ \hskip 1.0cm   \
\parbox[t]{7.5cm} {\epsfig{figure=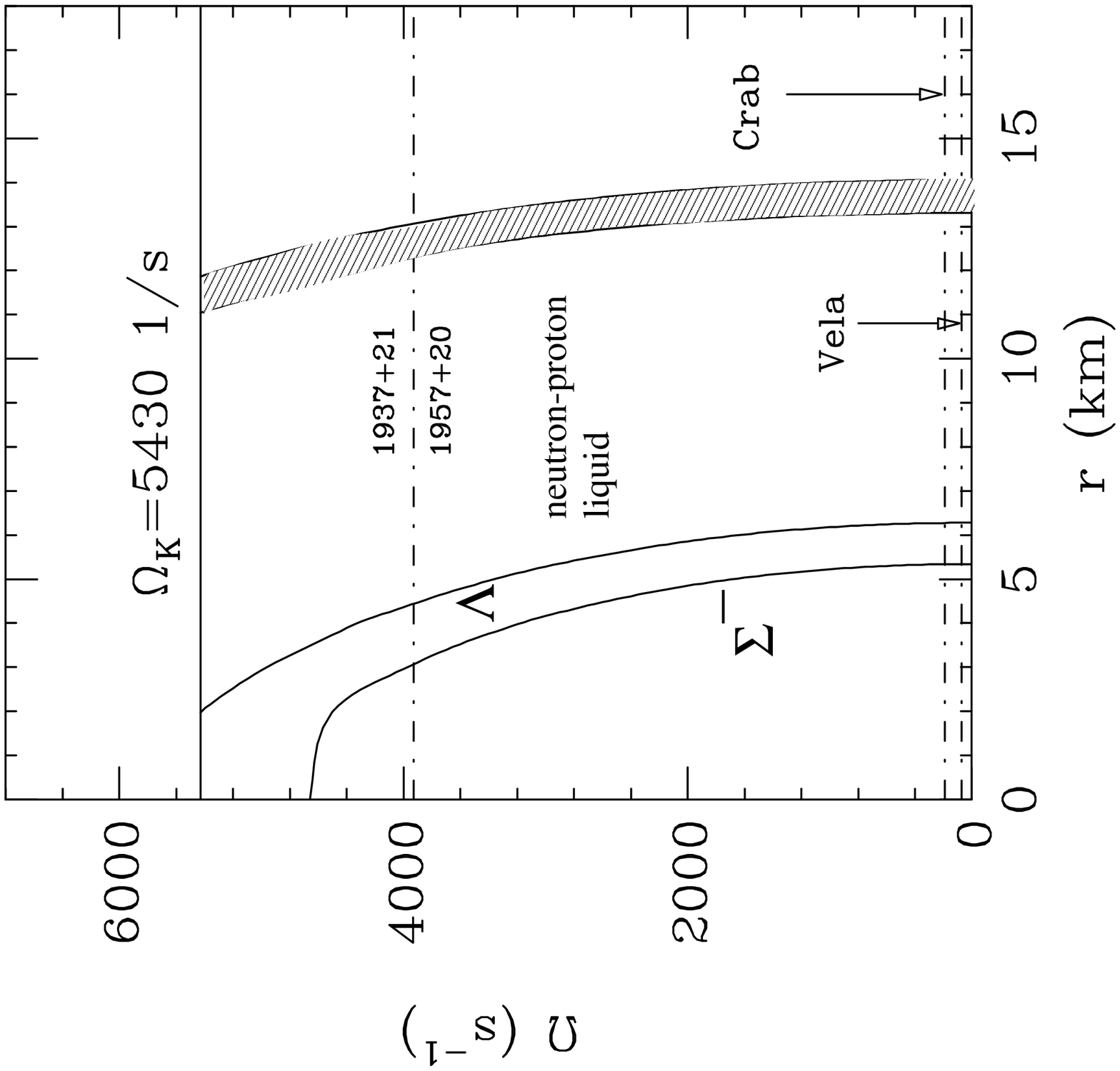,width=7.0cm,angle=-90}
{\caption{Same as Fig.\ \protect{\ref{fig:freqhp1}}, but in polar
direction~\protect{ \cite{weber99:book}}.}
\label{fig:freqhp2}}}
\end{center}
\vskip -0.4cm
\end{figure}
quark-hybrid star of Figs.\ \ref{fig:OkEq_1.42_G3B18} and
\ref{fig:OkPo_1.42_G3B18}.  The other curves correspond to a standard
hyperon star ($n,p,H$) constructed for $\KM$ and a standard neutron
star ($n,p$) where hyperons ($H$) have been ignored purposely. In
accordance with what has been said just above, the shrinkage of
quark-hybrid stars driven by the development of quark matter cores is
the less pronounced the smaller the quark matter cores which are being
built up in their centers.  Correspondingly, the dip in $I$ weakens
with decreasing star mass, as shown in Fig.~\ref{fig:moi1} for several
sample masses in the range $1.416 \leq M / \msun \leq 1.420$.  Model
calculations indicate that very strong reductions of $I$, such as
found for the $1.421 \,\msun$ model, for instance, may hardly be
obtainable for physical scenarios other than the hypothetical
quark-hadron phase transition \cite{weber99:book}.  Hyperon populations
alone as calculated in Ref.\  \cite{glen85:b} (Figs.\ \ref{fig:moi1},
\ref{fig:freqhp1} and \ref{fig:freqhp2}), for instance, appear to
modify the \eos way too little in order to cause significant changes
in $I$.  Nevertheless, there are models for the \eos of hyperonic
matter which also can strongly affect the spin evolution of isolated
neutron stars \cite{balberg99:a,zdunik04:a}.  As shown in these
references, depending on the nucleon-hyperon interaction and
hyperon-hyperon interaction in matter,
\begin{figure}[tb]
\begin{center}
\parbox[t]{7.5cm} 
{\epsfig{figure=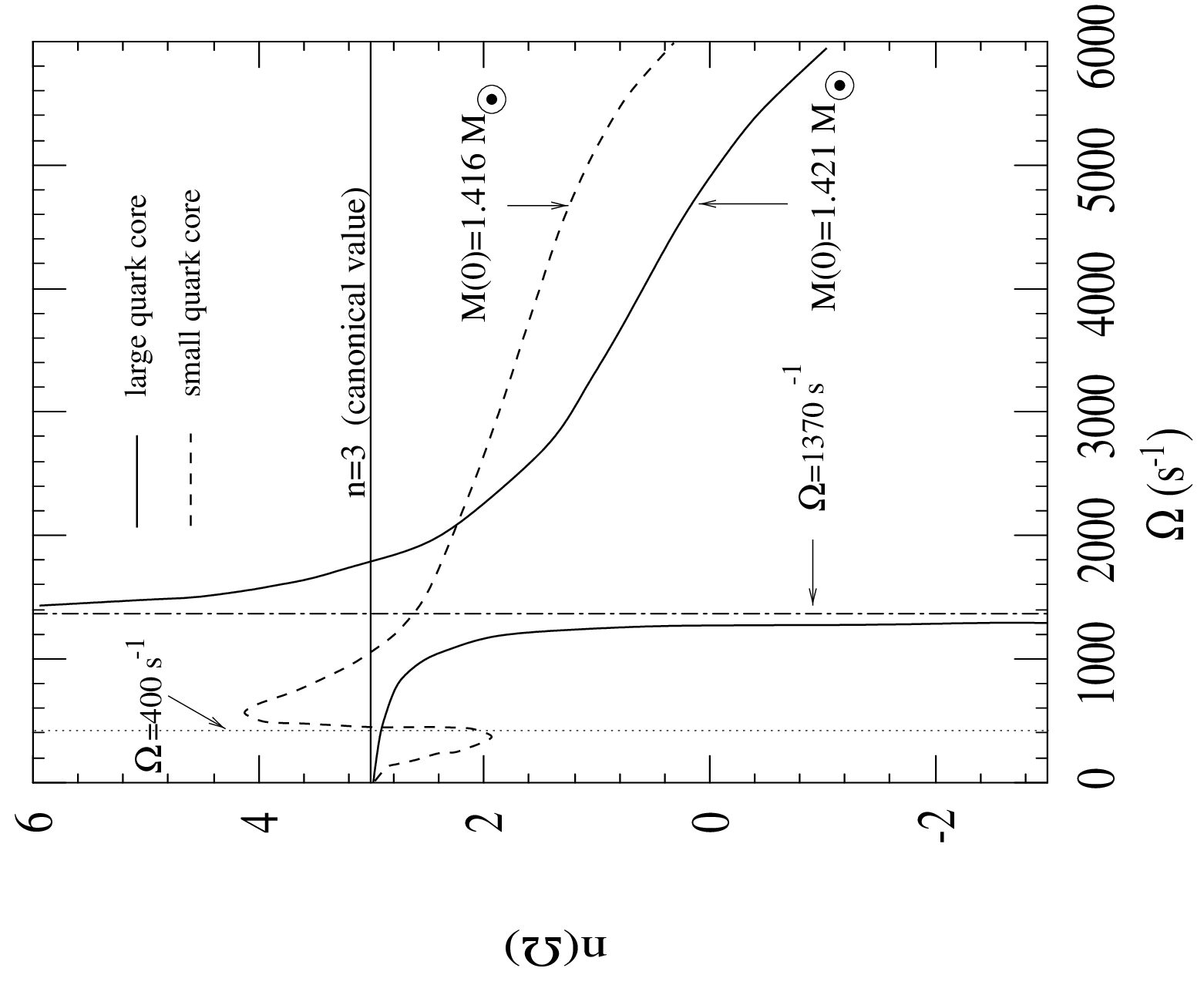,width=7.0cm,angle=-90}
{\caption{Braking index, $n$, of quark-hybrid stars of
Figs.\ \protect{\ref{fig:moi1}} and \protect{\ref{fig:moi2}}.  The
anomalies in $n$ at $\Omega \sim 400~\secm$ and $\Omega \sim
1370~\secm$ are caused by quark deconfinement. The overall reduction
of $n$ below 3 is due to rotation. (From Ref.\  \cite{weber99:book}.)}
\label{fig:n}}}
\ \hskip 1.0cm   \
\parbox[t]{7.5cm}
{\epsfig{figure=nvst_1_new.ps.bb,width=7.0cm,angle=-90}
{\caption{Braking index versus time for the quark-hybrid star of mass
$M=1.421\, \msun$ of Fig.~\protect{\ref{fig:n}}. The epoch over which
$n$ is anomalous because of quark deconfinement, $\sim 10^8$~years, is
indicated by the shaded area. (From Ref.\  \cite{weber99:book}.)}
\label{fig:nvst_1}}}
\end{center}
\end{figure}
hyperons can even cause backbending. The backbending episode can
terminate either unstably or through a stable continuous transition to
a standard spin-down behavior. The observation of backbending in the
timing behavior of isolated pulsars, or of spin clustering on
accreting neutron stars discussed in section \ref{sec:spin}, is
therefore not unambiguous evidence for quark deconfinement.  It is
also explored to which extent backbending may be caused by other
competing particle processes (cf.\ Figs.\ \ref{fig:cross} and
\ref{fig:ec1445fig}).

The decrease of the moment of inertia caused by the quark-hadron phase
transition, shown in Fig.~\ref{fig:moi1}, is superimposed on the
response of the stellar shape to a decreasing centrifugal force as the
star spins down due to the loss of rotational energy.  In order to
conserve angular momentum not carried off by particle radiation from
the star, the deceleration rate $\dot\Omega$~($< 0$) must respond
correspondingly by decreasing in absolute magnitude. More than that,
$\dot\Omega$ may even change sign, as shown in Fig.~\ref{fig:moi2}
 \cite{glen97:a}, which carries the important astrophysical information
that an isolated pulsar may spin up during a certain period of its
stellar evolution.  The situation may be compared with an ice skater
who spins up upon contraction of the arms before air resistance and
friction of the skate on the ice reestablishes spin-down again.  Such
an anomalous decrease of $I$ is analogous to the `backbending'
phenomenon known from nuclear physics, in which case the moment of
inertia of an atomic nucleus changes anomalously because of a change
in phase from a nucleon spin-aligned state at high angular momentum to
a pair-correlated superfluid phase at low angular momentum.  In the
nuclear physics case, the backbending in the rotational bands of
nuclei that was predicted by Mottelson and Valatin \cite{mottelson82:a}
and then observed years later by Stephens and
Simon \cite{stephens72:a}, and Johnson, Ride and
Hjorth \cite{johnson72:a}.  For neutron stars, the stellar backbending
of $I$ is shown in Fig.~\ref{fig:moi2}. Stars evolving from `b' to `a'
are rotationally accelerated ($\dot\Omega > 0$), while stars evolving
from `a' to `b', which could be part of the evolutionary track of
pulsars accreting matter from companions (see section \ref{sec:spin}),
are rotationally decelerated ($\dot\Omega < 0$).  As we shall see
next, the structure in the moment of inertia and, specifically, the
backbending phenomenon dramatically modifies the timing structure of
pulsar spin-down, rendering the observation of quark matter in neutron
stars accessible to radio astronomy.  Pulsars are identified by their
periodic signal believed to be due to a strong magnetic field fixed in
the star and oriented at an angle from the rotation axis. The period
of the signal is therefore that of the rotation of the star.  The
angular velocity of rotation decreases slowly but measurably over
time, and usually the first and occasionally second time derivative
can also be measured. Various energy loss mechanisms could be at play
such as the magnetic dipole radiation, part of which is detected on
each revolution, as well as other losses such as ejection of charged
particles \cite{ruderman87:a}. If one assume that pulsar slow-down is
governed by a single mechanism, or several mechanisms having the same
power law, the energy balance equation can then be written in the form
\begin{equation}
  \frac{d E}{d t} = \frac{d}{d t} \, \left( \frac{1}{2}\,
  I(\Omega) \, \Omega^2 \right) = - \, C \ \Omega^{n+1} \, .
  \label{eq:engloss}
\end{equation} In the case of magnetic dipole radiation, the constant $C$
is equal to $C= \frac{2}{3} \mu^2 \sin^2 \alpha$ where $\mu$ denotes the
star's magnetic dipole moment. The quantity $n$ in Eq.\
(\ref{eq:engloss}) is called braking index. It is $n=3$ if $I$ is kept
constant during spin-up (down).  If, as is customary, the star's
angular velocity $\Omega$ is regarded as the only time-dependent
quantity, one obtains the usual formula for the rate of change of
pulsar frequency, given by
\begin{equation}
\dot{\Omega} = - \ K \ \Omega^n \, ,
\label{eq:braking}
\end{equation} 
with $K=C/I$ a constant. With the braking formula (\ref{eq:braking})
one can define the spin-down age of a pulsar given by
\begin{equation}
  \tau = - \, (n-1)^{-1} \, \Omega / {\dot{\Omega}} \, ,
\label{eq:age_tau}
\end{equation}  
with $n=3$ for energy loss governed by magnetic dipole radiation.
However the moment of inertia is not constant in time but responds to
changes in rotational frequency, as shown in Figs.\ \ref{fig:moi1} and
\ref{fig:moi2}, more or less in accord with the softness or stiffness
of the equation of state and according to whether the stellar mass
is small or large.  This response changes the value of the braking
\begin{figure}[tb]
\begin{center} 
\epsfig{figure=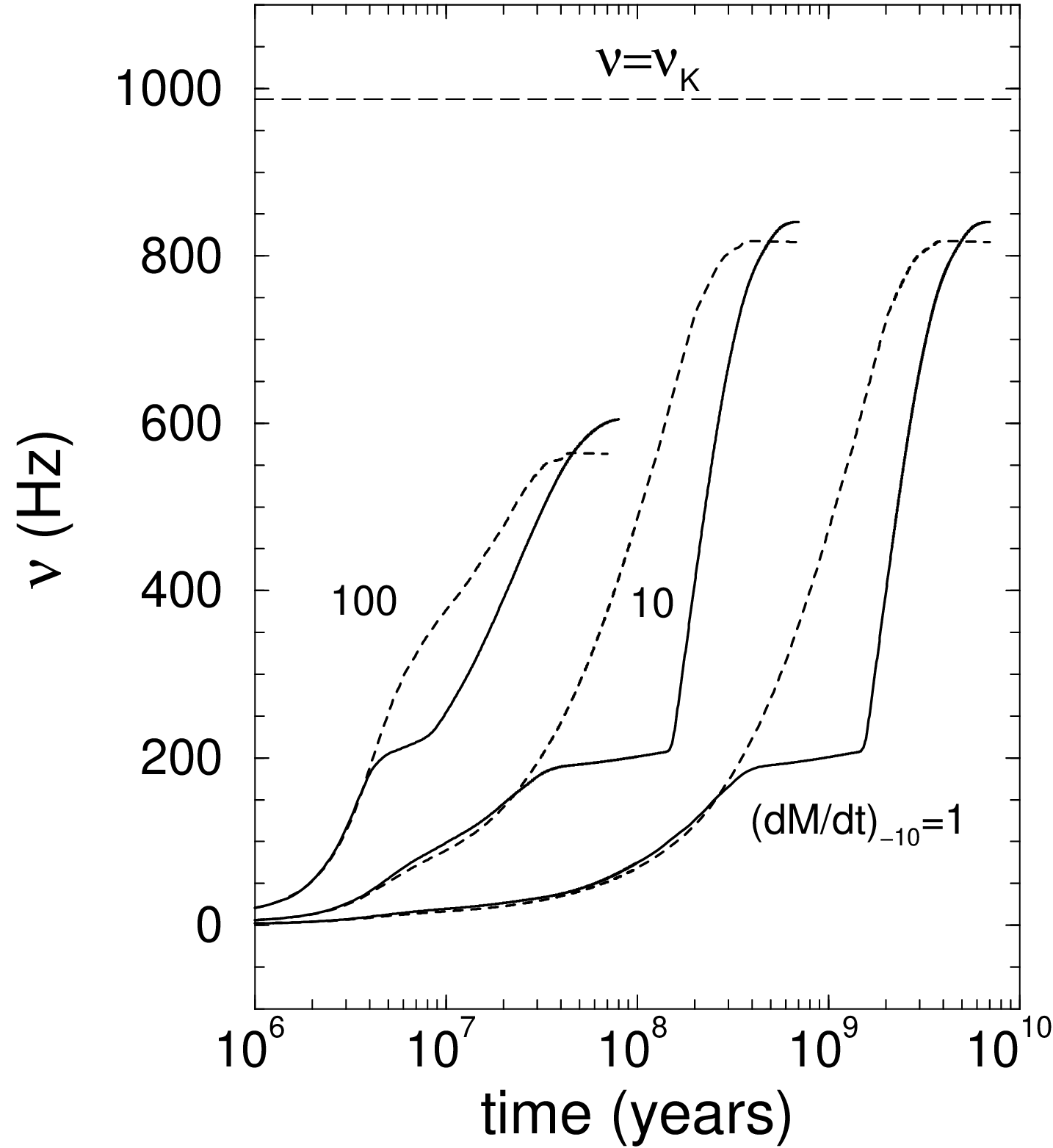,width=7.0cm}
\begin{minipage}[t]{16.5 cm}
\caption{Evolution of spin frequencies of accreting X-ray neutron
stars with (solid curves) and without (dashed curves) quark
deconfinement. The spin plateau around 200~Hz signals the ongoing
process of quark reconfinement in the stellar centers. (From Ref.\
 \cite{glen01:a}.)}
\label{fig:nue_t}
\end{minipage}
\end{center}
\vskip -0.4cm
\end{figure}
index in a frequency dependent manner, that is $n = n(\Omega)$, even
if the sole energy-loss mechanism were pure magnetic dipole, as
expressed in Eq.\ (\ref{eq:engloss}). Thus during any epoch of
observation, the braking index will be measured to be different from
it canonical value $n=3$ by a certain amount. How much less depends,
for any given pulsar, on its rotational frequency and for different
pulsars of the same frequency, on their mass and on their internal
constitution \cite{weber99:book,weber99:topr,chubarian00:a,blaschke99:trento,heiselberg98:a}.
When the frequency response of the moment of inertia is taken into
account, Eq.\ (\ref{eq:braking}) is replaced
with \cite{weber99:book,glen97:a}
\begin{equation}
  \dot{\Omega}= - 2 C \Omega^n \left( 2 I + \Omega (dI / d\Omega)
  \right)^{-1} \, .
\label{eq:braking2}
\end{equation} 
This explicitly shows that the frequency dependence of $\dot{\Omega}$
corresponding to any mechanism that absorbs (or deposits) rotational
energy cannot be a simple power law as given in Eq.\
(\ref{eq:braking}) (with $K$ a constant). It must depend on the mass
and internal constitution of the star through the response of the
moment of inertia to rotation as expressed in
(\ref{eq:braking2}). Equation (\ref{eq:braking2}) can be represented
in the form of (\ref{eq:braking}), but now with a frequency dependent
prefactor, by evaluating
\begin{equation}  
  n(\Omega) \equiv \frac{\Omega\, \ddot{\Omega} }{\dot{\Omega}^2} = 3
  - \left( 3 \, { {d I}\over{d \Omega} } \, \Omega + { {d^2 I}\over{d
      \Omega^2} } \, \Omega^2 \right) \left( 2\, I + { {d I}\over{d
      \Omega} } \, \Omega \right)^{-1} \, .
\label{eq:index}
\end{equation} One sees that this braking index depends
explicitly and implicitly on $\Omega$.  This relation reduces to the
canonical expression $n=3$ only if $I$ is independent of frequency,
which may not the case, as seen above, if there are compositional
changes driven by a varying star frequency.  As an example, we show in
Fig.\ \ref{fig:n} the variation of the braking index with frequency
for two selected quark-hybrid stars of Figs.\ \ref{fig:moi1} and
\ref{fig:moi2}. For illustrational purposes we assume dipole
radiation. Because of the response of the moment of inertia to quark
deconfinement, the braking index deviates dramatically from the
canonical value $n=3$ at rotational frequencies where quark
deconfinement leads to the built-up of pure quark matter cores in the
centers of these stars.  Such anomalies in $n(\Omega)$ are not
obtained for conventional neutron stars or hyperon stars because their
moments of inertia increase smoothly with $\Omega$, as known from
Fig.~\ref{fig:moi1}.  The observation of such an anomaly in the timing
structure of pulsar spin-down could thus be interpreted as a signal of
quark deconfinement in the centers of pulsars. Of course, because of
the extremely small temporal change of a pulsar's rotational period,
one cannot measure the shape of the curve which is in fact not
necessary. Just a single anomalous value of $n$ that differed
significantly from the canonical value of $n=3$ would
suffice \cite{glen97:a,glen97:b}.

Carried over to the observed pulsar data of $\Omega$ and $\dot\Omega$,
it appears that the change in centrifugal force over the life of a
canonical, slowly-rotating pulsar could eventually be too meager to
span a significant change.  The significant braking anomaly,
therefore, could be restricted to millisecond pulsars. For them, the
phase change may occur only in such millisecond pulsars that rotate
near the maximum-mass peak determined by the underlying
\eosp. Otherwise the fraction of pure quark matter in their centers
may not be sufficient to cause the required shrinkage. The phase
change itself may be first (as in our example) or second order.  Both
orders will cause a signal as long as quark deconfinement causes a
sufficient softening of the \eos and quark matter is generated at the
center of the star at a sufficiently high rate. On the observational
side, a serious drawback may be that the braking indices of
millisecond pulsars are very hard to measure, because of timing noise
which renders the determination of $\ddot\Omega$ very complicated.  As
a final but very important point on the subject of quark
deconfinement, we estimate the typical duration over which the braking
index is anomalous if quark deconfinement is well pronounced, as for
the quark-hybrid star of mass $M(0) = 1.421\, \msun$. The time span
can be estimated from $ \Delta T \simeq - \, {{\Delta\Omega}} /
{\dot\Omega} = {\Delta P} / {\dot{P}} \, , $ where $\Delta \Omega$ is
the frequency interval of the anomaly.  The range over which
$n(\Omega)$ is smaller than zero and larger than six
(Fig.~\ref{fig:n}) is $\Delta \Omega \approx - \, 100~\secm$, or
$\Delta P \approx - 2 \pi \Delta \Omega / \Omega^2 \approx 3 \times
10^{-4}$~s at $\Omega=1370~\secm$.  Hence, for a
\begin{figure}[tb] 
\begin{center}
\epsfig{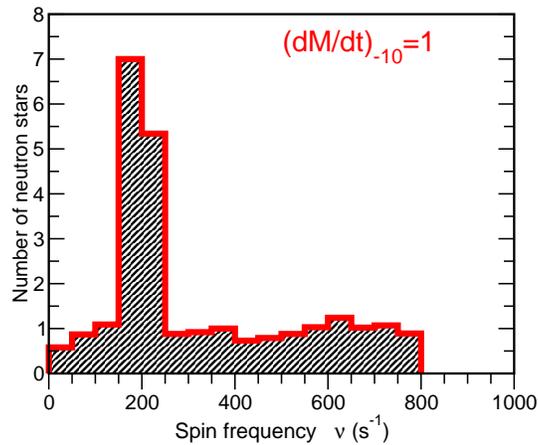}
\begin{minipage}[t]{16.5 cm}
\caption{Spin distribution of X-ray neutron stars. The spike
corresponds to the spinout of quark matter. Otherwise the spike would
be absent.}
\label{fig:histo}
\end{minipage}
\end{center}
\vskip -0.4cm
\end{figure}
millisecond pulsar whose period derivative is typically $\dot{P}\simeq
10^{-19}$ one has $\Delta T \simeq 10^8$~years, as graphically
illustrated in Fig.~\ref{fig:nvst_1}. The dipole age of such pulsars
is about $10^9$~years. So, as a rough estimate we expect about 10\% of
the $\sim 30$ presently known solitary millisecond pulsars to be in the
transition epoch during which pure quark matter cores are gradually
being built up in their centers. These pulsars could be signaling the
ongoing process of quark deconfinement in their cores.  Last but not
least we note that the spin-up time (region $b$--$a$ in
Fig.~\ref{fig:moi2}) is about 1/5 of the time span $\Delta T$, or
about 1/50 of the dipole age.  To avoid confusion, we point out that
the spin-up has nothing to do with the minuscule spin-ups known as
pulsar glitches. In the latter case the relative change of the moment
of inertia is very small, $\Delta I/I \simeq - \, \Delta\Omega/\Omega
\simeq 10^{-6}$ or smaller, and approximates closely a continuous
response of the star to changing frequency on any time scale that is
large compared to the glitch and recovery interval. Excursion of such
a magnitude as quoted would fall within the thickness of the line in
Fig.~\ref{fig:moi2}.

\goodbreak
\subsection{\it Accreting Neutron Stars}\label{sec:spin}

The signal of quark deconfinement described in section
\ref{sec:isolpuls} is computed for isolated neutron stars, where
deconfinement is driven by the gradual stellar contraction as the star
spins down.  The situation is reversed in neutron stars in binary
systems, which experience a spin-up torque because of the transfer of
angular momentum carried by the matter picked up by the star's
magnetic field from the surrounding accretion
disk \cite{glen01:a,chubarian00:a,blaschke99:trento,glen00:b,glen00:d}.
The spin-up torque causes a change in the stars' angular momentum that
can be expressed as \cite{glen01:a}
\begin{eqnarray}
{{dJ}/ {d t}} = {\dot M} {\tilde l}(r_{\rm m}) - N(r_{\rm c}) \,
,
\label{eq:dJdt}
\end{eqnarray}
where $\dot{M}$ denotes the accretion rate and
\begin{eqnarray}
{\tilde l}(r_{\rm m}) = \sqrt{M r_{\rm m}} 
\label{eq:l}
\end{eqnarray}
is the angular momentum added to the star per unit mass of accreted
matter. The quantity $N$ stands for the magnetic plus viscous torque
term,
\begin{equation}
N(r_{\rm c}) = \kappa \, \mu^2 \, r_{\rm c}^{-3} \, ,
\label{eq:N}
\end{equation}
with $\mu \equiv R^3 B$ the star's magnetic moment.  The quantities
$r_{\rm m}$ and $r_{\rm c}$ denote the radius of the inner edge of the
accretion disk and the co-rotating radius, respectively, and are given by
\begin{equation}
  r_{\rm m} = \xi \, r_{\rm A} \, , \qquad r_{\rm c} = \left( M
  \Omega^{-2} \right)^{1/3} \, ,
\label{eq:r_m+r_c}
\end{equation}
with $(\xi \sim 1)$. The Alfv\'en radius $r_{\rm A}$  is defined by
\begin{equation}
r_{\rm A} = \left( { {\mu^4} \bigl( {2 M \dot{M}^2} \bigr)^{-1} }
\right)^{1/7} \, .
\label{eq:r_A}
\end{equation}
Accretion will be inhibited by a centrifugal barrier if the neutron
star's magnetosphere rotates faster than the Kepler frequency at the
magnetosphere. Hence $r_{\rm m} < r_{\rm c}$, otherwise accretion onto
the star will cease.  The rate of change of a star's angular frequency
$\Omega$ then follows from Eq.\ (\ref{eq:dJdt}) as
\begin{equation}
  I(t) {{d\Omega(t)} \over {d t}} = {\dot M} {\tilde l}(t) - \Omega(t)
  {{dI(t)}\over{dt}} - \kappa \mu(t)^2 r_{\rm c}(t)^{-3} \, ,
\label{eq:dOdt.1}
\end{equation} with the explicit time dependences as indicated.
There are two terms on the right-hand-side of Eq.\ (\ref{eq:dOdt.1})
that grow linearly respectively quadratically with $\Omega$.  Ignoring
the linear term shows that mass transfer can spin up a neutron star to
an equilibrium period of \cite{heuvel91:a}
\begin{equation}
P_{\rm eq} = 2.4~{\rm ms} \left({{\dot M}\over{\dot M_{\rm
Edd}}}\right)^{-3/7} \left({{M}\over{M_{\odot}}}\right)^{-5/7} ~
R_6^{15/7} ~ B_9^{6/7} \, ,
\label{eq:equil.1}
\end{equation}
where $R_6$ and $B_9$ are the star's radius and its magnetic field in
units of $10^6$~cm and $10^9$~G, respectively.  $\dot{M}_{\rm Edd}$ in
Eq.\ (\ref{eq:equil.1}) denotes the maximum possible accretion rate,
defined by the Eddington limit (see Eq.\ (\ref{eq:Eddington})), at which
the accretion luminosity equals the luminosity at which the radiation
pressure force on ionized
hydrogen plasma near the star balances the gravitational acceleration
force exerted by the star on the plasma.  This condition leads to an
Eddington accretion rate of $\dot{M}_{\rm Edd} = 1.5\times 10^{-8} R_6
M_\odot~{\rm yr}^{-1}$.  For a typical accretion rate of ${\dot
M}_{-10} \equiv {\dot M}/(10^{-10} M_{\odot}~ {\rm yr}^{-1})$, the
Eddington rate can be expressed as ${\dot M}_{\rm Edd} = 150 \, R_6 \,
{\dot M}_{-10}^{-1} \, {\dot M}$.  The low-mass X-ray binaries (LMXBs)
observed with the RXTE are divided into Z sources and A(toll) sources,
which accrete at rates of $\dot{M}_{-10} \sim 200$ and $\dot{M}_{-10}
\sim 2$, respectively \cite{klis00:a}.

The solution of Eq.\ (\ref{eq:dOdt.1}) in combination with the
expression for the moment of inertia derived in Eq.\
(\ref{eq:11.71bk}) for the quark-hybrid model $\KBt$ ($M(0)=1.42\,
\msun$) is shown in Fig.~\ref{fig:nue_t}. The magnetic field is
assumed to evolve according to
\begin{eqnarray}
B(t) = B(\infty) + \bigl( B(t=0) - B(\infty) \bigr) \, e^{-t/t_{\rm d}}
\, ,
\label{eq:bevolution}
\end{eqnarray} 
with $t=0$ at the start of accretion, $B(t=0)=10^{12}$~G,
$B(\infty)=10^8$~G, and $t_{\rm d}=10^6$~yr.  Such a decay to an
asymptotic value seems to be a feature of some treatments of the
magnetic field evolution of accreting neutron stars
\cite{konar99:a}. Moreover, it expresses the fact that canonical
neutron stars have high magnetic fields, $\sim 10^{12}$~G, and
millisecond pulsars have low fields of $\sim 10^8$~G.  The result for
the spin-up of the quark-hybrid stars is most striking. One sees that
quark matter remains relatively dormant in the stellar core until the
star has been spun up to frequencies at which the central density is
about to drop below the threshold density at which quark matter is
predicted to exist for this model. As known from Fig.\ \ref{fig:moi2},
this manifests itself in a significant increase of the star's moment
of inertia. The angular momentum added to a neutron star during this
phase of evolution is therefore consumed by the star's expansion,
inhibiting a further spin-up until the entire quark matter core has
been spun out of the center, leaving the star with a mixed phase of
quarks and hadrons made up of hadrons and quarks surrounded by
ordinary nuclear matter (see Figs.\ \ref{fig:OkEq_1.42_G3B18} and
\ref{fig:OkPo_1.42_G3B18}).  Such accreters, therefore, tend to spend
a much greater length of time in the critical frequencies than
otherwise. There will be an anomalous number of accreters that appear
near the same frequency, as shown in Fig.~\ref{fig:histo}.  Evidence
that accreting neutron stars pile up at certain frequencies, which are
well below the mass shedding limit, is provided by the spin
distribution of accreting millisecond pulsars in 57 Tuc and neutron
stars in low mass X-ray binaries observed with the Rossi X-ray Timing
Explorer. The proposed limiting mechanisms responsible for this
behavior could be gravity-wave emission caused by the r-mode
instability, or a small stellar mass quadrupole moment
\cite{bildsten98:a,andersson00:a,chakrabarty03:a}. As shown here,
quark reconfinement may be linked to this phenomenon as well
\cite{glen01:a,chubarian00:a,glen00:b,poghosyan01:a}

\goodbreak
\section{Summary}\label{sec:summary}

The tremendous pressures in the cores of neutron stars might be able
to break neutrons, protons plus other hadronic constituents in the
centers of neutron stars into their quark constituents, creating a new
state of matter known as quark matter which is being sought
at the most powerful colliders. If quark matter exists in the cores of
neutron stars, it will be a color superconductor whose complex
condensation pattern is likely to change with density inside the
star. The exploration of the numerous astrophysical facets of (color
superconducting) quark matter is therefore of uppermost importance and
is pursued by physicists from different, complementing fields of
physics. Their joint scientific efforts, surveyed in this review,
provides most valuable information about the phase diagram of nuclear
matter at high baryon number density but low temperature, which is not
accessible to relativistic heavy ion collision experiments, and may
ultimately provide us with a glimpse on the kind of matter that
filled our universe just milliseconds after is was born.



\section*{Acknowledgments}

I thank the Institute for Nuclear Theory at the University of
Washington for its hospitality and the Department of Energy for
partial support during the completion of this work. Mark Alford
deserves my special thanks for discussions and communications.



\end{document}